\documentclass[aps,10pt,prd,citeautoscript,longbibliography,superscriptaddress,twocolumn]{revtex4-1}
\usepackage[T1]{fontenc}
\usepackage{graphicx}
\usepackage{times}
\usepackage{amsmath,amssymb,amsfonts,bm,dsfont,units}
\usepackage{hyperref}
\usepackage{braket}
\usepackage{multirow}
\usepackage[normalem]{ulem} 
\usepackage[caption=false]{subfig} 

\newcommand{\norder}[1]{ {\mkern1mu\colon\mkern-4mu{#1}\colon\mkern-3mu} }

\newcommand{\eqnref}[1]{Eq.~(\ref{#1})}
\newcommand{\eqsref}[1]{Eqs.~(\ref{#1})}
\newcommand{\figref}[1]{Fig.~\ref{#1}}
\newcommand{\figsref}[1]{Figs.~\ref{#1}}
\newcommand{\sfigref}[2]{Figs.\,\hyperref[#1]{\ref{#1}(#2)}}
\newcommand{\tabref}[1]{Tab.~\ref{#1}}
\newcommand{\secref}[1]{Sec.~\ref{#1}}
\newcommand{\appref}[1]{Appendix~\ref{#1}}

\usepackage{color}
\definecolor{dkgreen}{rgb}{0,0.5,0}
\definecolor{midnightblue}{rgb}{0.39,0.58,0.93}

\definecolor{kspink}{RGB}{200,0,200}

\newcommand{\comment}[1]{}{}

\hypersetup{
  colorlinks = true,
  citecolor=blue,
  linkcolor=blue,
  urlcolor = blue,
  pdfstartview=FitH,
  pdfauthor = {Kevin Slagle, David Aasen, Hannes Pichler, Roger S. K. Mong, Paul Fendley, Xie Chen, Manuel Endres, Jason Alicea},
  pdftitle = {Slagle et al. - 2021 - Microscopic characterization of Ising CFT in Rydberg chains}
}

\begin{document}

\title{Microscopic characterization of Ising conformal field theory in Rydberg chains}

\author{Kevin Slagle}
\affiliation{Department of Physics and Institute for Quantum Information and Matter, California Institute of Technology, Pasadena, CA 91125, USA}
\affiliation{Walter Burke Institute for Theoretical Physics, California Institute of Technology, Pasadena, CA 91125, USA}
\author{David Aasen}
\affiliation{Microsoft Quantum, Microsoft Station Q, University of California, Santa Barbara, California 93106 USA}
\affiliation{Kavli Institute for Theoretical Physics, University of California, Santa Barbara, California 93106, USA}
\author{Hannes Pichler}
\affiliation{Department of Physics and Institute for Quantum Information and Matter, California Institute of Technology, Pasadena, CA 91125, USA}
\affiliation{Institute for Theoretical Physics, University of Innsbruck, 6020 Innsbruck, Austria}
\affiliation{Institute for Quantum Optics and Quantum Information, Austrian Academy of Sciences, 6020 Innsbruck, Austria}
\author{\mbox{Roger S. K. Mong}}
\affiliation{Department of Physics and Astronomy and Pittsburgh Quantum Institute, University of Pittsburgh, Pittsburgh, PA 15260, USA}
\author{Paul Fendley}
\affiliation{All Souls College and Rudolf Peierls Centre for Theoretical Physics,
University of Oxford, Oxford OX1 3PU, UK}
\author{Xie Chen}
\affiliation{Department of Physics and Institute for Quantum Information and Matter, California Institute of Technology, Pasadena, CA 91125, USA}
\affiliation{Walter Burke Institute for Theoretical Physics, California Institute of Technology, Pasadena, CA 91125, USA}
\author{Manuel Endres}
\affiliation{Department of Physics and Institute for Quantum Information and Matter, California Institute of Technology, Pasadena, CA 91125, USA}
\author{Jason Alicea}
\affiliation{Department of Physics and Institute for Quantum Information and Matter, California Institute of Technology, Pasadena, CA 91125, USA}
\affiliation{Walter Burke Institute for Theoretical Physics, California Institute of Technology, Pasadena, CA 91125, USA}

\date{\today}

\begin{abstract}
Rydberg chains provide an appealing platform for probing conformal field theories (CFTs) that capture universal behavior in a myriad of physical settings.
Focusing on a Rydberg chain at the Ising transition separating charge density wave and disordered phases, we establish a detailed link between microscopics and low-energy physics emerging at criticality.
We first construct lattice incarnations of primary fields in the underlying Ising CFT \emph{including chiral fermions}---a nontrivial task given that the Rydberg chain Hamiltonian does not admit an exact fermionization.
With this dictionary in hand, we compute correlations of microscopic Rydberg operators, paying special attention to finite, open chains of immediate experimental relevance.  
We further develop a method to quantify how second-neighbor Rydberg interactions tune the sign and strength of four-fermion couplings in the Ising CFT.
Finally, we determine how the Ising fields evolve when four-fermion couplings drive an instability to Ising tricriticality.
Our results pave the way to a thorough experimental characterization of Ising criticality in Rydberg arrays,
  and can inform the design of novel higher-dimensional phases based on coupled critical chains.
\end{abstract}

\maketitle

\section{Introduction}
\label{intro}

Conformal field theory (CFT) plays a vital role in many branches of physics including condensed matter, statistical mechanics, high energy and quantum gravity \cite{AppliedCFT,IntroCFT}.  
CFTs describe systems that enjoy invariance under conformal spacetime transformations that strongly constrain physical properties.   These constraints are particularly powerful in one-dimensional quantum and two-dimensional classical systems, allowing universal behavior to be extracted from algebraic relations. In many cases of interest, the CFT here is `rational' and can be characterized by a finite set of `primary' fields and states \cite{AppliedCFT}. All other fields/states are found by acting with the generators of conformal and other symmetries. On the experimental front, CFTs capture low-energy physics in a wide variety of platforms ranging from quantum-critical spin chains (e.g., Refs.~\onlinecite{ColdeaE8,FavaE8}) to edge states of topological phases of matter (e.g., Ref.~\onlinecite{MooreRead}). 

Laser-excited Rydberg atoms trapped in optical tweezer arrays offer a route towards investigating CFTs with unprecedented depth via analog quantum simulation \cite{Browaeys2020,Morgado2021}.
These systems benefit from exceptional coherence, exquisite tunability, configurable atom array geometry , and site-resolved readout.
Moreover, Rydberg atoms exhibit strong induced dipole-dipole interactions that catalyze a rich set of accessible phases and transitions \cite{Lukin2017,Scholl2021,Lukin256};
  indeed, even the simplest linear chain architecture features quantum phase transitions described by Ising and tricritical Ising CFTs \cite{Fendley2004,Lesanovsky2012,Rader2019} (as well as a $\mathbb{Z}_3$ transition \cite{Samajdar2018,Whitsitt2018}).
Initial experimental forays into Rydberg-array quantum criticality have focused on Kibble-Zurek effects \cite{Kibble1976,Zurek1985} that describe excitations created upon dynamically sweeping across a quantum phase transition,
  revealing critical exponents of the associated universality classes \cite{RydbergKibbleZurek,Ebadi2021}.

Interrogating Rydberg arrays tuned precisely to criticality promises to reveal the more complete structure of CFTs.
For instance, can one directly measure critical correlations of fields that capture low-energy physics, and in doing so read off their scaling dimensions?
How do edge terminations---naturally relevant for experiment---impact correlations of microscopic quantities?
Do irrelevant perturbations away from `pure' CFT fixed-point theories produce measurable signatures?
Aside from fundamental interest, this line of inquiry can provide valuable benchmarking for quantum simulation, inform blueprints for exotic phases of matter based on coupled critical chains \cite{wiresKane,wiresTeo,wireTCI}, and perhaps even advance formal understanding of CFTs (e.g., in the realm of non-equilibrium dynamics or their connection \cite{YaoScarCriticality} to scar states \cite{Lukin2017,TurnerScar}).  

Addressing such questions requires understanding how physical microscopic Rydberg degrees of freedom map to emergent low-energy CFT fields.  
In some models, deriving such a correspondence can be straightforward. 
The canonical transverse-field Ising model---which as the name indicates displays a quantum critical point described by the Ising CFT---provides a classic example: 
Jordan-Wigner fermionization exposes the free-fermion nature of the problem and facilitates a precise mapping between microscopic spins and low-energy fermions that famously emerge in the Ising CFT.

\begin{figure}
  \centering
  \includegraphics[width=\columnwidth]{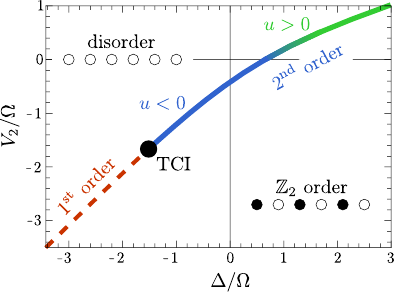} \\
  \caption{%
    Phase diagram of the Rydberg-chain Hamiltonian [\eqnref{H_FSS}] with $V_1 = \infty$.  
   Upon decreasing $V_2$, the 2nd order Ising transition (solid line) separating the disordered and $\mathbb{Z}_2$ CDW phases evolves into a tricritical Ising (TCI) point before turning first order (dashed line).
    We color the 2nd order line green and blue to respectively indicate positive and negative $T\bar{T}$ coefficients $u$ that characterizes four-fermion interactions in the Ising CFT; see \figref{fig:spectrum} for more details.
      Additional phases appear beyond the range of parameters displayed but are not relevant for this paper.
  }\label{fig:phases}
\end{figure}

In this paper we pursue an analogous dictionary for the Ising CFT governing the phase transition between charge density wave and trivial phases in a Rydberg chain; see the phase diagram in Fig.~\ref{fig:phases}.  
Finding lattice counterparts of operators in the field theory is not so simple here for two deeply intertwined reasons.
Unlike the transverse-field Ising problem, the Rydberg chain does not admit a known exact mapping to a local fermion model. 
At and near this transition, the chain is not even integrable, much less free-fermion.
Moreover, the $\mathbb{Z}_2$ symmetry spontaneously broken in the ordered phase is not a simple internal symmetry (again unlike transverse-field Ising), but rather corresponds to translation by a single site. 
A similar situation occurs in the antiferromagnetic Ising chain with both transverse and longitudinal fields \cite{OvchinnikovIsingXZ}. The longitudinal field explicitly breaks the internal $\mathbb{Z}_2$ symmetry, thus spoiling the usual Jordan-Wigner mapping and leaving single-site translation as the symmetry that sharply distinguishes ordered and disordered phases.

We show how to overcome these difficulties for a Rydberg chain.
In particular, we analytically construct microscopic incarnations of both bosonic and fermionic Ising CFT fields and verify the mappings using exact diagonalization.
As a bonus, our techniques allow us to identify lattice operators that map to fermion fields in the antiferromagnetic Ising model as well.

Armed with this dictionary, we develop a microscopic characterization of Ising criticality in Rydberg chains from several angles.  
First, although our mappings immediately predict long-distance power-law behavior of microscopic Rydberg operators in \emph{periodic} chains, edge effects operative in more experimentally accessible open chains can (and do in this case) strongly modify correlations.
We use results from CFT with fixed boundary conditions to quantify open-chain correlations of microscopic operators, providing key input for near-term experiments.
Second, the continuous Ising transition populates a one-dimensional line in the two-dimensional phase diagram from Fig.~\ref{fig:phases}.  
We show how moving along this line (by modulating the external parameters at which Ising criticality appears) tunes the sign and strength of four-fermion interactions in the CFT.  
Moreover, we argue that these interactions, while formally irrelevant at weak coupling, have a visible effect on finite-size open-chain correlations, providing an experimental window into quantifying perturbations to vanilla CFT theories.
Third, the continuous Ising transition line eventually terminates at a quantum critical point described by a tricritical Ising CFT (driven by strong four-fermion interactions).  
We track the evolution of Ising CFT fields upon approaching the tricritical Ising point and establish a partial dictionary linking microscopic Rydberg operators to tricritical Ising fields. 
We anticipate that our results will pave the way to detailed experimental characterization of Ising criticality, and quantum criticality more broadly, in Rydberg arrays.  

The remainder of the paper is organized as follows.
Section~\ref{sec:model} reviews the Rydberg-chain model that we study throughout.
Section~\ref{sec:IsingCFT} surveys the Ising CFT, Sec.~\ref{sec:operators} develops the dictionary linking microscopic Rydberg operators to Ising CFT fields, and Sec.~\ref{sec:AFMising} briefly discusses implications for the antiferromagnetic Ising chain.
Section~\ref{TTbar_sec} explores the microscopic origin of four-fermion interactions in the field-theory action.
In Sec.~\ref{sec:open_chain} we quantify microscopic Rydberg correlations in an open chain, where edge effects play a pivotal role, and propose a scheme for locating the critical point using finite open chains. 
Section~\ref{sec:tricriticality} studies the approach to the tricritical Ising CFT driven by four-fermion interactions, and,  
finally, Sec.~\ref{discussion} provides a summary and experimental outlook.

\section{Model and phase diagram}
\label{sec:model}

We consider a Rydberg chain governed by the Hamiltonian
\begin{equation}
  H = \sum_j\left[\frac{\Omega}{2}(b_j + b_j^\dagger) -\Delta n_j + V_1 n_j n_{j+1} + V_2 n_j n_{j+2}\right].
  \label{H_FSS}
\end{equation}
Here $b_j$ is a canonical hard-core boson operator and $n_j = b_j^\dagger b_j$ is the associated number operator;
$n_j = 0$ and $1$ respectively correspond to the ground state and Rydberg excited state for the atom at site $j$.
The first two terms in $H$ describe (within the rotating-wave approximation) atoms driven at Rabi frequency $\Omega$ with detuning $\Delta$ from the Rydberg state.  
The $V_1$ term encodes nearest-neighbor induced dipole-dipole interactions.
Unless specified otherwise, we take $V_1 \rightarrow +\infty$.
This limit energetically enforces the nearest-neighbor Rydberg blockade constraint $n_j n_{j+1} = 0$,  precluding two nearest-neighbor atoms from simultaneously entering the Rydberg state.
Finally, the $V_2$ term in Eq.~\eqref{H_FSS} encodes subdominant induced dipole-dipole interactions among second-nearest neighbors.
We allow $V_2$ to take either sign in our analysis.
Although the most natural physical regime corresponds to $V_2>0$,
  negative $V_2<0$ may also be realizable (as discussed in \secref{discussion}).
Rydberg chains naturally host a rapidly decaying $V_r \propto r^{-6}$ interaction;
  we drop interactions beyond next-nearest-neighbor.
The full Hamiltonian preserves bosonic time reversal $\mathcal{T}$ and, with suitable boundary conditions, translation by one lattice site $T_x$  and reflection $R_x$ about a site.  
Table~\ref{SymmetriesTable} (upper rows) specifies the action of these symmetries on microscopic Rydberg-chain operators.

The model in Eq.~\eqref{H_FSS} was introduced by Fendley, Sengupta, and Sachdev \cite{Fendley2004} as a quantum-chain limit of Baxter's hard-square model \cite{Baxter1982}. It was  further explored via large-scale numerics in Refs.~\onlinecite{Samajdar2018,Chepiga2019,Chepiga2019b,Giudici2019} (see also Refs.~\onlinecite{Whitsitt2018,Rader2019,IsingLongitudinal}).
Figure~\ref{fig:phases} reproduces the phase diagram as a function of $\Delta/\Omega$ and $V_2/\Omega$ over the range of couplings relevant for this paper.
Two phases appear: The first is a disordered, symmetric gapped state that smoothly connects to the trivial boson vacuum with no Rydberg excitations.  Negative detuning ($\Delta <0$) and repulsive second-neighbor interactions ($V_2>0$) naturally favor such a state.  
The second is a two-fold-degenerate $\mathbb{Z}_2$-ordered charge density wave (CDW) promoted by either positive detuning ($\Delta >0$) or second-neighbor attraction ($V_2<0$)---both of which favor maximal packing of Rydberg excitations subject to nearest-neighbor Rydberg blockade.  
Each of the two CDW ground states accordingly exhibits enhanced Rydberg-excitation probability on every other site, quantified by
\begin{equation}
  \langle n_j\rangle = a_0 + a_\pi (-1)^j
  \label{CDWorder}
\end{equation}
for non-universal $a_0,a_\pi$ constants.  
Importantly, the CDW ground states are exchanged under $T_x$ but preserve $T_x^2$.
The broader phase diagram features additional phases (not shown) including a three-fold-degenerate charge density wave and incommensurate order; see Refs.~\onlinecite{Fendley2004,Samajdar2018,Chepiga2019,Chepiga2019b,Giudici2019}.

The nature of the transition separating the disordered and CDW phases evolves nontrivially as one moves along the phase boundary in Fig.~\ref{fig:phases}.
The solid line---which includes the physically relevant $V_2>0$ regime---corresponds to a continuous Ising transition, with translation $T_x$ playing the role of the global $\mathbb{Z}_2$ spin-flip symmetry familiar from the Ising model.  
We determined the location of this portion of the phase boundary via a standard scaling collapse of the rescaled energy gap $L E_\text{gap}$ vs $\Delta$ (here and below $L$ denotes the number of sites) obtained from exact diagonalization of a Rydberg chain with periodic boundary conditions \cite{Fendley2004}.
At 
\begin{equation} 
  \frac{V_2}{\Omega} = \mathcal{V}_{\rm TCI} \equiv -\frac{1}{2}\left(\frac{1+\sqrt{5}}{2}\right)^{5/2}
  \label{TCIpoint}
\end{equation} 
the continuous transition evolves into a tricritical Ising point (labeled `TCI' in Fig.~\ref{fig:phases}). The location of the tricritical point is known exactly because the chain is integrable here \cite{Baxbook}; its Hamiltonian can be expressed in terms of the Temperley-Lieb algebra and is sometimes known as the golden chain \cite{Feiguin2007}. The transition at still more negative $V_2/\Omega$ becomes first order (dashed line in Fig.~\ref{fig:phases}) \cite{Fendley2004}. Its location is also known from integrability to be at
\begin{equation}
  \frac{V_2}{\Omega} = \frac{1}{2}\left[\frac{\Delta}{\Omega}-\sqrt{\left(\frac{\Delta}{\Omega}\right)^2+1}\right]~~~~{\text{(first-order line)}}
  \label{firstorder}
\end{equation}
for $V_2/\Omega< \mathcal{V}_{\rm TCI}$.

\section{Operator dictionary at Ising criticality}
\label{sec:dictionary}

We here begin an in-depth exploration into the continuous Ising transition separating the disordered phase from the $\mathbb{Z}_2$-ordered CDW along the solid line in \figref{fig:phases}.
In this section we first review the Ising CFT, then derive a mapping between CFT fields and microscopic Rydberg operators, and finally comment on implications of this mapping for the antiferromagnetic transverse field Ising model.

\subsection{Ising CFT Review}
\label{sec:IsingCFT}

The continuous Ising transition line is described by a CFT with central charge $c = 1/2$ \cite{Cardy_1984,CARDY1986186}.
In the CFT, the local $\mathbb{Z}_2$ CDW order parameter that condenses on the ordered side of the transition corresponds to a `spin field' $\sigma$.  The CFT exhibits a Kramers-Wannier duality as does the Ising lattice model, with the dual of the spin field known as the disorder field $\mu$. The disordered phase on the other side of the transition can be understood as arising from condensation of this disorder field, which is non-local in terms of the original spin field.
Both $\sigma$ and $\mu$ are Hermitian fields of scaling dimension $1/8$ that satisfy
\begin{align}
  \sigma(x)\mu(x') &= {\rm sgn}(x-x')\mu(x')\sigma(x).
  \label{sigma_mu_commutator}
\end{align}
  where $x$ and $x'$ are spatial coordinates, and ${\rm sgn}(x-x') = 1$ if $x>x'$ and $-1$ if $x<x'$.
Right- and left-moving emergent Majorana fermions $\gamma_{R/L}$ with dimension $1/2$ follow upon combining order and disorder fields via the
  operator product expansion (with spatially dependent coefficients omitted)
\begin{equation}
  \sigma \, \mu \sim \gamma_R + \gamma_L + \dots,
  \label{FermionOPE}
\end{equation}
where the ellipsis denotes descendant operators.
Consistent with Eq.~\eqref{sigma_mu_commutator}, $\sigma$ and $\mu$ enact sign changes on the fermions:
\begin{gather}
\begin{aligned}
  \sigma(x')\gamma_{R/L}(x) &= {\rm sgn}(x'-x)\gamma_{R/L}(x) \sigma(x')\\
  \mu(x')\gamma_{R/L}(x) &= {\rm sgn}(x-x')\gamma_{R/L}(x) \mu(x').
\end{aligned} \label{signchange}
\end{gather}

\begin{table}
\centering
\begin{tabular}{|c | c | c | c | c|} 
 \hline
  & $T_x$ & $\mathcal{T}$ & $R_x$ & $\mathbb{Z}_2^{\rm dual}$ \\ 
 \hline
 $n_j \rightarrow$ & $n_{j+1}$ & $n_j$ & $n_{-j}$ & {\rm N/A} \\ 
 $b_j \rightarrow$ & $b_{j+1}$ & $b_j$ & $b_{-j}$ & {\rm N/A} \\
 \hline
 $\sigma \rightarrow$   & $-\sigma$   & $\sigma$   & $\sigma$                  & $ \sigma$ \\
 $\mu \rightarrow $     & $ \mu$      & $\mu$      & $\mu(\infty) \mu$         & $-\mu$ \\
 $\gamma_R \rightarrow$ & $-\gamma_R$ & $\gamma_L$ & $-i \gamma_L \mu(\infty)$ & $-\gamma_R$ \\ 
 $\gamma_L \rightarrow$ & $-\gamma_L$ & $\gamma_R$ & $+i \gamma_R \mu(\infty)$ & $-\gamma_L$ \\ 
 \hline
\end{tabular}
\caption{Symmetry properties of the microscopic Rydberg chain operators (upper rows) and CFT fields (lower rows) that describe low-energy physics at Ising criticality.  For brevity we suppressed the position coordinate $x$ for the CFT fields; note, however, that $R_x$ additionally sends $x\rightarrow -x$.
}\label{SymmetriesTable}
\end{table}

The above fields and their descendants capture the low-energy physics at and near Ising criticality.
In particular, in terms of the dimension-1 Majorana-fermion mass term 
\begin{equation}
  \varepsilon = i \gamma_R \gamma_L
  \label{epsilon}
\end{equation} 
and dimension-2 kinetic energies 
\begin{equation}
  T = -i \norder{ \gamma_R \partial_x \gamma_R } , \qquad \overline{T} = i \norder{ \gamma_L \partial_x \gamma_L } ,
\end{equation}
where colons indicate normal ordering,
the low-energy Hamiltonian can be written as
\begin{align}
  \mathcal{H} = \int_x\left[m\, \varepsilon + v(T + \overline{T}) + u\,T \overline{T}\right].
  \label{H_CFT}
\end{align}
The pure CFT Hamiltonian corresponds to setting $m=u=0$.
The $T \overline{T}$ operator has dimension 4, and is the least irrelevant operator that preserves the self-duality and $\mathbb{Z}_2$ symmetry of the Ising CFT \footnote{%
  When $u<0$, $T \overline{T}$ is dangerously irrelevant in the sense that the long-distance IR physics is sensitive to the UV cutoff.}.
Ising criticality thus persists for sufficiently small $u$ while keeping $m = 0$.
Section~\ref{TTbar_sec} discusses in detail the effects of including this term.
Resurrecting $m \neq 0$ shifts the system into either the disordered phase or $\mathbb{Z}_2$ CDW depending on the sign of $m$.

\subsection{Lattice Operators}
\label{sec:operators}

Next we pursue a dictionary linking microscopic operators to the CFT fields defined above.  
In the canonical transverse-field Ising model, exact solvability aided by the Jordan-Wigner transformation enables a straightforward algorithmic identification of microscopic order and disorder operators as well as fermions.  
For a brief review see Appendix~\ref{IsingReview}.
An exact solution to Eq.~\eqref{H_FSS} at the continuous Ising transition is, by contrast, unknown.  
We can nevertheless obtain the desired dictionary using analytic arguments bolstered by numerics.

\begin{figure}
  \subfloat[\label{fig:sigmaScaling}]{\includegraphics[width=.8\columnwidth]{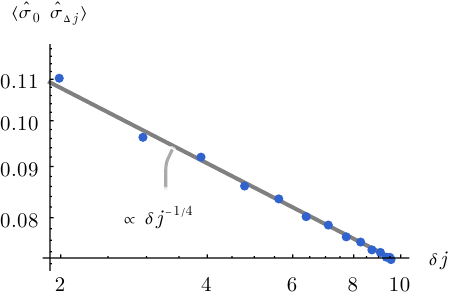}} \\
  \subfloat[\label{fig:nnScaling}]{\includegraphics[width=.8\columnwidth]{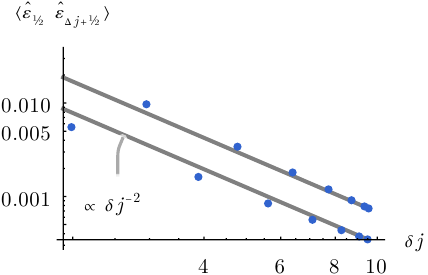}} \\
  \caption{%
    Numerical verification of power-law scaling for the microscopic operators (a) $\hat\sigma_j$ and (b) $\hat\varepsilon_{j+1/2}$ predicted by \eqsref{sigma_lattice} and (\ref{epsilon_lattice}), respectively.  
    Data were obtained using exact diagonalization on a periodic length $L=30$ chain at Ising criticality with $V_2 = 0$.
    Gray lines are fits to the expected power-law, $c \, (\delta j)^{-2\Delta}$,
      where $c$ is the only fitting parameter and $\Delta$ is the expected scaling dimension.
    The horizontal axis corresponds to the adjusted lattice distance between the operators: $\delta j = \frac{L}{\pi} \sin \frac{\pi \Delta j}{L} \sim \Delta j$ \cite{Afoot:deltaj}.
  }\label{fig:sigma nnScaling}
\end{figure}

First we expand the boson number operator at criticality as \footnote{%
  The Hermitian operator $b_j + b_j^\dagger$ has identical symmetry properties to $n_j$, and thus exhibits a low-energy expansion of the same form (of course with different coefficients).  
  The Hermitian operator $i(b_j-b_j^\dagger)$ is odd under time reversal but even under $R_x$.  
  In the Heisenberg picture, we therefore obtain $i[b_j(t)-b_j^\dagger(t)] \sim (-1)^j \partial_t \sigma + \cdots$, implying $i(b_j-b_j^\dagger)\sim i(-1)^j[\sigma,H]$ for the Schrodinger picture that we typically employ in this paper.
  We focus on the number operator rather than creation and annihilation operators due to ease of measurement.}
\begin{equation}
  n_j \sim \langle n\rangle + c_\sigma(-1)^j \sigma + c_\varepsilon\varepsilon + \cdots,
  \label{n_expansion}
\end{equation}
where $\langle n\rangle$ is the (generically non-zero) ground-state expectation value of $n_j$, $c_{\sigma,\varepsilon}$ are constants, and the ellipsis denotes subleading terms with higher scaling dimension.
The $c_\sigma$ term reflects the fact that condensing $\sigma$ generates $\mathbb{Z}_2$ CDW order [recall Eq.~\eqref{CDWorder}].
As for $c_\varepsilon$, observe that adding a term $\propto \sum_j n_j$ to the critical Hamiltonian [i.e., shifting $\Delta$ in Eq.~\eqref{H_FSS}] moves the system off of criticality; thus $n_j$ must contain the fermion bilinear $\varepsilon$ in its low-energy expansion.
We can isolate $\sigma$ as the leading contribution by defining 
\begin{equation}
  \hat\sigma_{j} \equiv (-1)^j(n_j - \langle n\rangle) \sim c_\sigma \sigma + \cdots
  \label{sigma_lattice}
\end{equation}
and similarly isolate $\varepsilon$ through
\begin{equation}
  \hat \varepsilon_{j+1/2} \equiv (n_j + n_{j+1})-2 \langle n \rangle \sim 2c_\varepsilon \varepsilon + \cdots.
  \label{epsilon_lattice}
\end{equation}
Exact diagonalization numerics plotted in Fig.~\ref{fig:sigma nnScaling} support the identifications in Eqs.~\eqref{sigma_lattice} and \eqref{epsilon_lattice}
  by demonstrating power-law correlations consistent with the
  $1/8$ and $1$ scaling dimensions for the CFT fields $\sigma$ and $\varepsilon$, respectively.  
  [The even-odd effect in Fig.~\ref{fig:sigma nnScaling}(b) arises from a $(-1)^j \partial_x \sigma$ term (with scaling dimension $9/8$) allowed in the ellipsis from Eq.~\eqref{epsilon_lattice}.]

For a microscopic counterpart of the disorder field $\mu$,
  we introduce a non-local operator $\hat\mu_{j}$ that flips $\mathbb{Z}_2$ CDW order to the left of site $j$
  via a partial translation:
\begin{equation}
  \hat\mu_{j} \, n_i \, \hat\mu_{j}^\dagger = \begin{cases}
    n_{i+1}    & i < j \\
    \braket{n} & i = j \\
    n_i        & i > j
    \end{cases} \label{eq:mu action}.
\end{equation}
In effect, $\hat\mu_{j}$ removes the site $j$ to accommodate the translated sites;
mapping the number operator $n_j$ to its expectation value $\braket{n}$ makes this action as non-violent as possible.
This definition presumes an infinite number of sites, though we explain below how to treat a finite system size. 

To precisely define  $\hat \mu_j$, we introduce operators $S_{j+1/2}$ that swap sites $j$ and $j+1$,
\begin{equation}
  S_{j+1/2} \ket{n_j n_{j+1}} = \ket{n_{j+1} n_j},
\end{equation}
along with an operator
\begin{align}
  \zeta_{j} &= \ket{0_{j}} \bra{\psi_{j}}, \cr
  \ket{\psi_{j}} &= \sqrt{1-\langle n \rangle} \ket{0_{j}} - \sqrt{\langle n \rangle} \ket{1_{j}}
\end{align}
that implements the $i = j$ case in \eqnref{eq:mu action}. 
In particular, $\zeta_j$ disentangles site $j$ from the rest of the chain by first projecting onto the `typical' quantum state $\ket{\psi_j}$, which has a sign structure favored by $\Omega>0$ and an average occupation number $\braket{n}$, and then parking the disentangled site into the $n_j = 0$ state.
Putting these pieces together, we arrive at
\begin{equation}
  \hat\mu_{j} = \cdots S_{j-5/2} \, S_{j-3/2} \, S_{j-1/2}  \, \zeta_{j} 
  \label{mu_expansion}
\end{equation}
We then can define the operator $\hat\mu_\infty$ to enact a single-site translation to the right. 

\begin{figure}[t]
  \subfloat[\label{fig:muScaling}]{\includegraphics[width=.8\columnwidth]{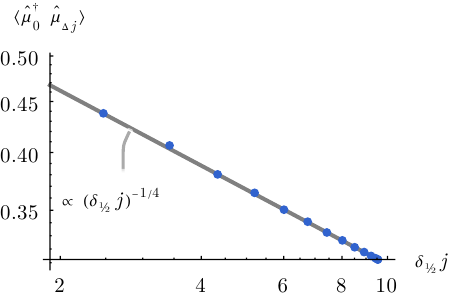}} \\
  \subfloat[\label{fig:gammaPowerLaw}]{\includegraphics[width=.8\columnwidth]{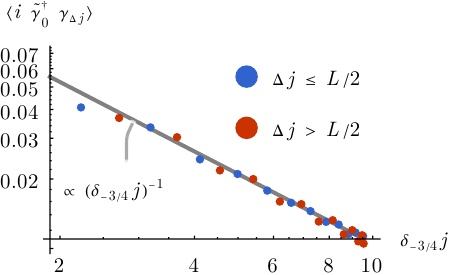}} \\
  \subfloat[\label{fig:gammaDecay}]{\includegraphics[width=.8\columnwidth]{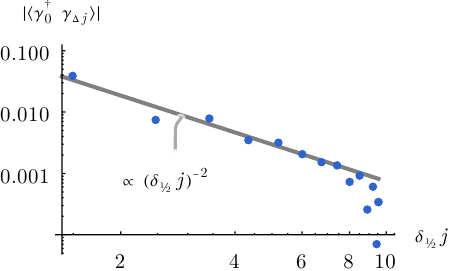}}
  \caption{%
    Correlation functions (a) $\langle \hat\mu^\dagger_0 \hat \mu_{\Delta j}\rangle$, (b) $\langle i \tilde \gamma^\dagger_0\gamma_{\Delta j}\rangle$, and (c) $\langle \gamma_0^\dagger\gamma_{\Delta j}\rangle$ obtained for a periodic $L = 30$ chain at Ising criticality with $V_2 = 0$.
        The horizontal axes are adjusted lattice distances: $\delta_\epsilon j = \frac{L}{\pi} \sin \frac{\pi}{L} (\Delta j + \epsilon) \sim \Delta j$ \cite{Afoot:deltaj}.
        Data from (a) and (b) verify the power-law scaling predicted by Eqs.~\eqref{mu_expansion}, \eqref{gammaj}, and \eqref{gamma_def}.
        }\label{fig:gammaScaling}
\end{figure}

We expect that with these definitions, lattice and CFT operators are related via
\begin{align}
\hat\mu_j \sim c_\mu \mu + \cdots\ .
\end{align}
(Although $\hat\mu_j$ is not Hermitian, time-reversal symmetry requires that the prefactor $c_\mu$ is real.)  
Exact diagonalization results shown in \figref{fig:muScaling} confirm that $\hat\mu_{j}$ indeed exhibits power-law correlations consistent with the CFT field $\mu$.  We measure the combination
  $\hat\mu^\dagger_i \hat\mu_j = \zeta_i^\dagger S_{i+1/2} \cdots S_{j-1/2} \zeta_j$.
This combination does make sense on finite lattices, and so here we utilize periodic boundary conditions for $L=30$ sites.

The product of lattice order and disorder operators $\hat\mu_i \hat \sigma_j$ exhibits the following simple off-site commutation relations:
\begin{equation}
\hat \mu_i \hat\sigma_j = \begin{cases}
    - \hat\sigma_{j+1} \hat\mu_i, & j<i \\
    + \hat\sigma_j     \hat\mu_i & j>i
  \end{cases}.
\end{equation}
Let us denote the on-site commutator as
\begin{equation}
  \tilde \gamma_j \equiv i\big[\hat\mu_j,\hat \sigma_j\big] = i\hat\mu_j(\hat\sigma_{j-1} + \hat\sigma_{j}),
  \label{gammaj}
\end{equation}
and further define
\begin{equation}
  \gamma_j \equiv i R_x \tilde \gamma_{-j} R_x T_x = \hat\mu_{j-1}(\hat\sigma_{j-1} + \hat\sigma_j).
  \label{gamma_def}
\end{equation}
Here we used $T_x R_x = R_x T_x^\dagger$ and the decomposition $T_x = \hat\mu_\infty$.
As the notation suggests, $\gamma_j,\tilde \gamma_j$ constitute lattice counterparts of the CFT fermion fields $\gamma_{R/L}$ that arise from products of order and disorder operators.

Symmetry partially constrains the form of this UV-IR relation.  
Time reversal swaps $\gamma_R \leftrightarrow \gamma_L$ in the CFT, implying 
\begin{gather}
\begin{aligned}
  \gamma_j &\sim e^{i \alpha} \gamma_R + e^{-i\alpha} \gamma_L + \cdots \\
  \tilde \gamma_j &\sim e^{i \beta} \gamma_R - e^{-i\beta} \gamma_L + \cdots 
\end{aligned} \label{gamma_exp1}
\end{gather} 
for real $\alpha,\beta$.
Recalling that $R_x$ also swaps right- and left-movers and identifying $T_x \sim \mu(\infty)$, we can insert Eq.~\eqref{gamma_exp1} into the left and middle parts of Eq.~\eqref{gamma_def} to infer that 
\begin{gather}
\begin{aligned}
  R_x \gamma_R(-x)R_x &= -ie^{-i(\alpha+\beta)} \gamma_L(x) \mu(\infty)
  \\
  R_x \gamma_L(-x)R_x &= ie^{i(\alpha+\beta)} \gamma_R(x)\mu(\infty).
\end{aligned}
\end{gather} 
Reflections must preserve Hermiticity of $\gamma_{R/L}$; since $\mu(\infty)$ anticommutes with $\gamma_{L/R}(x)$, this condition requires $e^{i\beta} = s e^{-i \alpha}$ for some sign $s = \pm 1$.  
Equation~\eqref{gamma_exp1} then reduces to
\begin{gather}
\begin{aligned}
  \gamma_j &\sim e^{i \alpha} \gamma_R + e^{-i\alpha} \gamma_L + \cdots \\
  \tilde \gamma_j &\sim s\left(e^{-i \alpha} \gamma_R - e^{i\alpha} \gamma_L\right) + \cdots.
\end{aligned} \label{gamma_exp2}
\end{gather} 
The lattice operators $\gamma_j,\tilde \gamma_j$ on the left side are not Hermitian, and so it appears that general arguments do not enable determination of the remaining parameters $s,\alpha$.

Nevertheless, Eq.~\eqref{gamma_exp2} implies that $\langle i \tilde \gamma^\dagger_j \gamma_{j'}\rangle$ generically exhibits power-law correlations with scaling dimension 1/2, whereas for $\langle i \tilde \gamma^\dagger_j \tilde \gamma_{j'}\rangle$ and $\langle i \gamma^\dagger_j \gamma_{j'}\rangle$ the leading power-law contributions from right- and left-moving pieces exactly cancel.   Numerics presented in \sfigref{fig:gammaScaling}{b,c} indeed show that $\langle i \tilde \gamma^\dagger_j \gamma_{j'}\rangle$ obeys the predicted power-law correlations (decay exponent of 1) while $\langle i \tilde \gamma^\dagger_j \tilde \gamma_{j'}\rangle$ and $\langle i \gamma^\dagger_j \gamma_{j'}\rangle$ decay with a subleading power law (decay exponent of 2).  
We attribute the observed subleading power law to terms represented by the ellipses of Eq.~\eqref{gamma_exp2} involving $\partial_x \gamma_{R/L}$.   Other fermion correlation functions are given by the exact microscopic relations
      $\langle i \tilde{\gamma}_0^\dagger \gamma_{\Delta j} \rangle = \langle i \gamma_1^\dagger \tilde{\gamma}_{L-\Delta j} \rangle$ and
      $\langle \gamma_0^\dagger \gamma_{\Delta j} \rangle = - \langle \tilde{\gamma}_1^\dagger \tilde{\gamma}_{L-\Delta j} \rangle$.

The lower rows of Table~\ref{SymmetriesTable} summarize the symmetry transformations for the CFT fields $\sigma, \mu,$ and $\gamma_{R/L}$ that are compatible with the preceding dictionary.  
In the final column we include a dual $\mathbb{Z}_2$ symmetry---labeled $\mathbb{Z}_2^{\rm dual}$---preserved by the CFT, which sends $\mu \rightarrow -\mu$ but leaves $\sigma$ invariant.

\subsection{Application to the antiferromagnetic transverse-field Ising model}
\label{sec:AFMising}

The preceding analysis also has interesting implications for the antiferromagnetic transverse field Ising model.  
Upon setting $\Delta = V_1$ and $V_2 = 0$ and identifying Pauli matrices 
\begin{align}
  Z_j = 2 n_j - 1,~~~X_j = b_j + b_j^\dagger\ .
\end{align} 
the Rydberg Hamiltonian in Eq.~\eqref{H_FSS} reduces to an \emph{antiferromagnetic} transverse-field Ising model,
\begin{equation}
  H_{\rm TFIM} = \sum_j \left(J Z_j Z_{j+1} - h_x X_j\right),
  \label{Hising}
\end{equation}
with $h_x = -\Omega/2$ and $J = V_1/4$.  
In this fine-tuned limit the Hamiltonian preserves a $\mathbb{Z}_2$ Ising spin-flip symmetry that sends $Z_j \rightarrow -Z_j$ as well as $T_x, R_x$, and $\mathcal{T}$. The antiferromagnetic ordered phase appearing at $J>h_x$ spontaneously breaks both the Ising spin-flip \emph{and} translation symmetries.  

For any choice of couplings $H_{\rm TFIM}$ can be written exactly as a bilinear in the familiar Jordan-Wigner fermions 
assembled from order and disorder operators associated with the local Ising  $Z_j \to - Z_j$ spin-flip symmetry. Explicit expressions are given in \eqnref{gammaIsing}. 
At Ising criticality, these fermions map onto continuum CFT fields $\gamma_{R/L}$, as described in thousands of papers (which for compactness we will not reference). 
Because the antiferromagnetically ordered state also breaks translation symmetry, \emph{so do the microscopic operators $\gamma_j, \tilde \gamma_j$ constructed in Eqs.~\eqref{gammaj} and \eqref{gamma_def}}.
We have indeed verified that the power-law correlations shown in \sfigref{fig:gammaScaling}{b,c} persist with parameters appropriate for the Ising model.
The antiferromagnetic transverse-field Ising chain thus admits two sets of microscopic fermions, one associated with local Ising symmetry, and the other with translation symmetry. 
Both map to equivalent continuum fermions at criticality.  

The interesting wrinkle is that the well-known Jordan-Wigner fermions become confined when supplementing Eq.~\eqref{Hising} with a uniform longitudinal-field term $-h_z\sum_j Z_j$, as arises when $\Delta \neq V_1$ in Rydberg language. Such a term explicitly breaks Ising spin-flip symmetry.  A sharp continuous Ising transition nevertheless survives (at a value of $h_x/J$ changing with $h_z)$ because the Hamiltonian continues to preserve the spontaneously broken translation symmetry \cite{OvchinnikovIsingXZ}. Thus even though the longitudinal field is relevant at the ferrogmagnetic transition, it is irrelevant at the antiferromagnetic one.
In the presence of this term, the Jordan-Wigner fermions are confined, because their strings do not commute with the longitudinal field. The Hamiltonian cannot even be written locally in terms of the Jordan-Wigner fermions. 
Our microscopic $\gamma_j, \tilde \gamma_j$ operators, however, generate the `correct' power-law-correlated 
 low-energy fermions at the transition even when $h_z\ne 0$.

\section{Four-fermion interactions at Ising criticality} 
\label{TTbar_sec}

Here we will discuss four-fermion interactions encoded by the $u T\overline{T}$ term in Eq.~\eqref{H_CFT}, assuming a critical Rydberg chain with $m = 0$.
In particular, we determine how the strength of the four-fermion interaction changes as one moves along the critical Ising line.

One gains valuable intuition by writing 
\begin{equation}
  u \int_x T \overline{T} \approx \frac{u}{\delta x^2} \int \varepsilon(x+\delta x)\varepsilon(x) + \cdots,
  \label{u_rewriting}
\end{equation}
where $\delta x$ is a microscopic length, $\varepsilon$ is the fermion bilinear from Eq.~\eqref{epsilon}, and the ellipsis represents fermion bilinears and an unimportant constant.
The derivation of Eq.~\eqref{u_rewriting} follows by expanding  $\varepsilon(x+\delta x) = i \gamma_R(x+\delta x) \gamma_L(x+\delta x)$ to $O(\delta x^2)$.
From the form on the right side, it is clear that turning on sufficiently large $u<0$ catalyzes an instability with $\langle \varepsilon\rangle \neq 0$---in turn gapping the critical theory by \emph{spontaneously} generating a nonzero mass $m \neq 0$ with arbitrary sign.  
Since the sign of $m$ dictates whether the system enters the $\mathbb{Z}_2$ CDW or trivial phase, we conclude that large $u<0$ renders the continuous Ising transition first order, in harmony with the exact results in Eq.~\eqref{firstorder}.
Conversely, $u>0$ opposes mass generation.

We now argue that the sign and strength of $u$ are determined primarily by the second-neighbor interaction strength $V_2$ at which one accesses Ising criticality; i.e., $u$ can be varied by moving along the continuous Ising line in \figref{fig:phases}.
On a qualitative level, inserting Eq.~\eqref{n_expansion} into the $V_2$ interaction naturally recovers the $u$ term as written on the right side of Eq.~\eqref{u_rewriting}.  
We can alternatively exploit the identity
\begin{equation}
  V_2 \sum_j n_j n_{j+2} = \frac{V_2}{2(1-\langle n\rangle)\langle n\rangle}\sum_j\big(i \tilde \gamma_{j-1}^\dagger \gamma_j\big)\big(i \tilde \gamma_{j}^\dagger \gamma_{j+1}\big),
  \label{V2}
\end{equation}
which follows from Eqs.~\eqref{gammaj} and \eqref{gamma_def} upon dropping terms that are trivial due to the nearest-neighbor Rydberg blockade,
to recover the $u$ term as written on the left side of Eq.~\eqref{u_rewriting}.
Indeed, expanding $\gamma_j, \tilde \gamma_j$ in terms of $\gamma_{R/L}$ in Eq.~\eqref{V2} yields $T \overline{T}$ as the leading four-fermion interaction.
This analysis suggests that moving along the continuous Ising transition line in the $V_2<0$ direction realizes Eq.~\eqref{H_CFT} with increasingly large $u<0$, eventually giving way to a first-order transition consistent with the established phase diagram \cite{Fendley2004} reproduced in Fig.~\ref{fig:phases}.  
Moving along the Ising transition line in the physically relevant $V_2>0$ direction instead yields Eq.~\eqref{H_CFT} with increasingly large $u>0$ that disfavors spontaneous mass generation.
Note, however, that $u$ is generically non-zero even with $V_2 = 0$ since the chain remains interacting due to nearest-neighbor Rydberg blockade.

\begin{figure}
  \includegraphics[width=.9\columnwidth]{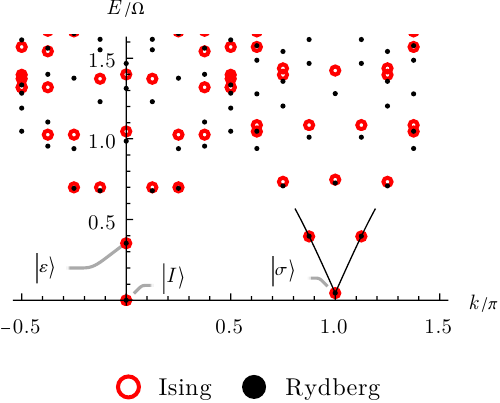} \\
  \caption{%
    Energy spectrum versus momentum for a periodic $L = 16$ Rydberg chain tuned to Ising criticality with $V_2=0$. 
    For comparison we also show the corresponding spectrum for the critical antiferromagnetic transverse-field Ising model, which realizes the non-interacting limit of the Ising CFT with $u = 0$ [Eq.~\eqref{H_CFT}].  
    As described in the text, the slight reduction in the Rydberg energies near the middle of the plot suggests that the critical Rydberg chain at $V_2 = 0$ retains weak four-fermion interactions with $u<0$.
  }\label{fig:spectrum0}
\end{figure}

Although $u T\overline{T}$ has scaling dimension 4 and is therefore irrelevant under RG,
  this interaction still influences the physics at finite energy density or finite system sizes.
For a more quantitative treatment, we examine the excitation spectrum versus momentum $k$ for an $L = 16$ site chain (with periodic boundary conditions) tuned to the continuous Ising transition line at various $V_2$ values.
Black dots in Fig.~\ref{fig:spectrum0} present exact diagonalization data for a Rydberg chain at $V_2 = 0$.
Accompanying red dots represent simulations for the critical antiferromagnetic transverse-field Ising model [Eq.~\eqref{Hising} with $J = h_x$, denoted hereafter by $H_{\rm TFIM}^{J = h_x}$]---which provides an illuminating comparison given that the latter realizes a free-fermion theory with $u = 0$.
Both the Rydberg chain and transverse-field Ising model admit a unique ground state $\ket{I}$ carrying zero momentum and a first excited state $\ket{\sigma}$ that follows from acting the CFT field $\sigma$ on $\ket{I}$ and thus carries momentum $\pi$.  
Here and below the spectrum for $H_{\rm TFIM}^{J = h_x}$ has been shifted and rescaled to match the energies of the $\ket{I}$ and $\ket{\sigma}$ states for the Rydberg chain.

Consider for the moment the non-interacting limit of the CFT---i.e., with $u = 0$---realized by $H_{\rm TFIM}^{J = h_x}$.
There, low-energy excitations about the states $\ket{I}$ and $\ket{\sigma}$ follow simply by adding an even number of free-fermion modes.
Fermions added to the ground state $\ket{I}$ obey anti-periodic boundary conditions, yielding momenta quantized to $\frac{2\pi}{L}\times \left(\mathbb{Z}+\frac{1}{2}\right)$.
Fermions added to $\ket{\sigma}$ obey periodic boundary conditions [which stems from Eq.~\eqref{signchange}] and instead exhibit momenta quantized to $ \frac{2\pi}{L}\times \mathbb{Z}$.
Starting from either $\ket{I}$ or $\ket{\sigma}$, adding a pair of fermions carrying appropriately quantized momenta $k_1$ and $k_2$ adds energy $v(|k_1|+|k_2|)$ and momentum $k_1+k_2$, where positive and negative momenta respectively correspond to right- and left-movers. 
Importantly, \emph{turning on $u T \overline{T}$ interactions shifts the excitation energy for counter-propagating fermion pairs}: their energy increases for $u>0$ and decreases for $u<0$ by an amount dependent on the chiral fermion kinetic energies $(K_L,K_R)$.
The energy shift (in first order perturbation theory) for descendants of $\ket{I}$ and $\ket{\sigma}$ are
\begin{align}
	\delta E &\propto u \begin{cases} (K_L-1/48) (K_R-1/48) & \text{desc.\ of $\ket{I}$}, \\ (K_L+1/24) (K_R+1/24) & \text{desc.\ of $\ket{\sigma}$}. \end{cases}
	\label{eq:deltaEu}
\end{align}
Here $K_{L/R}$ are the kinetic energy contributions to a state from the left/right-moving fermion modes, in units of $2\pi v/L$.
For instance, the $\ket{\varepsilon}$ state (labeled in Fig.~\ref{fig:spectrum0}) contains right- and left-moving fermions with energies $K_L=K_R=1/2$ and are thus susceptible to energy shifts $\propto 0.23 u$.
By contrast, the states near $k = \pi$ connected to $\ket{\sigma}$ by solid lines in Fig.~\ref{fig:spectrum0} involve one chiral fermion with unit energy;
  these states have $(K_L,K_R) = (1,0)$ or $(0,1)$ and are only very weakly affected by $u$.  
Figure~\ref{fig:spectrum0} thus indicates that at $V_2 = 0$, the critical Rydberg chain retains weak four-fermion interactions with $u<0$.  

\begin{figure*}[ht]
  \subfloat[\label{fig:spectrum pos}]{\includegraphics[width=.61\columnwidth]{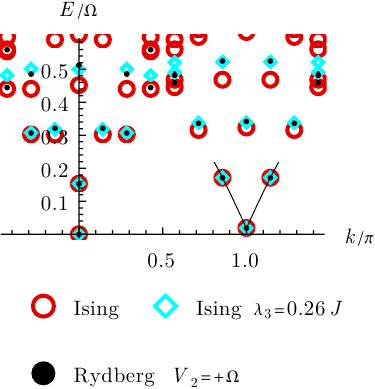}} 
  ~~~~~~~\subfloat[\label{fig:spectrum zero}]{\includegraphics[width=.61\columnwidth]{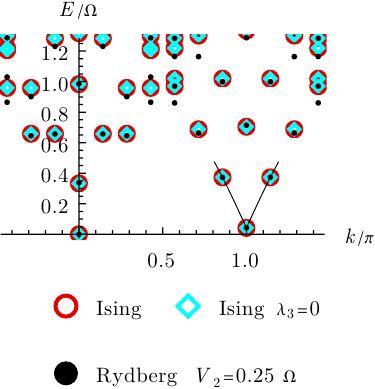}} 
  ~~~~~~~\subfloat[\label{fig:spectrum neg}]{\includegraphics[width=.61\columnwidth]{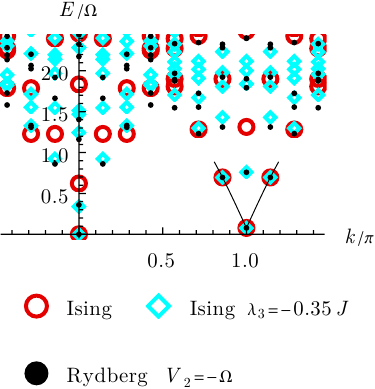}}
  \caption{%
    Excitation spectrum versus momentum for a periodic $L = 16$ Rydberg chain at Ising criticality with (a) $V_2/\Omega = 1$, (b) $V_2/\Omega = 0.25$, and (c) $V_2/\Omega = -1$.
    Overlaid for comparison are the spectra for the critical transverse-field antiferromagnetic Ising model from Eq.~\eqref{interactingTFIM} without (red circles) and with (blue diamonds) three-spin interactions, i.e.\ $\lambda_3=0$ and $\lambda_3\ne 0$ respectively. The three-spin interactions produce $T \overline{T}$ interactions in the Ising CFT with coefficient $u \propto \lambda_3$; choosing $\lambda_3$ that best matches the low-energy part of the Rydberg spectrum allows one to infer the evolution of $T \overline{T}$ interactions in the critical Rydberg chain.  Blue diamonds were obtained with the optimal $\lambda_3$ and indicate that the Rydberg chain exhibits four-fermion interactions with $u>0$ in (a), $u \approx 0$ in (b), and $u< 0$ in (c), as illustrated by the color coding of the continuous Ising line in Fig.~\ref{fig:phases}.
  }\label{fig:spectrum}
\end{figure*}

Figure~\ref{fig:spectrum}, black dots, shows the excitation spectrum for a critical Rydberg chain with (a) $V_2 = \Omega$, (b) $V_2 = 0.25 \Omega$, and (c) attractive $V_2 = -\Omega$.  
Red dots once again correspond to the critical antiferromagnetic transverse-field Ising model, $H_{\rm TFIM}^{J = h_x}$.
Comparing the black and red spectra near 0 and $\pi$ momentum in (a), we see that the excitation energies are enhanced for the Rydberg chain relative to the non-interacting Ising model, as expected if the repulsive $V_2$ delivers a $uT \overline{T}$ interaction with $u>0$.  
In (b) the two spectra agree fairly well, suggesting $u \approx 0$, while in (c) the Rydberg chain excitation energies are reduced, as expected for $u<0$.  

To probe further, we perturb the critical TFIM via
\begin{align}
 H= H_{\rm TFIM}^{J = h_x} + \lambda_3 \sum_j (X_{j-1} Z_j Z_{j+1} + Z_{j-1} Z_j X_{j+1}).
 \label{interactingTFIM}
\end{align}
Ref.~\onlinecite{Fendley2018} introduced the ferromagnetic counterpart of Eq.~\eqref{interactingTFIM}, motivated in part by connections to supersymmetry.
Despite the rather different underlying microscopics, this interacting model and the Rydberg-chain Hamiltonian are expected to display common low-energy properties.
The interaction term preserves self-duality---thereby precluding explicit mass generation---but, upon expanding in terms of low-energy Majorana-fermion fields, produces a $uT \overline{T}$ term in the CFT with $u = 512 \lambda_3$ \cite{Aasen2020}. Indeed, for a suitable value of $\lambda_3$, one recovers the tricritical Ising point. 
We can thereby quantitatively estimate the strength of $u T\overline{T}$ interactions in the critical Rydberg chain by deducing the $\lambda_3$ coupling strength that yields good agreement between the low-energy spectra for the two models.

The green data points in Fig.~\ref{fig:spectrum} were obtained from Eq.~\eqref{interactingTFIM}  using (a) $\lambda_3 = 0.26J$, (b) $\lambda_3 = 0$, and (c) $\lambda_3 = -0.35J$.
In all three cases the low-energy parts of the spectra indeed match that of the corresponding Rydberg chain quite well.  
As the energy increases, departures become more significant. The discrepancies can be attributed to additional irrelevant interactions that we did not consider, e.g., corrections to linear dispersion. 
Figure~\ref{fig:spectrum} thus substantiates the qualitative arguments provided earlier: Moving along the critical Ising line engenders $u T \overline{T}$ interactions with $u>0$ along the repulsive $V_2>0$ direction and $u<0$ along the attractive $V_2 < 0$ direction, with $u$ vanishing near $V_2 = 0.25 \Omega$.  
The color coding of the second-order line in Fig.~\ref{fig:phases} illustrates this dependence.

\begin{figure}[t]
  \includegraphics[width=.8\columnwidth]{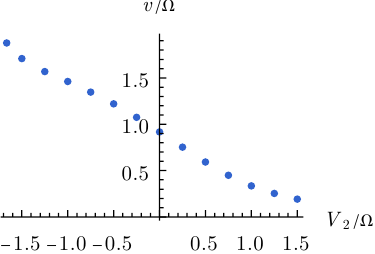} \\
  \caption{%
    Fermion velocity $v$ in the low-energy Ising CFT Hamiltonian [Eq.~\eqref{H_CFT}] versus $V_2$ for a critical Rydberg chain.  
    The velocity is calculated from the slope of the lines near the $\ket{\sigma}$ state in \figref{fig:spectrum},
      but for a larger system size with $L=28$ sites
      (for which finite-size effects are negligible.)
  }\label{fig:velocity}
\end{figure}

The dimensionless interaction strength in the Ising CFT is set by $\tilde u \equiv u \Lambda^2/v$, where $\Lambda$ is a momentum cutoff and $v$ is the fermion velocity.  
When $\tilde u$ becomes of order unity, the nominally irrelevant four-fermion interactions can induce non-perturbative effects (as indeed happens upon approaching the tricritical Ising point).  
If microscopic terms responsible for $uT \overline{T}$ interactions \emph{also} sharply enhance the velocity $v$, then $\tilde u$ can remain small even with superficially `strong' interactions.  
Such a scenario plays out in certain interacting self-dual Majorana chains reviewed in Ref.~\onlinecite{Rahmani2019}, for which dramatic upward velocity renormalization suppresses interaction effects except at extremely strong microscopic fermion interaction strengths \cite{Aasen2020}.
Downward renormalization of $v$ would instead promote non-perturbative interaction effects.  
To investigate velocity renormalization effects in the Rydberg chain, we extract $v$ from the lowest-lying excitations near momentum $\pi$ at various $V_2$ values along the continuous Ising transition line.  
More precisely, in Figs.~\ref{fig:spectrum0} and \ref{fig:spectrum}, $v$ follows from the slope of the solid lines emanating from $k = \pi$; as noted above, the associated energies are not influenced by $u$, and thus this procedure backs out the velocity present in the non-interacting part of the Hamiltonian.  
Figure~\ref{fig:velocity} shows the resulting velocity $v$ as a function of $V_2$.  
Over the $V_2$ range shown, $v$ varies by roughly an order of magnitude.  
Perhaps most notably, the reduction in $v$ at $V_2>0$ is expected to boost $u T\overline{T}$ interaction effects in the physically relevant repulsive regime.

\section{Open Rydberg chains}
\label{sec:open_chain}

\subsection{Critical correlations induced by open boundary conditions}
\label{sec:open_chain_correlations}

In previous sections, we either assumed an infinite chain or (in our numerics) invoked periodic boundary conditions.
Although periodic boundary conditions could be realized by arranging the atoms in a circle,
  finite chains with open boundary conditions are more naturally accessible to Rydberg array experiments.  
Our goal here is to quantify how open boundaries affect correlations of microscopic Rydberg chain operators at Ising criticality.

Open boundaries explicitly break the translation symmetry $T_x$ that distinguishes the CDW and trivial phases; i.e., the edges act as $\mathbb{Z}_2$ symmetry-breaking fields.  
Thus only time reversal $\mathcal{T}$ and reflection remain as good symmetries. 
The latter is site-centered ($R_x$) for $L$ odd and bond-centered ($R_x' \equiv R_x T_x^{-1}$) for $L$ even, leading to a pronounced even-odd effect in system size as we will see below.  
This reduction in symmetry injects considerable nuance into the problem.  
Edge effects cause $\hat \sigma_j$ and the field $\sigma$ to acquire a nonzero, position-dependent expectation value in the ground state even along the continuous Ising transition line in Fig.~\ref{fig:phases}.
Moreover, CFT self-duality changes the boundary conditions and therefore is broken here. Since this duality swaps $\sigma\leftrightarrow \mu$ and sends $\varepsilon \rightarrow - \varepsilon$, its breaking implies that $\varepsilon$ also takes on a nonzero, position-dependent ground-state expectation value.  
The loss of translation symmetry generically renders expectation values of the fermion kinetic energies $T, \overline{T}$ (among other operators) position-dependent as well.

Open boundary conditions further non-universally amend the link between lattice operators and CFT fields.
Under the appropriate reflection, $R_x$ or $R_x'$, Table~\ref{SymmetriesTable} implies that the fields $\sigma, \varepsilon$ transform as
\begin{align}
  {\rm Reflection}: \quad 
  \begin{aligned} \sigma(x) &\rightarrow (-1)^{L+1} \sigma(L-x)\ ,
 \cr
  \varepsilon(x) &\rightarrow\ \varepsilon(L-x)\ .
  \end{aligned}
 \label{Reflection_sigma} 
\end{align}
Enforcing only $\mathcal{T}$ and reflection symmetries, we obtain the following generalization of Eq.~\eqref{n_expansion}:
\begin{equation}
  n_j \sim c_{I,j} + c_{\sigma,j}(-1)^j  \sigma + c_{\varepsilon,j} \varepsilon + \cdots,
  \label{n_general}
\end{equation}
where all coefficients are real and satisfy $c_{\alpha, j} = c_{\alpha,L-j+1}$ for $\alpha = I,\sigma,\varepsilon$.
Sufficiently far from the edges, these position-dependent coefficients must tend to uniform values appropriate for a translation-invariant system.  
Here we will boldly postulate that the $c_{\alpha,j}$'s are uniform \emph{throughout} the chain and simply replace $c_{\alpha,j} \rightarrow c_\alpha$ in what follows.  
Equation~\eqref{n_general} then reduces to the form in Eq.~\eqref{n_expansion}; however, $c_I$ should not be interpreted as the mean Rydberg occupation number since $\sigma, \varepsilon$ take on nonuniform expectation values.  
To isolate $\sigma$ or $\varepsilon$ in this case, it is useful to consider variations on Eqs.~\eqref{sigma_lattice} and \eqref{epsilon_lattice} that do not reference the (now position-dependent) mean Rydberg occupation number.   
In particular, we utilize a bond-centered CDW order parameter
\begin{align}
  \hat \sigma^\text{bond}_{j+1/2} \equiv (-1)^j(n_{j} - n_{j+1}) \sim 2c_\sigma \sigma + \cdots
  \label{sigmaprime}
\end{align}
and define 
\begin{align}
  \hat \varepsilon^\text{bare}_{j+1/2} &\equiv n_{j} + n_{j+1} \sim 2c_I + 2 c_\varepsilon \varepsilon +  c_\sigma (-1)^j \partial_x \sigma + \cdots.
  \label{epsilonprime}
\end{align}
We displayed the subleading $\partial_x \sigma$ term since including it substantially improves agreement with the numerics below.
Note that to isolate $\varepsilon$ to leading order, we need to consider $\hat \varepsilon^\text{bare}_{j+1/2} - 2c_I$.
With Eqs.~\eqref{sigmaprime} and \eqref{epsilonprime} in hand, computing correlation functions in the CFT allows us to back out physical correlations of microscopic Rydberg-chain operators.

\begin{table*}[ht]
\centering
\begin{tabular}{c | c} 
 \hline
  CFT correlators for odd $L$ & CFT correlators for even $L$ \\ 
 \hline
 $\braket{\sigma(x)}_{(+)} = (2/\sin x)^{1/8}$ &
    $\braket{\sigma(x)}_{(-)} = \cos x \braket{\sigma(x)}_{(+)}$\\
    $\braket{\sigma(x) \sigma(y)}_{(+)}
      =  \left( \rho^{1/4} + \rho^{-1/4} \right)^{1/2}/(4 \sin x \sin y)^{1/8}$ &
    $\braket{\sigma(x) \sigma(y)}_{(-)}
      = \left( 1 - |\cos x - \cos y| \right) \braket{\sigma(x) \sigma(y)}_{(+)}$\\    
      $\braket{\varepsilon(x)}_{(+)} = 1/\left( 2\sin x \right)$ &
    $\braket{\varepsilon(x)}_{(-)}
      = \braket{\varepsilon(x)}_{(+)} - \braket{\varepsilon(x)}_{(+)}^{-1}$ \\
      $\braket{\varepsilon(x) \varepsilon(y)}_{(+),\text{c}}
      = (\sin x \sin y)/(\cos x - \cos y)^2$ &
    $\braket{\varepsilon(x) \varepsilon(y)}_{(-),\text{c}}
      = \braket{\varepsilon(x) \varepsilon(y)}_{(+),\text{c}} - 4 \sin x \sin y$ \\
      \hline
\end{tabular}
 \caption{%
   One-point and equal-time two-point CFT $\sigma$ and $\varepsilon$ correlation functions for boundary conditions appropriate for odd $L$ [left column, labeled $(+)$] and even $L$ [right column, labeled $(-)$].  
   In the second row $ \rho(x,y) = \left[ \sin\left(\frac{x+y}{2}\right)/\sin\left(\frac{x-y}{2}\right) \right]^2$, and in the last row the `$c$' subscript indicates a connected correlator, e.g.,  $\braket{\varepsilon(x) \varepsilon(y)}_{(+),\text{c}}
      \equiv \braket{\varepsilon(x) \varepsilon(y)}_{(+)}
            - \braket{\varepsilon(x)}_{(+)} \braket{\varepsilon(y)}_{(+)}$.  
  }\label{tab:boundaryCFT}
\end{table*}

Open boundaries act as $\mathbb{Z}_2$ symmetry breaking fields, as noted above, that impose fixed boundary conditions
\begin{equation}
  \big\langle \sigma(x = 0)\big\rangle = (-1)^{L+1} \big\langle \sigma(x = L)\big\rangle \neq 0.
  \label{BC}
\end{equation}
The $(-1)^{L+1}$ factor on the right side follows from reflection symmetry [Eq.~\eqref{Reflection_sigma}].
In \appref{app:CFT}, we review the CFT calculation for one-point and equal-time two-point $\sigma$ and $\varepsilon$ correlation functions subject to fixed boundary conditions; \tabref{tab:boundaryCFT} summarizes the results.
For convenience, the correlators listed there are evaluated with space rescaled such that the chain lives on the interval $0 \leq x \leq \pi$.
When the positions are close to the middle of the chain,
  then the connected two-point correlators reproduce the periodic lattice correlators to leading order:
  $\braket{\sigma(x) \sigma(y)}_{(\pm),c} \approx (\delta x - \delta y)^{-1/4}$ and
  $\braket{\epsilon(x) \epsilon(y)}_{(\pm),c} \approx (\delta x - \delta y)^{-2}$
  where $x=\frac{\pi}{2} + \delta x$ and $y=\frac{\pi}{2} + \delta y$
  with $|\delta x| \ll 1$ and $|\delta y| \ll 1$
  (which can only be achieved in the long chain limit).

\begin{figure}
  \includegraphics[width=\columnwidth]{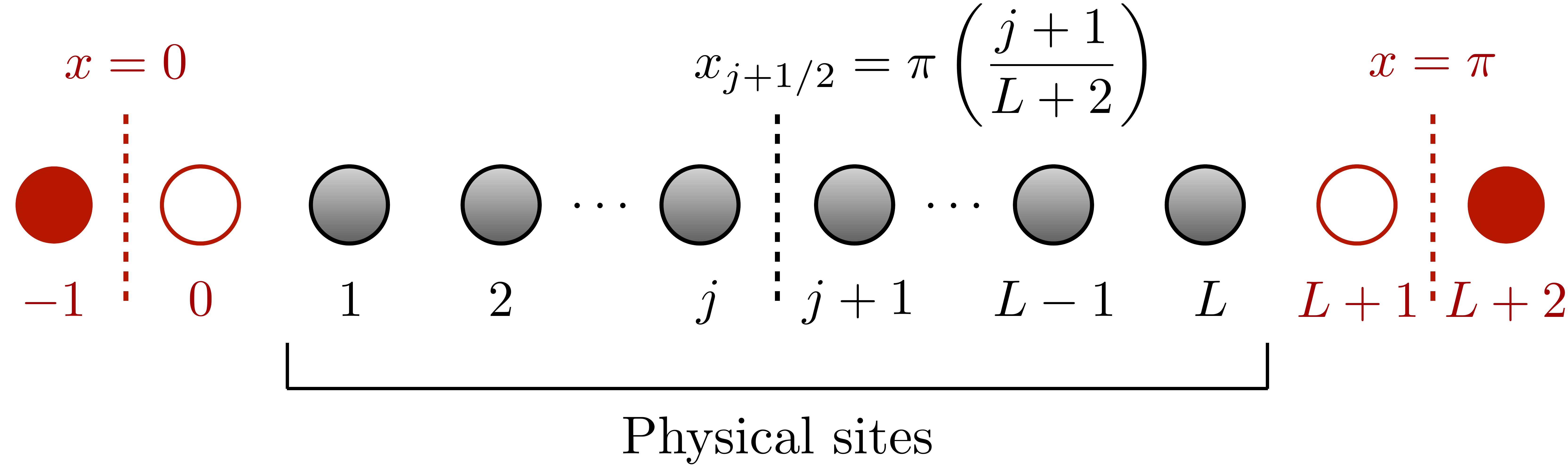}
  \caption{%
    Physical sites of an open Rydberg chain augmented by a pair of fictitious boundary sites (red) on each end.      
    Solid (open) augmented sites are parked into $n = 1$ ($n = 0$) states, thereby seeding CDW order into the chain from the edges.  
    The rescaled boundary coordinates $x = 0$ and $x = \pi$ in the CFT respectively correspond to the left and right fictitious bond; the continuum coordinate $x_{j+1/2}$ associated with bond $j+1/2$ is then given by Eq.~\eqref{eq:xi}.
  }\label{fig:boundaries}
\end{figure}

To compare these CFT results with lattice numerics, we must relate the continuum position $x$ used in the CFT to lattice coordinates.  A subtlety occurs for 
the leftmost and rightmost bonds of the chain. They cannot correspond to the positions $x=0$ and $\pi$ since, according to \tabref{tab:boundaryCFT}, $\langle\sigma\rangle$ diverges there.
We therefore augment each end of the open chain with an extra (ficitious) pair of sites---labeled $j = -1,0$ on the left side and $j = L+1,L+2$ on the right---that seed CDW order into the system from the edges.  
We park these auxiliary sites into fixed configurations $n_{-1} = 1, n_{0} = 0$ and $n_{L+1} = 0, n_{L+2} = 1$ as illustrated in \figref{fig:boundaries}.
Importantly, this assignment preserves reflection symmetry, and in the $V_2=0$ limit does not affect the Hamiltonian for the physical sites.
Continuum coordinates $x = 0, \pi$ are then associated with the outermost bonds of the \emph{enlarged} $(L+4)$-site system.  
The physical bond $j+1/2$ (with $j = 1,\ldots,L-1$) of the chain thereby corresponds to a continuum coordinate
\begin{equation}
  x_{j+1/2} = \pi \left(\frac{j+1}{L+2}\right). \label{eq:xi}
\end{equation}
Note that this change of coordinates rescales the CFT fields in \tabref{tab:boundaryCFT} by $\phi \to \phi \big(\frac{\pi}{L+2}\big)^{-\Delta_\phi}$
  [\eqnref{eq:conformalTransformation}],
  where $\Delta_\sigma=1/8$ and $\Delta_\varepsilon = 1$ for $\phi=\sigma,\varepsilon$.
This rescaling is necessary for the coefficients $c_\phi$ with $\phi=I,\sigma,\varepsilon$ to asymptote to a constant as $L \to \infty$.

\begin{figure*}
  \subfloat[\label{fig:sigmaOdd}]{\includegraphics[width=.73\columnwidth]{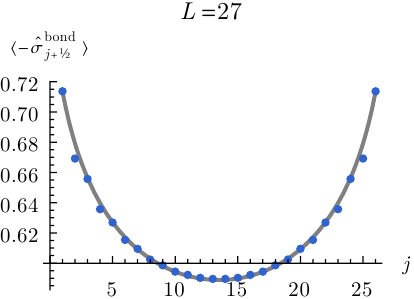}} \hspace{1cm}
  \subfloat[\label{fig:sigmaEven}]{\includegraphics[width=.73\columnwidth]{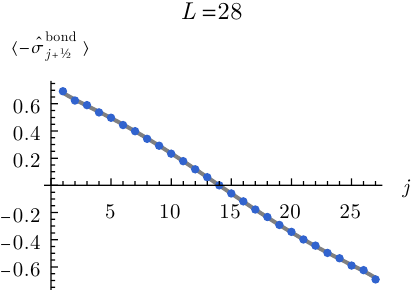}} \\
  \subfloat[\label{fig:sigmaSigmaOdd}]{\includegraphics[width=.73\columnwidth]{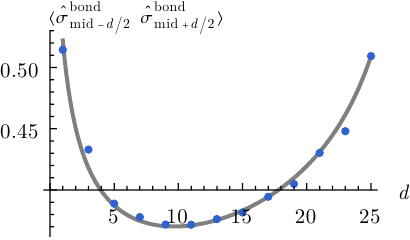}} \hspace{1cm}
  \subfloat[\label{fig:sigmaSigmaEven}]{\includegraphics[width=.73\columnwidth]{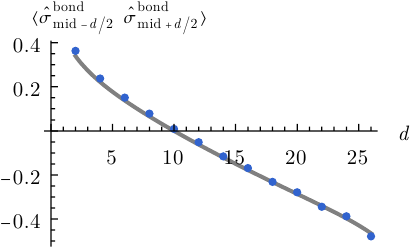}} \\
  \subfloat[\label{fig:sigmaSigmaCOdd}]{\includegraphics[width=.73\columnwidth]{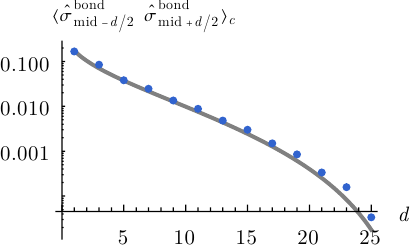}} \hspace{1cm}
  \subfloat[\label{fig:sigmaSigmaCEven}]{\includegraphics[width=.73\columnwidth]{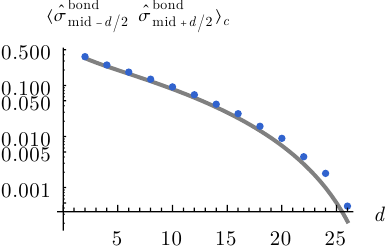}} \\
  \caption{%
    (a,b) One-point and (c-f) two-point correlators of the microscopic bond-centered CDW order parameter $\hat \sigma^{\rm bond}_{j+1/2}$ [Eq.~\eqref{sigmaprime}] for open Rydberg chains at Ising criticality with $V_2 = 0$.
    Panels (e,f) display connected two-point correlators.       
    Blue points result from exact diagonalization of open $L=27$ (left column) and $L=28$ (right column) Rydberg chains.
    Gray lines are obtained from the low-energy expansion in \eqnref{sigmaprime} and the CFT expressions in \tabref{tab:boundaryCFT},
      with a single fitting parameter $c_\sigma$ for each $L$.
  }\label{fig:sigmaBoundary}
\end{figure*}

\begin{figure*}
  \subfloat[\label{fig:epsOdd}]{\includegraphics[width=.73\columnwidth]{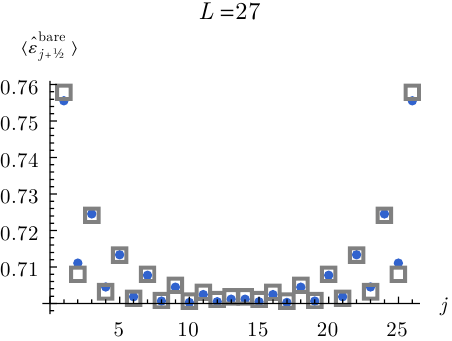}} \hspace{.5cm}
  \subfloat[\label{fig:epsEven}]{\includegraphics[width=.73\columnwidth]{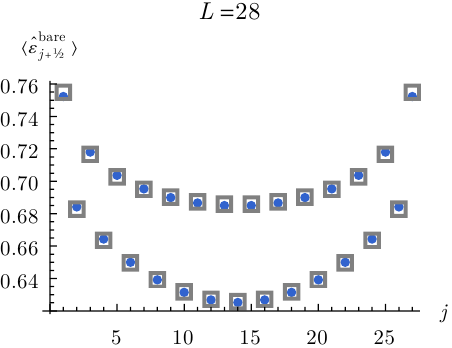}} \\ 
  \subfloat[\label{fig:epsEpsCOdd}]{\includegraphics[width=.73\columnwidth]{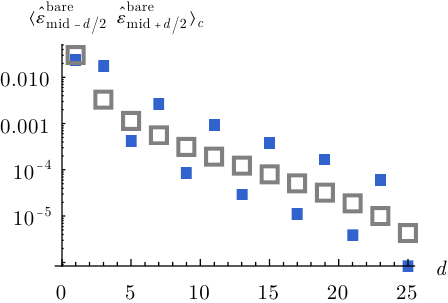}} \hspace{.5cm}
  \subfloat[\label{fig:epsEpsCEven}]{\includegraphics[width=.73\columnwidth]{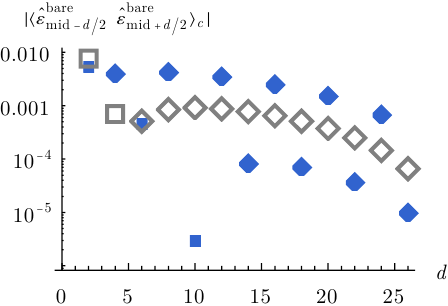}}
  \caption{%
    (a,b) One-point and (c,d) connected two-point correlators of the microscopic operator $\hat \varepsilon^{\rm bare}_{j+1/2}$ [Eq.~\eqref{epsilonprime}] for open Rydberg chains at Ising criticality with $V_2 = 0$.
    Blue points are exact diagonalization data for $L=27$ (left column) and $L=28$ (right column) Rydberg chains, and gray squares
    are fits obtained from the low-energy expansion in Eq.~\eqref{epsilonprime} and the CFT expressions in \tabref{tab:boundaryCFT}.
    For the fits in (a,b), we use the same $c_\sigma$ as in \figref{fig:sigmaBoundary} along with two additional free parameters $c_I$ and $c_\varepsilon$ for each $L$.
    For (c,d) we use the same $c_I,c_\varepsilon$ fitting parameters from (a,b) but set $c_\sigma = 0$ for simplicity; we thus do not capture the oscillatory structure, though the fits nevertheless track the exact diagonalization data fairly well.
    In (d), the blue and gray diamonds (squares) are used to specify that $\braket{\varepsilon \varepsilon}_c$ is negative (positive).
  }\label{fig:epsBoundary}
\end{figure*}

\begin{figure*}
  \subfloat[\label{fig:sigmaOddV}]{\includegraphics[width=.8\columnwidth]{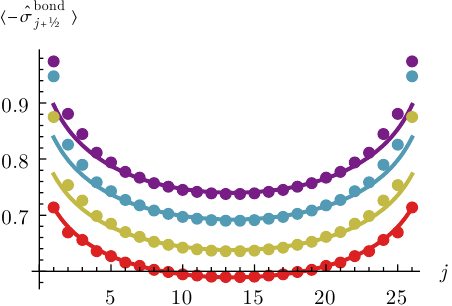}} 
  \subfloat[\label{fig:sigmaSigmaCOddV}]{\includegraphics[width=\columnwidth]{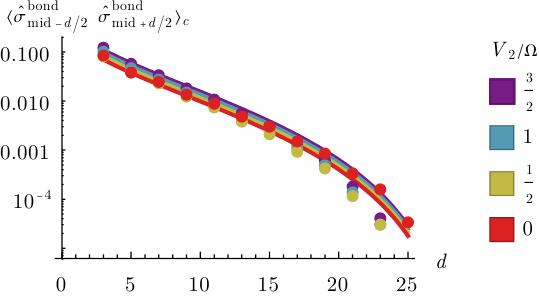}}
  \caption{%
  (a) One-point and (b) two-point connected correlators of $\hat \sigma^{\rm bond}_{j+1/2}$ [Eq.~\eqref{sigmaprime}] for an open $L = 27$ Rydberg chains at Ising criticality with varying second-neighbor repulsion $V_2\geq 0$.
    Dots are exact diagonalization data while lines are fits to CFT predictions, similar to \figref{fig:sigmaBoundary}, with a single free parameter $c_\sigma$ for each $V_2$.
  Second-neighbor repulsion boosts the one-point correlator and produces a sharper upturn at the edges, but the data nevertheless continue to qualitatively (and for the most part quantitatively) track CFT predictions.
  }\label{fig:sigmaBoundaryV}
\end{figure*}

We are now in position to evaluate correlators of microscopic Rydberg operators.  
Blue data points in Figs.~\ref{fig:sigmaBoundary} and \ref{fig:epsBoundary} present $\sigma^\text{bond}_{j+1/2}$ and $\hat \varepsilon^\text{bare}_{j+1/2}$ correlators obtained using exact diagonalization for $L = 27$ (left columns) and $L = 28$ (right columns) with $V_2 = 0$.  
These lattice results can now be compared with the CFT results using Eqs.~(\ref{sigmaprime}) and (\ref{epsilonprime}) and replacing $x$ with $x_{j+1/2}$ given in \eqnref{eq:xi}; for example, $\hat\sigma^\text{bond}_{j+1/2} \sim 2 c_\sigma \sigma(x_{j+1/2})$.
Overlaid in gray in Figs.~\ref{fig:sigmaBoundary} and \ref{fig:epsBoundary} are fits to the corresponding CFT formulas with $c_{I,\varepsilon,\sigma}$ as three fitting parameters, one set for each system size.   
$c_\sigma$ is obtained by fitting to $\langle \hat{\sigma}^\text{bond}_{j+1/2} \rangle$
  in \sfigref{fig:sigmaBoundary}{a,b} (separately for $L=27$ and $L=28$).
The same $c_\sigma$ is used to also obtain $c_I$ and $c_\varepsilon$ by fitting to $\langle \hat{\varepsilon}^\text{bare}_{j+1/2} \rangle$
  in \sfigref{fig:epsBoundary}{a,b}.
For the connected two-point $\hat \varepsilon^\text{bare}_{j+1/2}$ correlators,
  we set $c_\sigma = 0$ for simplicity since we do not have CFT expressions for $\langle \sigma \varepsilon \rangle$;
  we thus do not capture the  $\langle \sigma \varepsilon \rangle$ cross terms that are responsible for the zigzagging of blue data points in \sfigref{fig:epsBoundary}{c,d}.

The agreement with CFT predictions is rather striking and supports the validity of our treatment that approximated the coefficients in Eq.~\eqref{n_general} as position independent.
Notice that edge effects induce $O(1)$ expectation values for both the CDW order parameter $\hat \sigma^\text{bond}_{j+1/2}$ and $\hat \varepsilon^\text{bare}_{j+1/2}$.  
As Fig.~\ref{fig:sigmaBoundaryV} illustrates for $L = 27$, turning on second-neighbor repulsion ($V_2>0$) in the open chain further boosts the CDW order parameter and yields a sharper upturn at the edges.  
The fits represented by solid lines nevertheless continue to demonstrate good agreement between numerics and CFT predictions for both one-point and two-point $\hat \sigma^\text{bond}_{j+1/2}$ correlators, provided the operators are not within a few lattice sites of the boundary.

\begin{figure}
  \includegraphics[width=.95\columnwidth]{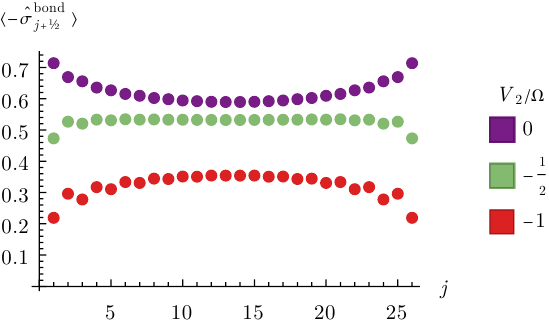}
  \caption{%
  Exact diagonalization results for $\langle -\hat \sigma^{\rm bond}_{j+1/2}\rangle$ [Eq.~\eqref{sigmaprime}] in an open $L = 27$ Rydberg chains at Ising criticality with varying second-neighbor attraction $V_2 \leq 0$.
  The downward curvature appearing at $V_2<0$ disagrees qualitatively with CFT predictions for $\langle \sigma\rangle$, signaling nonuniversal boundary physics induced by second-neighbor attraction.  These boundary effects can be offset by modifying the detuning on the outermost sites; see main text and Fig.~\ref{fig:boundary addV}.  
  }\label{fig:boundary negV}
\end{figure}

While less physically relevant, it is instructive to explore the effects of second-neighbor attraction ($V_2<0$) on open-chain correlations.  
Figure~\ref{fig:boundary negV} shows the evolution of the microscopic CDW order parameter for $L = 27$ with increasing second-neighbor attraction.
Two trends appear: attraction suppresses $\langle-\hat \sigma^\text{bond}_{j+1/2}\rangle$ throughout the chain and produces a \emph{downturn} in the expectation value at the edges.  
The latter feature stands in stark contrast to the upturn present both in our simulations with $V_2\geq 0$ and in the CFT calculation of $\langle \sigma \rangle$ with fixed boundary conditions---suggesting the emergence of nonuniversal boundary physics.

Revisiting the enlarged $(L+4)$-site open chain provides further insight into this boundary conundrum.  
We expect that the $L$ physical sites in the center conform best to fixed-boundary-condition CFT predictions when one starts from an enlarged open chain governed by a uniform Hamiltonian and \emph{then} freezes the outer auxiliary sites to seed CDW order.  
Displaying only the $\Delta$ and $V_2$ terms for the first three sites in the enlarged chain, the uniform Hamiltonian is
\begin{equation}
  H_{\rm enlarged} = -\Delta(n_{-1} + n_0 + n_1) + V_2 n_{-1} n_1 + \cdots.
\end{equation}
Projection of the auxiliary sites to $n_{-1} = 1$ and $n_0 = 0$ yields (up to a constant)
\begin{equation}
  H_{\rm enlarged} \rightarrow -(\Delta-V_2)n_1 + \cdots. \label{eq:addV}
\end{equation}
When $V_2 = 0$, we see that the resulting effective Hamiltonian for the $L$ physical sites is unmodified by the frozen auxiliary sites as noted earlier.  
When $V_2 \neq 0$, however, the outermost frozen auxiliary sites shift the detuning on physical sites $1,L$ from $\Delta$ to $\Delta - V_2$. 
This line of reasoning suggests that the CFT analysis more naturally describes a chain with nonuniform detuning given by $\Delta$ in sites $2,\ldots,L-1$ and $\Delta - V_2$ in sites $1,L$.
As one sanity check, simulations of a chain with uniform detuning (as we carried out above) would overshoot the optimal $\Delta$ in the outer sites for $V _2>0$ but undershoot the optimal $\Delta$ for $V_2<0$.
Figure~\ref{fig:phases} implies that overshooting and undershooting moves the edges locally toward the ordered and disordered phases, respectively; one would then expect enhanced edge CDW order for $V_2>0$ but suppressed edge CDW order for $V_2<0$, precisely as seen in Figs.~\ref{fig:sigmaBoundaryV} and \ref{fig:boundary negV}.  

\begin{figure*}[t]
  \subfloat[\label{fig:sigmaOddAddV}]{\includegraphics[width=.8\columnwidth]{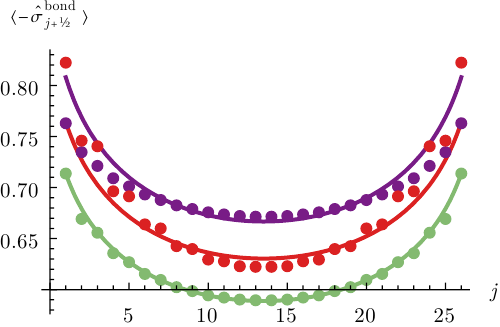}} \hspace{.9cm}
  \subfloat[\label{fig:epsOddAddV}]{\includegraphics[width=\columnwidth]{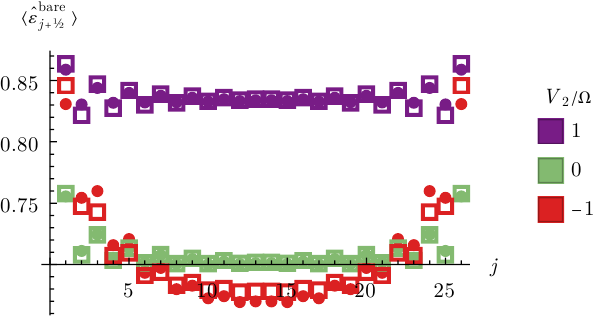}}
  \caption{%
   One-point (a) $\hat \sigma^{\rm bond}_{j+1/2}$ and (b) $\varepsilon^{\rm bare}_{j+1/2}$ correlators for an open, critical $L = 27$  Rydberg chain with detuning shifted from $\Delta\rightarrow \Delta-V_2$ on sites $1$ and $L$.
   Dots represent exact diagonalization data; lines in (a) and open squares in (b) are fits to CFT predictions, similar to Figs.~\ref{fig:sigmaBoundary}-\ref{fig:sigmaBoundaryV}, with three fitting parameters ($c_\sigma$, $c_I$, $c_\varepsilon$) for each $V_2$.
  Panel (a) demonstrates that the shifted detuning on the outermost sites counteracts the nonuniversal boundary effects visible in \figref{fig:boundary negV}, resulting in upward curvature for all $V_2$ shown.
  Reasonable agreement with CFT predictions then follows in both (a) and (b).  
The data in (b) flatten considerably upon varying $V_2$ from attractive to repulsive; see also Fig.~\ref{cVsV}.
  }\label{fig:boundary addV}
\end{figure*}

For additional support, Fig.~\ref{fig:boundary addV} (dots) presents one-point $\hat \sigma^\text{bond}_{j+1/2}$ and $\hat \varepsilon^\text{bare}_{j+1/2}$ correlators for $L = 27$ with detuning for sites $1,L$ shifted to $\Delta-V_2$.  
The characteristic upturn in the CDW order parameter predicted by the CFT is now evident for both repulsive \emph{and} attractive $V_2$.
Moreover, the numerical data can be reasonably fit to the CFT for both signs of $V_2$ as demonstrated by the solid lines [Fig.~\ref{fig:boundary addV}(a)] and squares [Fig.~\ref{fig:boundary addV}(b)].
Still better fits may be possible if one treats the detuning on sites $1,L$ as adjustable parameters, though we will not go down that route for the sake of simplicity.

\begin{figure}[ht]
  \includegraphics[width=0.8\columnwidth]{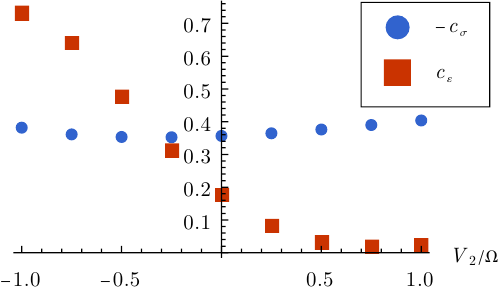}
  \caption{%
    Optimal $c_\sigma$ and $c_\varepsilon$ fitting parameters (as obtained in \figref{fig:boundary addV})
     for an open, critical $L=27$ Rydberg chain with detuning shifted from $\Delta\rightarrow \Delta-V_2$ on sites $1$ and $L$.
     Whereas $c_\sigma$ varies modestly over the $V_2/\Omega$ window shown, $c_\varepsilon$ varies by more than a factor of 30.  
     The latter variation relates to the flattening of the curves in Fig.~\ref{fig:boundary addV}(b) as $V_2$ varies from attractive to repulsive, and reflects four-fermion interactions in the Ising CFT (Sec.~\ref{TTbar_sec}).  
  }\label{cVsV}
\end{figure}

Finally, inspection of Fig.~\ref{fig:boundary addV}(b) reveals a curious feature: Upon changing $V_2$ from attractive to repulsive, the edge-induced $\hat \varepsilon^\text{bare}_{j+1/2}$ expectation value flattens considerably [contrary to the CDW order parameter in Fig.~\ref{fig:boundary addV}(a)].  
In fact at $V_2/\Omega = +1$ the dominant source of spatial variation by far originates from the $c_\sigma(-1)^j \partial_x\sigma$ contribution to Eq.~\eqref{epsilonprime} rather than the more relevant $2c_\varepsilon \varepsilon$ piece.
For a deeper look, Fig.~\ref{cVsV} plots the optimal $c_\sigma$ and $c_\varepsilon$ fitting parameters for second-neighbor interaction ranging from $V_2/\Omega = -1$ to $V_2/\Omega = +1$.  
While $c_\sigma$ varies modestly over this range, $c_\varepsilon$ \emph{changes by more than a factor of 30}.
Thus second-neighbor interactions effectively freeze out the contribution from the CFT field $\varepsilon$ for $V_2>0$ but enhance its contribution for $V_2<0$.

This behavior arises naturally from the four-fermion interactions---$uT \overline{T}$ in Eq.~\eqref{H_CFT}---analyzed in Sec.~\ref{TTbar_sec}. This analysis suggests that the one-point $\varepsilon$ correlator presented in Table~\ref{tab:boundaryCFT} would be reduced by four-fermion interactions generated with second-neighbor repulsion but enhanced with second-neighbor attraction.  
Namely, as expressed on the right side of Eq.~\eqref{u_rewriting}, $uT \overline{T}$ interactions clearly either promote or suppress the one-point $\langle \varepsilon\rangle$ correlator induced by fixed boundary conditions in the CFT, depending on the sign of $u$.
Suppose that $\langle \varepsilon\rangle_{\rm exact}$ denotes the exact correlator including $u T \overline{T}$ effects.  
Let us further assume that $\langle \varepsilon\rangle_{\rm exact} = \kappa \langle \varepsilon\rangle_{\rm free}$, where $\langle \varepsilon\rangle_{\rm free}$ is the result from Table~\ref{tab:boundaryCFT} that neglected interactions, and $\kappa$ is a scale factor that varies along the continuous Ising line.
Equation~\eqref{epsilonprime} would then ideally yield 
\begin{align}
\big\langle \hat \varepsilon^\text{bare}_{j+1/2}\big\rangle -2c_I &\approx 2c_\varepsilon \big\langle \varepsilon\big\rangle_{\rm exact} \nonumber\\
  &= 2c_\varepsilon\kappa \big\langle \varepsilon\big\rangle_{\rm free} \\
 &\equiv 2c^{\rm eff}_\varepsilon \big\langle \varepsilon\big\rangle_{\rm free}. \nonumber
\end{align}
Crucially, the effective parameter $c^{\rm eff}_\varepsilon$ extracted based on a fit to $\langle \varepsilon \rangle_{\rm free}$---as we pursued in this section---implicitly contains the scale factor $\kappa$ reflecting interaction effects.  
(In the notation from this paragraph, Fig.~\ref{cVsV} actually displays $c^{\rm eff}_\varepsilon$.)
The dramatic evolution of $\langle \hat \varepsilon^\text{bare}_{j+1/2}\rangle$ observed in our open-chain simulations thus can be viewed as an interaction effect in the effective CFT description given the variation of $u$ with $V_2$ along the Ising transition line \footnote{Technically, $c^{\rm eff}_\varepsilon = c_{\varepsilon} \kappa$ varies with $V_2$ due to a combination of changes in $\kappa$ \emph{and} $c_\varepsilon$.  Variation in $c_\varepsilon$ can have a trivial origin unrelated to interactions, e.g., the lattice operator $\hat \varepsilon^\text{bare}_{j+1/2}$ can have a smaller overlap with the CFT field $\varepsilon$ as $V_2$ increase simply due to curvature in the phase boundary of Fig.~\ref{fig:phases}.  We expect, however, that the latter effect is $O(1)$, in contrast to the dramatic change in $c^{\rm eff}_\varepsilon$ (again, by more than a factor of 30!) evident in Fig.~\ref{cVsV}.}.

\subsection{Locating the critical point}
\label{sec:Delta c}

In this subsection,
  we address how one could experimentally determine the critical detuning $\Delta_c$ to reach criticality.
In classical simulation of critical systems,
  critical points are typically located via a scaling collapse or curve-crossing of some rescaled observable.
For systems with periodic boundary conditions,
  popular observables include a susceptibility, correlation length, or Binder cumulant.
That is, one of these observables is plotted versus a tuning parameter (e.g., temperature or the detuning $\Delta$) for different system sizes;
  the data is then rescaled such that it collapses (for a range of tuning parameters) or crosses (at the critical point) for different large system sizes \cite{SandvikComputational}.

For open Rydberg chains, the edges explicitly break translation symmetry---yielding a charge density wave order parameter that is pinned near the boundaries and slowly decays into the bulk as seen in Fig.~\ref{fig:sigmaBoundaryV}.
It is therefore useful to consider a scheme that is optimized for open Rydberg chains.
We propose to locate the critical point by measuring a curve-crossing of an appropriately rescaled order parameter at the midpoint of an odd-$L$ chain, $\braket{\hat\sigma^\text{bond}_{L/2}}$.  
[For even-$L$ chains the order parameter vanishes by symmetry in the center; recall Fig.~\ref{fig:sigmaBoundary}(b).]
This approach leverages translation symmetry-breaking by the boundaries as a feature:
  it allows us to locate the critical point using a simple one-point correlator that is diagonal in the number basis and thus easy to measure.
  
\begin{figure}[ht]
  \subfloat[]{\includegraphics[width=0.95\columnwidth]{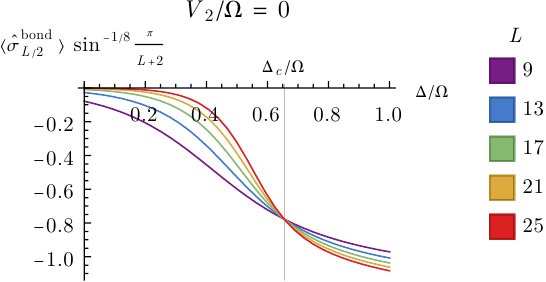}} \vspace{.5cm}\\
  \subfloat[]{\includegraphics[width=0.95\columnwidth]{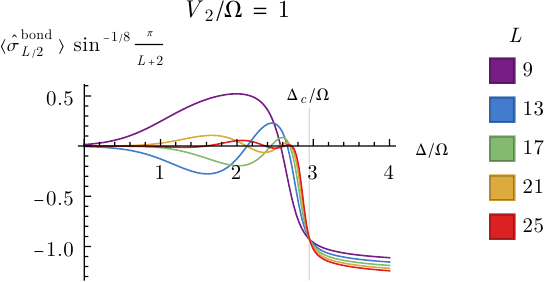}}
  \caption{%
    The midpoint correlator from  \eqnref{eq:midpointCorrelator} 
    for increasing odd-integer length $L$ versus detuning $\Delta$ for open Rydberg chains
      with (a) $V_2=0$ and (b) $V_2=\Omega$. 
    The correlator is rescaled so that the curves cross at the critical detuning $\Delta_c$ (vertical dashed lines)
  }\label{fig:scalingCollapse}
\end{figure}

For large odd-integer chain lengths $L \to \infty$,
  the midpoint correlator scales at criticality as $\braket{\hat\sigma^\text{bond}_{L/2}} \sim L^{-1/8}$.
A crude curve-crossing can therefore be obtained from $\braket{\hat\sigma^\text{bond}_{L/2}} L^{1/8}$.
This result follows because the left-most nontrivial bond,
  between the virtual site $j=0$ and physical site $j=1$,
  is fixed to an $O(1)$ value at the boundary.
From \tabref{tab:boundaryCFT}, the CFT correlator for this bond is $\braket{\sigma(x_{1/2})} = (2/\sin x_{1/2})^{1/8}$.
A more accurate curve-crossing therefore can be obtained from the following rescaling:
\begin{equation}
  \braket{\sigma^\text{bond}_{L/2}} \sin^{-1/8}(x_{1/2}), \label{eq:midpointCorrelator}
\end{equation}
where $x_{1/2} = \frac{\pi}{L+2}$ from \eqnref{eq:xi}.
In \figref{fig:scalingCollapse},
  we verify that the curves of different odd-$L$ chains with $V_2 \geq 0$ indeed cross near the critical point (vertical dashed lines)
  calculated from scaling collapses on periodic chains; notably, the crossings hold for both (a) $V_2/\Omega = 0$ and (b) $V_2/\Omega = 1$. 
We consider only length increments by 4 (rather than 2)
  because of a very slight `even-odd' effect between $L \equiv 1$ and $L \equiv 3$ (mod 4) chain lengths.


\section{Approach to tricriticality} 
\label{sec:tricriticality}

Figures~\ref{fig:spectrum}(c) and \ref{fig:boundary addV} revealed a pronounced deformation of the critical Rydberg chain spectrum induced by second-neighbor attraction ($V_2<0$) and the accompanying $uT \overline{T}$ interactions in the CFT. 
As we elucidate below, this deformation reflects proximity to the tricritical Ising (TCI) point in Fig.~\ref{fig:phases}.
Tracking the spectral evolution upon approaching tricriticality yields useful insight into the relation between Ising and TCI theories, both at the CFT and microscopic levels.

The TCI point is described by a $c = 7/10$ CFT with six primary fields of chiral dimensions 0, 3/80, 7/16, 1/10, 3/5 and 3/2 \cite{Friedan1984,Lassig1990}.
Which combinations of left- and right-moving fields are realized in the low-energy limit of a critical lattice model varies with model. We find that that the combinations of right- and left-movers appearing in the Rydberg chain must have spins---given by the difference in right- and left-moving scaling dimensions---that are either integer (yielding local bosons) or half-integer (yielding fermions).
Table~\ref{tab:isingTCI}, left column, lists the scaling dimensions of some of those we identify in the finite-size spectrum below. There are four spinless bosonic fields listed there. The   $\sigma$ field is the analog of the Ising spin field, with dimension $3/40=3/80+3/80$. The $\sigma'$ field is a less-relevant operator also breaking the $\mathbb{Z}_2$ symmetry, and is of dimension $7/8=7/16+7/16$.  The field $\varepsilon$ of dimension $1/5=1/10+1/10$ is the lowest-dimension nontrivial operator invariant under the $\mathbb{Z}_2$ symmetry. Perturbing by it moves away from the transition lines in \figref{fig:phases}, as it is odd under the CFT self-duality. The operator $\varepsilon'$ of dimension $6/5=3/5+3/5$ is self-dual and $\mathbb{Z}_2$ invariant, so perturbing by it drives the system along the transition lines, with different signs corresponding to the different directions. The fermionic field $\psi$ is the TCI analog of the Ising fermion, but a key distinction is that it is not a purely chiral operator, being of dimension $7/10 =1/10+3/5$. The other fermionic field $G$ of dimension $3/2$ is purely chiral or antichiral, although presumably what is observed on the lattice is a sum of the two. 
[See Ref.~\onlinecite{Fendley2018} for a more in-depth discussion for the Hamiltonian in \eqnref{interactingTFIM}.] 
The chiral and antichiral parts generate the left-and right-moving supersymmetries in the CFT. 

\begin{table}
  \centering
  \setlength{\tabcolsep}{0em}
  \begin{tabular}{ll|lll}
    \multicolumn{2}{c|}{TCI} & \multicolumn{3}{c}{Ising} \\ \hline
    $\Delta_\sigma$&$=3/40\;$      &$\;$& $\Delta_\sigma$&$=1/8$ \\
    $\Delta_\varepsilon$&$=1/5$    &$\;$& $\Delta_\varepsilon$&$=1$ \\
    $\Delta_\psi$&$=7/10$               &$\;$& $\Delta_\psi$&$=1/2$ \\
    $\Delta_{\sigma'}$&$=7/8$      &$\;$& $\Delta_{\sigma~{\rm desc}}$&$=1/8+2$ \\
    $\Delta_{\varepsilon'}$&$=6/5$ &$\;$& $\Delta_{T \overline{T}}$&$=4$ \\
    $\Delta_{G}$&$=3/2$                &$\;$& $\Delta_{\psi~{\rm desc}}$&$=1/2+1$
   \end{tabular}
  \caption{%
    Correspondence between tricritical Ising (TCI) and Ising CFT fields, along with their scaling dimensions.
    That is, upon moving along the continuous Ising line in Fig.~\ref{fig:phases} towards the TCI point, Ising fields in the right column evolve into the TCI fields in the same row of the left column.
    This correspondence enables us to obtain a partial dictionary linking microscopic Rydberg operators and tricritical Ising CFT as described in the main text.
  }\label{tab:isingTCI}
\end{table}

\begin{figure*}
  \vspace{1cm} 
  \subfloat[$L=28$\label{fig:evenSpec}]{\includegraphics[width=2\columnwidth]{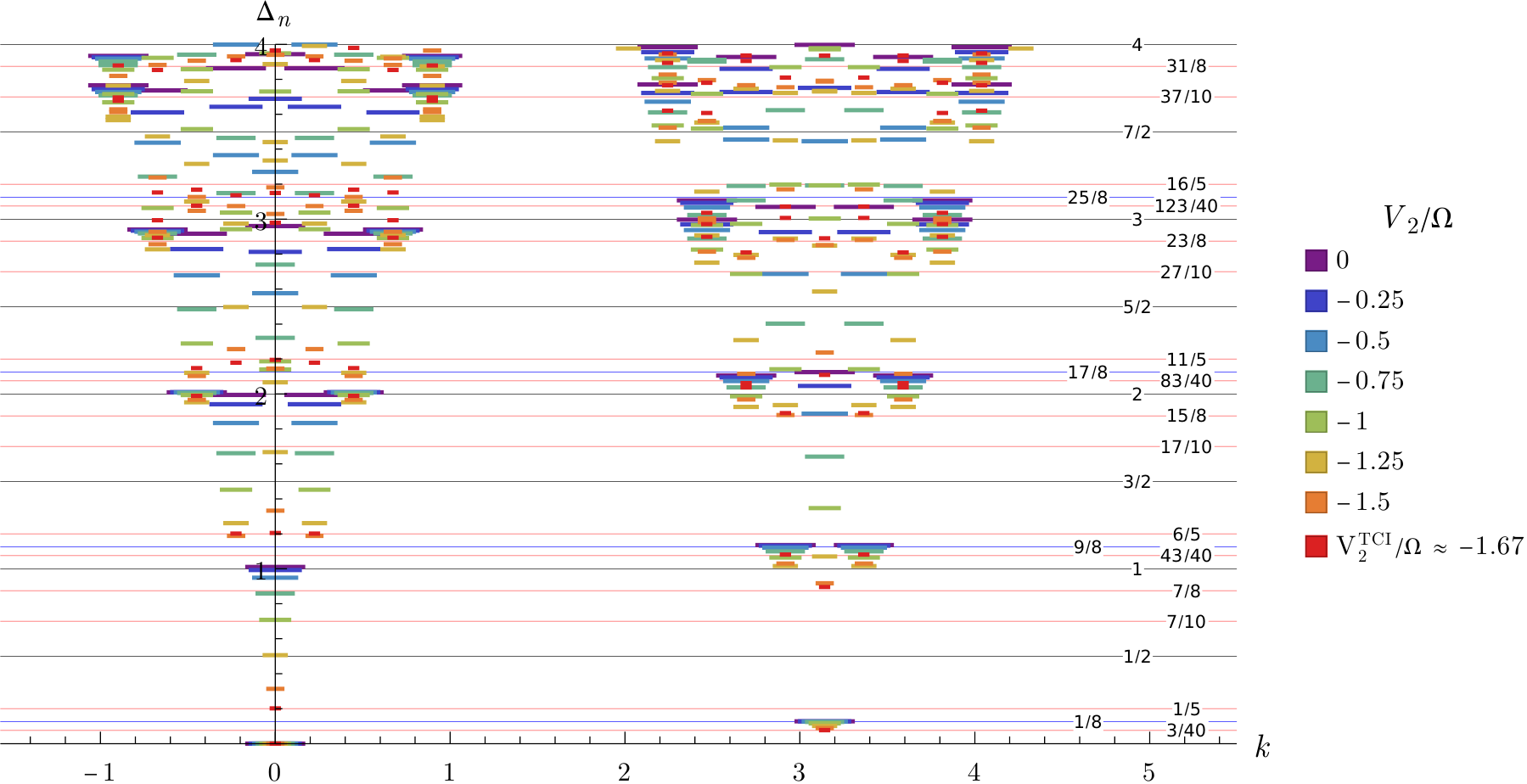}} \\
  \subfloat[$L=27$\label{fig:oddSpec}]{\includegraphics[width=2\columnwidth]{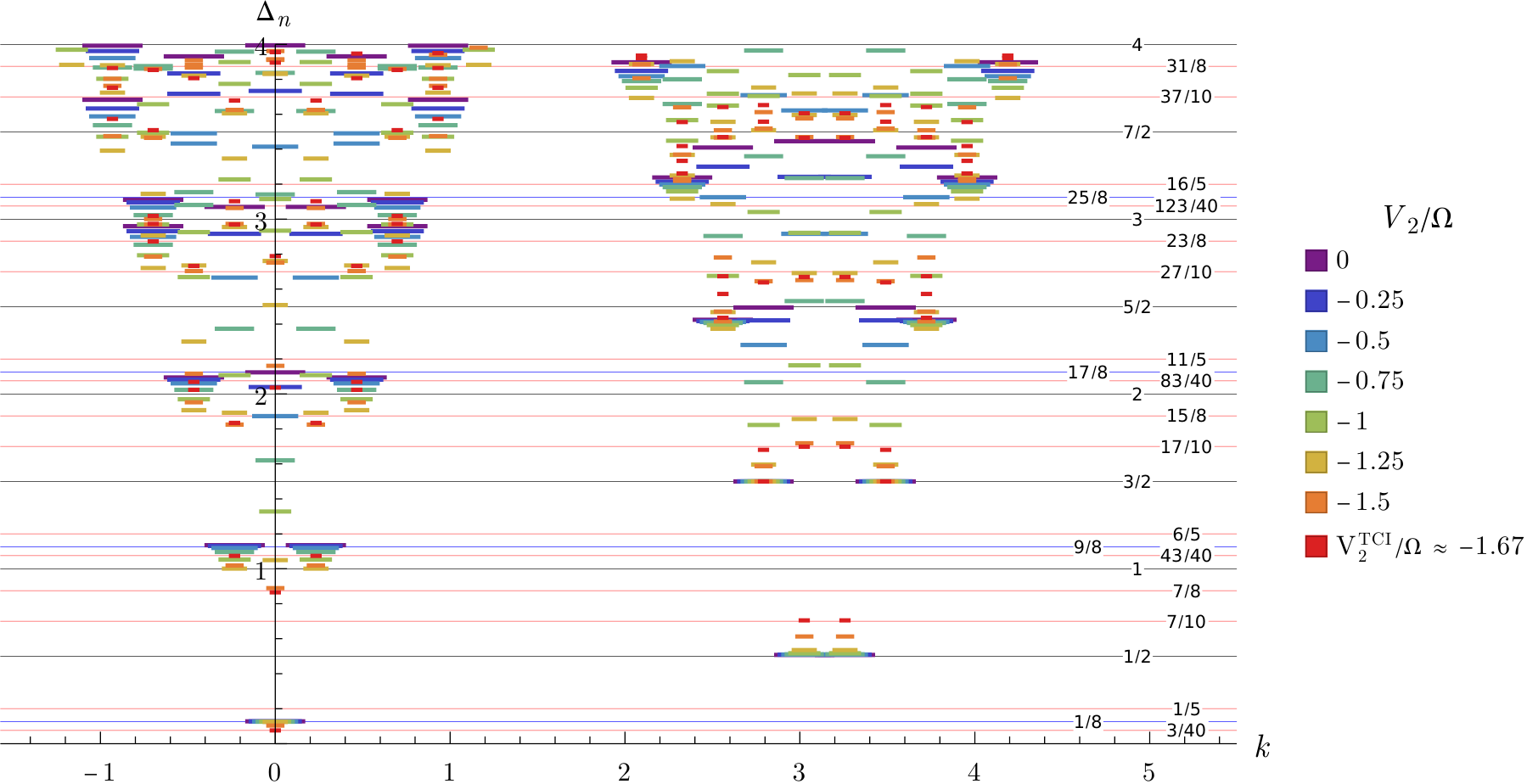}}
  \caption{%
   Rescaled energy spectrum $\Delta_n$ from Eq.~\eqref{En} versus momentum along the critical Ising line, starting from $V_2 = 0$ (long purple bars)
     and terminating at the tricritical Ising (TCI) point $V_2 = V_2^{\rm TCI} \approx -1.67 \Omega$ (short red bars).
   See text for how we fix parameters in \eqnref{En}. 
   The data were obtained from exact diagonalization of periodic Rydberg chains with (a) even length $L = 28$ and (b) odd length $L = 27$.
   Since $\Delta_n$ is the dimension of the CFT field associated with a given energy level, these plots reveal how Ising CFT fields morph into TCI fields upon approaching tricriticality, leading to the correspondence summarized in \tabref{tab:isingTCI}.
  }\label{fig:isingTCISpec}
\end{figure*}

\subsection{Connection to Ising CFT}

One of the many profound consequences of conformal symmetry in two spacetime dimensions is that the spectrum of the associated 1d quantum Hamiltonian is determined exactly by the scaling dimension of the operators creating the states. This fact allows a direct probe of the CFT from the lattice. Namely, for a length-$L$ periodic chain described in the low-energy limit by some CFT with central charge $c$, the energies $E_n$ are given approximately by \cite{Bloete1986,Affleck1986,Feiguin2007}
\begin{equation}
  E_n = e_0 L + \frac{2\pi v}{L} \left(\Delta_n - \frac{c}{12}\right).
  \label{En}
\end{equation}
The universal quantity $\Delta_n$ is the scaling dimension of the CFT field that yields the corresponding energy eigenstate
  (labeled by $n$)
  when acting on the ground state.
($\Delta_n$ should not be confused with the detuning $\Delta$ in the Rydberg Hamiltonian).
The other quantities are non-universal:  $e_0$ is an energy density, while $v$ is a velocity.
A critical Rydberg chain at $V_2 = 0$ conforms well to the $c = 1/2$ Ising CFT with only small $u T \overline{T}$ corrections to the energies for finite system sizes,
  as shown in Fig.~\ref{fig:spectrum}(b).
Equation~\eqref{En} allows us to associate energy levels in that limit with the constituent Ising CFT fields.
At the TCI point [Eq.~\eqref{TCIpoint}] occurring at $V_2/\Omega \approx -1.67$, Eq.~\eqref{En} instead relates the energy levels to tricritical Ising fields.
As $V_2/\Omega$ is tuned from 0 to the TCI point,
  for finite-sized systems the $u T \overline{T}$ corrections increase
  and there is a crossover between the Ising and TCI CFT energy level predictions.
Monitoring the energy levels for the finite-sized critical chain as $V_2$ varies from $0$ to the TCI point thereby reveals the mapping between fields for the two CFTs.  

In particular, we track $\Delta_n$ as a function of $V_2$ by computing the energy levels using exact diagonalization and  then fitting to \eqnref{En}.
For the central charge, we set $c = 1/2$ for all $V_2$ along the continuous Ising line but set $c = 7/10$ exactly at the TCI point.
The constants $v$ and $e_0$ depend nonuniversally on $V_2$, but we can determine both using a pair of energies with known $\Delta_n$.
For the first energy we choose the state with momentum $k = 0$ and dimension $\Delta = 0$ for even $L$; for the second we choose the $k = \pi - 3\pi/L$ state with $\Delta = 3/2$ for odd $L$.
That is, we find $v$ and $e_0$ such that the lowest-energy state for those momenta and system sizes have the corresponding $\Delta_n$ value.
This choice is convenient since both the Ising and TCI theories exhibit fields with dimension $0$ and $3/2$---hence the values of $\Delta_n$ for the above pair of states can (and do) evolve trivially as the system marches toward tricriticality.

\begin{figure}
  \includegraphics[width=.8\columnwidth]{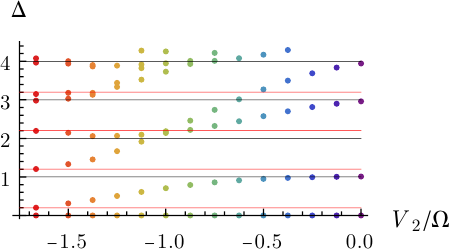}
  \caption{%
    Same as \figref{fig:evenSpec}, but versus coupling and only showing $k=0$ data.
    This view illustrates that the Ising fields associated with $\Delta_{T \overline{T}} = 4$ and $\Delta_n = 3$ respectively evolve into TCI fields associated with $\Delta_{\varepsilon'} =6/5$ and $\Delta_n =2+1/5$.
  }\label{fig:isingTCIk0}
\end{figure}

Figure~\ref{fig:isingTCISpec} shows the resulting values of $\Delta_n$ versus momentum for (a) $L = 28$ and (b) $L = 27$, with $V_2$ values ranging from 0 to the TCI point; Fig.~\ref{fig:isingTCIk0} displays the $L = 28$, $k = 0$ energies for additional clarity.
At $V_2 = 0$ (purple) one can clearly identify the Ising primary fields $I, \sigma, \psi$ as well as their descendants.
Similarly, at tricriticality (red) one can identify the fields from the left column of Table~\ref{tab:isingTCI} and their descendants, in agreement with those found in Ref.~\onlinecite{Feiguin2007}.
Data points at intermediate $V_2$ indicate that the fields morph into one another as follows (and summarized in Table~\ref{tab:isingTCI}):
The $\sigma, \psi,$ and $\varepsilon$ fields from the $c = 1/2$ line respectively evolve into $\sigma, \psi,$ and $\varepsilon$ from the TCI theory---evading a potential notational nightmare.
The Ising field $T \overline{T}$ evolves into the TCI field $\varepsilon'$.  
Notice that the former irrelevant perturbation thus becomes relevant in the TCI theory, as expected given that accessing the TCI point requires fine-tuning two relevant parameters rather than one.  
Finally, the TCI fields $\sigma'$ and $G$ evolve from descendants of $\sigma$ and $\psi$ in the Ising CFT.

\subsection{Lattice Operators}

\begin{figure*}
  \subfloat[]{\includegraphics[width=.67\columnwidth]{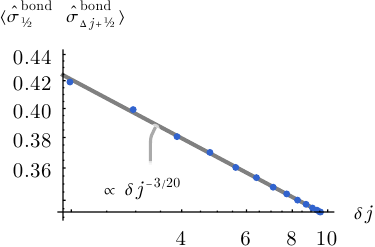}}
  \subfloat[]{\includegraphics[width=.67\columnwidth]{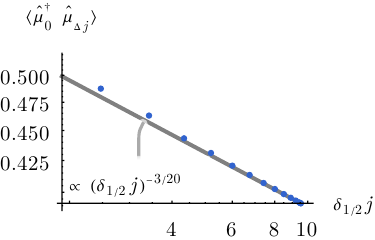}} 
  \subfloat[]{\includegraphics[width=.67\columnwidth]{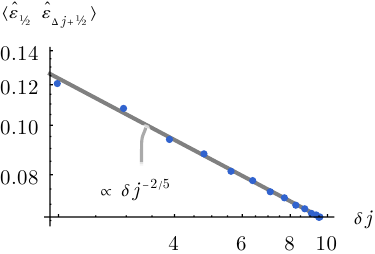}} \\
  \subfloat[]{\includegraphics[width=.8\columnwidth]{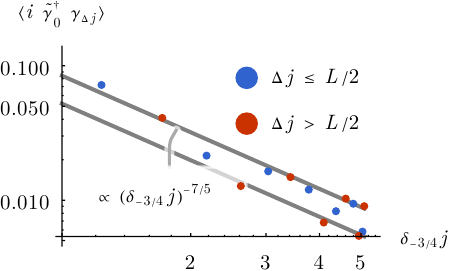}} \hspace{1cm}
  \subfloat[]{\includegraphics[width=.8\columnwidth]{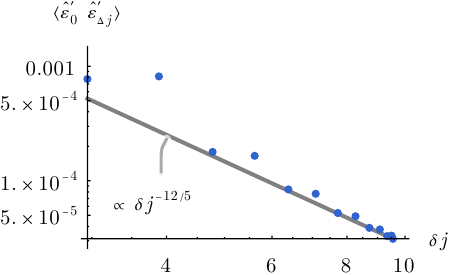}}
  \caption{%
  Correlation functions of various microscopic operators obtained from exact diagonalization of a periodic $L = 30$ Rydberg chain tuned to the tricritical Ising (TCI) point.
    Panels (a,b,c,e) verify power-law scaling of the microscopic operators $\hat\sigma^\text{bond}$, $\hat\mu$, $\hat\varepsilon$, and $\hat\varepsilon'$---respectively defined in \eqsref{sigmaprime}, \eqref{mu_expansion}, \eqref{epsilon_lattice}, and \eqref{epsilon'}---predicted by our mapping to $c = 7/10$ CFT fields.
    Panel (d) presents correlations for lattice operators defined in \eqsref{gammaj} and \eqref{gamma_def}, which are predicted to map onto TCI fermions with dimension $\Delta_\psi = 7/10$; here the data do not clearly exhibit such power-law behavior, possibly due to finite-size effects.  
    In the horizontal axes, $\delta j$ and $\delta_\epsilon j$ are defined as in Figs.~\ref{fig:sigma nnScaling} and \ref{fig:gammaScaling}.
  }\label{fig:TCI scaling}
\end{figure*}

The above Ising-TCI dictionary allows us to identify microscopic incarnations of some of the TCI fields.  
Precisely, the lattice operators that map onto the Ising fields $\sigma$ (and its dual $\mu$), $\varepsilon$, and $\psi$ yield the corresponding TCI fields when couplings are tuned to the TCI point [Eq.~\eqref{TCIpoint}].
Figure~\ref{fig:TCI scaling}(a-d) presents correlators of these lattice operators evaluated at tricriticality for an $L = 30$ chain with periodic boundary conditions.  
Panels (a), (b), and (c) respectively correspond to the bond-centered CDW order parameter $\hat \sigma^\text{bond}_{j+1/2}$ [Eq.~\eqref{sigmaprime}]
\footnote{We show results for the bond-centered CDW order parameter rather than Eq.~\eqref{sigma_lattice} since the latter exhibits a pronounced even-odd effect that muddies somewhat the power-law correlations arising from the $c = 7/10$ $\sigma$ field.}, $\hat \mu_j$ [Eq.~\eqref{mu_expansion}], and $\hat \varepsilon_{j+1/2}$ [Eq.~\eqref{epsilon_lattice}].
These cases confirm the power-laws, with scaling dimensions $\Delta_\sigma = \Delta_\mu = 3/40$ for (a,b) and $\Delta_\varepsilon = 1/5$ for (c), expected from the associated $c = 7/10$ fields.
Panel (d) presents $\langle i \tilde \gamma_0^\dagger \gamma_{\Delta j}\rangle$ [Eqs.~\eqref{gammaj} and \eqref{gamma_def}], which we predict displays power-law correlations with scaling dimension $\Delta_\psi = 7/10$ associated with the $c = 7/10$ $\psi$ fermion field.  
Here the data are less conclusive, however, presumably due to finite-size effects.
We speculate that $\sigma'$ and $G$ appear as subleading terms in the low-energy expansion of $\hat \sigma^\text{bond}_{j+1/2}$ and $\gamma_j,\tilde \gamma_j$, though we will not attempt to pinpoint their lattice counterparts.

The dimension-$6/5$ CFT field $\varepsilon'$ corresponds to a perturbation that moves the chain away from the TCI point and into an adjacent part of the phase boundary in Fig.~\ref{fig:phases}.
Since the exact first-order line is known from integrability [recall Eq.~\eqref{firstorder}], we can precisely determine a lattice operator with $\varepsilon'$ as its leading low-energy contribution.
Consider first
\begin{align}
  \hat\varepsilon^{(1)}_j &= - n_j + \braket{n_j} \nonumber\\
    &+ \left(\frac{\partial V_2}{\partial \Delta} \right)_{\rm TCI} \left(n_{j-1} n_{j+1} - \braket{n_{j-1} n_{j+1}} \right),
    \label{epsilon1}
\end{align}
where 
\begin{equation}
    \left(\frac{\partial V_2}{\partial \Delta} \right)_{\rm TCI} = \frac{9+5\sqrt{5}}{22} \approx 0.917
\end{equation}
is the derivative of Eq.~\eqref{firstorder} evaluated at the TCI point.
The sum $\sum_j \hat \varepsilon^{(1)}_j$ encodes the proper ratio of detuning and second-neighbor interaction that nudges a tricritical Rydberg chain into the first-order line.
Accordingly, the leading \emph{slowly varying} part of $\hat\varepsilon^{(1)}_j$ is $\varepsilon'$.  
The expansion of $\hat \varepsilon^{(1)}_j$ also, however, contains an oscillatory $(-1)^j \sigma$ term involving a field with much smaller scaling dimension. This term does not contribute to the sum, but will dominate correlation functions of the local operator.
We can distill this unwanted term away by coarse graining via 
\begin{align}
\hat \varepsilon_{j+1/2}^{(2)} = \tfrac12(\hat \varepsilon_j^{(1)} + \hat \varepsilon_{j+1}^{(1)})\ .
\end{align}
The expansion of this coarse-grained operator 
however involves an oscillatory $(-1)^j \partial_x \sigma$ term, and even with the extra derivative, $\partial_x\sigma$ still has a smaller scaling dimension than $\varepsilon'$. 
An additional coarse-graining step is thus needed to to isolate $\varepsilon'$ as the leading contribution, namely
\begin{equation}
  \hat \varepsilon_j' \equiv \tfrac12(\hat\varepsilon_{j-1/2}^{(2)} + \hat\varepsilon_{j+1/2}^{(2)}) \sim c_{\varepsilon'} \varepsilon' + \cdots \label{epsilon'}
\end{equation}
for some non-universal $c_{\varepsilon'}$ coefficient.
Figure~\ref{fig:TCI scaling}(e) demonstrates that $\hat\varepsilon_j'$ indeed exhibits power-law correlations with scaling dimension $6/5$, in line with this expansion.
Although the coefficient of the power-law fit is small,
  we have verified that the coefficient does not significantly depend on system size
  for any of the power-law decays in \figsref{fig:sigma nnScaling}, \ref{fig:gammaScaling}, or \ref{fig:TCI scaling}.

\section{Discussion}
\label{discussion}

Motivated in part by near-term experimental prospects, we have developed a detailed microscopic characterization of Ising criticality in Rydberg chains.
One of our main results was constructing a set of lattice operators that yield bosonic CFT fields $\sigma, \mu, \varepsilon$ and fermionic CFT fields $\gamma_{R/L}$ as the leading contribution to their low-energy expansions.
Devising microscopic counterparts of the disorder field $\mu$ and fermions $\gamma_{R/L}$ was particularly nontrivial given the non-on-site nature of the relevant $\mathbb{Z}_2$ symmetry combined with the lack of exact fermionizability for the Rydberg chain Hamiltonian.  

This dictionary enables CFT results to be readily translated into measurable predictions involving physical microscopic Rydberg operators.
These predictions become particularly clear-cut for Rydberg arrays defined on a ring, as realized in Refs.~\onlinecite{arrayJaewook,arrayBrowaeys}, thereby emulating periodic boundary conditions: 
Two-point correlation functions of microscopic operators yield power-laws associated with the leading CFT field in their expansion.
Such measurements would directly reveal the field content of the CFT and the associated scaling dimensions---arguably constituting a major achievement for quantum simulation.

Site-resolved measurements of the Rydberg occupation numbers $n_j$ would suffice for backing out correlations of the microscopic operators $\hat \sigma_j$ (or $\hat \sigma^{\rm bond}_{j+1/2}$) and $\hat \varepsilon_{j+1/2}$ that map to CFT fields $\sigma$ and $\varepsilon$, as these operators are local and diagonal in the $n_j$ basis.
Correlation functions of the non-local, off-diagonal operators $\hat \mu_j$ and $\gamma_j,\tilde \gamma_j$, which map to CFT fields $\mu$ and $\gamma_{R/L}$, could be measured using the classical shadow technique \cite{classicalShadow}.
This technique involves making measurements in the occupation number basis after applying a random unitary evolution \cite{classicalShadow,rydbergDesigns},
  from which the desired correlation functions can then be calculated.

Due to edge effects, linear Rydberg chains exhibit more nuanced critical behavior that we nevertheless showed could also be captured, with reasonable accuracy, using results from Ising CFT subject to fixed boundary conditions.
Even one-point correlators are rich here.
Translation symmetry breaking by the boundaries induces a nontrivial ground-state expectation value of the charge density wave order parameter $\hat \sigma^{\rm bond}_{j+1/2}$, which decays (slowly) into the bulk of the chain with a spatial profile governed by the CFT.
In \secref{sec:Delta c}, we showed how this edge effect can be utilized to experimentally determine the location of the critical point.
We further argued that the expectation value of the lattice operator $\hat \varepsilon^{\rm bare}_{j+1/2}$ manifests four-fermion interactions in the Ising CFT that can be tuned in both sign and strength by moving along the continuous Ising line in Fig.~\ref{fig:phases}.
Specifically, these interactions produce an effective enhancement (with attractive $V_2$) or suppression (with repulsive $V_2$) of contributions to the expectation value arising from the CFT field $\varepsilon$. 
This effect is pronounced even if one restricts to the physically natural $V_2\geq 0$ regime---recall Fig.~\ref{cVsV}---and can be probed by tracking the characteristic flattening of $\hat \varepsilon^{\rm bare}_{j+1/2}$ [Fig.~\ref{fig:boundary addV}(b)] upon accessing the Ising transition at progressively larger $V_2/\Omega$ values. 

Realizing these predictions in practice requires not only tuning Hamiltonian parameters to criticality, but \emph{also} initializing into the associated low-energy subspace.  
The most natural way of preparing target states in Rydberg experiments is to begin with a Hamiltonian whose ground state(s) can be easily prepared and then adiabatically deform to the desired final Hamiltonian \cite{Lukin2017}. 
In our context, one can initialize a Rydberg chain with all atoms in the $n_j=0$ configuration,
  which is the ground state for $H$ [\eqnref{H_FSS}] with $\Delta<0$, $\Omega=0$, and $V_r \geq 0$;
critical states can then be prepared by adiabatically tuning $\Omega$ and $\Delta$.
Since the gap at criticality scales like the inverse chain length, maintaining adiabaticity requires evolution times proportional to system size.
Our CFT predictions could be used to benchmark how well the Rydberg quantum simulator prepares critical ground states.

The $V_2<0$ regime may be realizable using an alternative adiabatic preparation scheme.  
Suppose that we again initialize the $n_j=0$ state,
  which is the \emph{highest-energy} Rydberg-constrained ($n_j n_{j+1}=0$) state of $H$ with $\Delta>V_2>0$ and $\Omega=0$.
This state is also the ground state of the Rydberg-constrained
  $H^\text{eff} = -H$ with $\Delta^\text{eff} = -\Delta < 0$ and $V_2^\text{eff} = -V_2 < 0$.
One could then prepare a critical state with $V_2^\text{eff} < 0$
  by adiabatically tuning $\Delta$ and $\Omega$.
However, the Rydberg constraint then becomes a dynamical constraint due to
  $V_1^\text{eff} \approx 2^6 V_2^\text{eff} < 0$ being large and negative.
That is, $H^\text{eff}$ has lower-energy states (than the desired critical state) that violate the Rydberg constraint,
  but the evolution into these unwanted states is slow in the $V_1^\text{eff} \to -\infty$ limit.
More work is necessary to determine the validity of this approximation.

Our work additionally paves the way to more forward-looking investigations of criticality in Rydberg chains.  
For instance, it would be interesting to develop a similar microscopic understanding of other quantum critical points in the phase diagram.
Real-time tunability further suggests tantalizing opportunities for exploring non-equilibrium dynamics in CFTs.
And finally, one can exploit insights gained here to study two-dimensional arrays assembled from coupled critical Rydberg chains---which we will pursue in a sequel to this work to uncover fractionalized phases relevant for fault-tolerant quantum computation.


\begin{acknowledgments}
It is a pleasure to thank Lesik Motrunich for stimulating conversations.
This work was supported by
	the Army Research Office under Grant Award W911NF-17-1-0323; 
	the U.S. Department of Energy, Office of Science, National Quantum Information Science Research Centers, Quantum Science Center;
	the National Science Foundation through grants DMR-1723367 (JA) and DMR-1848336 (RM); 
	the Caltech Institute for Quantum Information and Matter, an NSF Physics Frontiers Center with support of the Gordon and Betty Moore Foundation through Grant GBMF1250; 
	the Walter Burke Institute for Theoretical Physics at Caltech; 
	the ESQ by a Discovery Grant;
    the Gordon and Betty Moore Foundation's EPiQS Initiative, Grant GBMF8682;
    the AFOSR YIP (FA9550-19-1-0044);
    and the UK Engineering and Physical Sciences Research Council through grant EP/S020527/1 (PF).
\end{acknowledgments}

\appendix

\section{Operator-CFT field mapping in the transverse-field Ising model}
\label{IsingReview}

Here we briefly review the standard mapping between microscopic spin operators and CFT fields in the transverse-field Ising model:
\begin{align}
  H &= \sum_j (J Z_j Z_{j+1} - h X_j).
\end{align}
In terms of microscopic order and disorder operators
\begin{align}
  \hat\sigma_j &= Z_j \\
  \hat\mu_j    &= \cdots X_{j-2} X_{j-1} X_j,
\end{align}
exact microscopic Majorana fermion operators follow as
\begin{gather}
\begin{aligned}
  \tilde{\psi}_j &= \tfrac{1}{2} i [\hat\mu_j,\hat\sigma_j] = \cdots X_{j-2} X_{j-1} Y_j \\
         \psi_j  &= i R_x \tilde{\psi}_{-j} R_x \mathcal{C} = \cdots X_{j-2} X_{j-1} Z_j.
\end{aligned} \label{gammaIsing} 
\end{gather}
Here $R_x$ denotes $x\to -x$ reflection symmetry and $\mathcal{C} = \prod_j X_j$ implements the Ising spin-flip symmetry.
We have written the middle parts of Eq.~\eqref{gammaIsing} in a way that parallels our definition of microscopic operators that map to low-energy fermions in the Rydberg model; recall Eqs.~\eqref{gammaj} and \eqref{gamma_def}.
The Hamiltonian expressed in terms of Majorana fermions becomes quadratic,
\begin{align}
  H &= \sum_j(J i \tilde{\psi}_j \psi_{j+1} - h i \psi_j \tilde{\psi}_j),
\end{align}
and can therefore be solved exactly at any $h/J$.
At the Ising transition occurring when $h = J$, the microscopic operators above relate to Ising CFT fields according to the dictionary
\begin{gather}
\begin{aligned}
  \hat \sigma_j &\sim \sigma
  \\
  \hat \mu_j &\sim \mu
  \\
  \psi_j & \sim \gamma_L + \gamma_R
  \\
  \tilde \psi_j & \sim \gamma_L - \gamma_R.
\end{aligned}
\end{gather}

\section{Open boundary CFT calculations}
\label{app:CFT}

\begin{figure*}[t]
  \includegraphics[width=1.65\columnwidth]{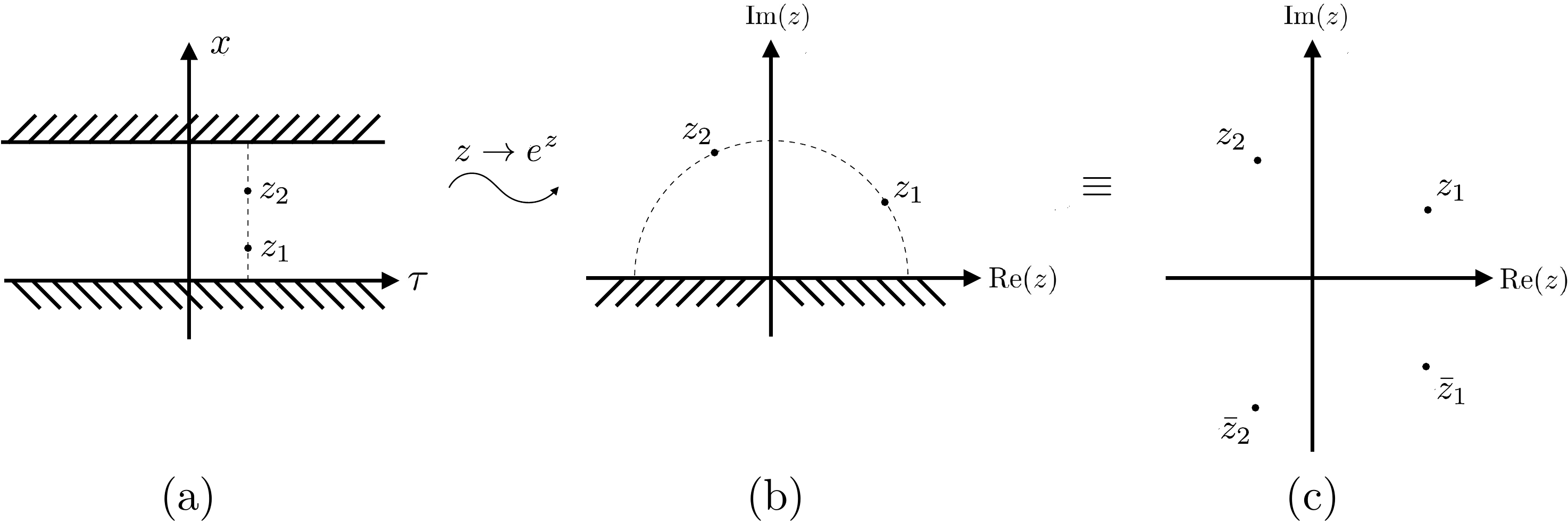}
  \caption{%
    (a) An infinite strip of spacetime is conformally mapped to (b) the upper half plane.
    (c) Two-point correlation functions in the upper half plane satisfy the same differential
equations as four-point correlation functions in the whole plane \cite{Cardy_1984}.
  }\label{fig:conformalMap}
\end{figure*}

This Appendix sketches the calculation of the open-boundary CFT correlation functions listed in \tabref{tab:boundaryCFT}.
After rescaling space so that the chain lives on the interval $0 \leq x \leq \pi$, the relevant spacetime $(\tau,x)$ populates an infinite strip, with imaginary time $\tau$ running from $-\infty$ to $+\infty$.
We label the boundary conditions for the infinite strip---specified by Eq.~\eqref{BC}---as $(+)$ for odd $L$ and $(-)$ for even $L$.  
We map the infinite strip to the upper-half plane via the conformal transformation
\begin{equation}
  \tau + i x = z \to w(z) = e^z \label{eq:conformalMap}
\end{equation}
illustrated in \sfigref{fig:conformalMap}{a-b}.
Under this transformation, the boundary lines at $x = 0$ and $x = \pi$ respectively map to the positive and negative real axis.
Equation~\eqref{BC} then dictates that the real axis exhibits a homogeneous boundary condition (which we label $[+]$) for odd $L$ but a piecewise homogeneous boundary condition with a jump at $x = 0$ (which we label $[-]$) for even $L$.
In two dimensions, conformal invariance requires that correlation functions transform under Eq.~\eqref{eq:conformalMap} as
\begin{align}
  \braket{\phi(z)}_{({\rm BC})} &=|w'(z)|^{\Delta_\phi} \braket{\phi(w(z))}_{[{\rm BC}]} \nonumber\\
    &= e^{\tau \Delta_\phi} \braket{\phi(e^{\tau + i x})}_{[{\rm BC}]} \label{eq:conformalTransformation}\\
  \braket{\phi(z_1)\phi(z_2)}_{({\rm BC})} &= |w'(z_1)|^{\Delta_\phi} |w'(z_2)|^{\Delta_\phi} 
  \nonumber \\
  &\times \braket{\phi(w(z_1)) \phi(w(z_2))}_{[{\rm BC}]} \nonumber\\
    &= e^{(\tau_1 + \tau_2) \Delta_\phi} \braket{\phi(e^{\tau_1 + i x_1}) \phi(e^{\tau_2 + i x_2})}_{[{\rm BC}]}, \nonumber
\end{align}
where $\Delta_\phi$ is the scaling dimension of field $\phi$.  
Subscripts $({\rm BC})$ and $[{\rm BC}]$ indicate that the correlator is evaluated with boundary conditions applicable for the infinite strip and upper-half plane, respectively.

The $n$-point correlation functions $\braket{\phi(z_1) \dots \phi(z_n)}$ on the upper-half plane  are the same as the $2n$-point correlation functions $\braket{\phi(z_1)\phi(\bar z_1)\dots \phi_n(z_n)\phi_n(\bar z_n)}$ on the infinite 2D plane [see \sfigref{fig:conformalMap}{b-c}] \cite{Cardy_1984}.
Results for the Ising CFT with fixed homogeneous and piecewise homogeneous boundary conditions
appear in Ref.~\onlinecite{Burkhardt1993}.
For fixed homogeneous boundary conditions appropriate for odd $L$, one-point and two-point $\sigma$ correlations read
\begin{gather}
\begin{aligned}
  \braket{\sigma(z)}_{[+]} &= 2^{1/4} [2{\rm Im}(z)]^{-1/8} \\
  \braket{\sigma(z_1) \sigma(z_2)}_{[+]} &=  \frac{\left(\rho^{1/4} + \rho^{-1/4} \right)^{1/2}}{[{\rm Im}(z_1){\rm Im}(z_2)]^{1/8}}
\end{aligned} \label{eq:sigmaPP}
\end{gather}
with
\begin{equation}
  \rho(z_1,z_2) = \left| \frac{z_1 - \overline{z}_2}{z_1 - z_2} \right|^2.
\end{equation}
General $n$-point $\varepsilon$ correlations take the compact form
\begin{equation}
  \braket{\varepsilon(z_1) \varepsilon(z_2) \cdots \varepsilon(z_n)}_{[+]} = i^n \operatorname{Pf}\mathopen\Big(\frac{1}{w_a-w_b}\Big),
\end{equation}  
where $\operatorname{Pf}$ denotes the Pfaffian of the $2n \times 2n$ matrix defined using
$(w_1,\ldots,w_{2n}) = (z_1, \overline{z}_1, \ldots, z_{n}, \overline{z}_n)$.
(The Pfaffian equals a square root of the determinant.)
Correlations for piecewise homogeneous boundary conditions appropriate for even $L$ are related to those above as follows. For spin fields,
\begin{align}
  \braket{\sigma(z)}_{[-]} &= \frac{{\rm Re}(z)}{|z|} \braket{\sigma(z)}_{[+]} \nonumber\\
  \braket{\sigma(z_1)\sigma(z_2)}_{[-]} &= \frac{1}{|z_1z_2|} \bigg[ {\rm Re}(z_1z_2) + \label{eq:sigmaPM}\\
    \frac{1}{2} |z_1-\bar{z}_2| \big(|z_1 & -\bar{z}_2|-|z_1-z_2|\big) \bigg] \braket{\sigma(z_1)\sigma(z_2)}_{[+]}, \nonumber
\end{align}
while defining $w'_j =w_j$ for $j\le 2n$ along with $w_{2n+1}=0$ and $w_{2n+2}=\zeta$ yields
\begin{align}
  \braket{\varepsilon(z_1) \varepsilon(z_2) \cdots \varepsilon(z_n)}_{[-]} &= i^n \lim_{\zeta \to \infty} \zeta^{-1} \operatorname{Pf}\mathopen\Big(\frac{1}{w'_a-w'_b}\Big).
\end{align}

Equations~\eqref{eq:conformalTransformation} allow extraction of correlation functions on the original infinite strip; for instance, we obtain
\begin{align}
  \braket{\sigma(\tau + ix)}_{(+)} &= e^{\tau/8} \braket{\sigma(e^{\tau + i x})}_{[+]} = \Bigg(\frac{2}{\sin x}\Bigg)^{1/8}
  \label{sigmaexample}
\end{align}
and similarly find the one- and equal-time two-point correlators provided in \tabref{tab:boundaryCFT}.
For brevity in \tabref{tab:boundaryCFT}, we set $\tau=0$ and
  abuse the notation to write e.g. $\braket{\sigma(x)}_{(+)}$ instead of $\braket{\sigma(\tau + ix)}_{(+)}$.


\hbadness=10000	
\bibliography{RydbergChain}

\begin{thebibliography}{52}%
\makeatletter
\providecommand \@ifxundefined [1]{%
 \@ifx{#1\undefined}
}%
\providecommand \@ifnum [1]{%
 \ifnum #1\expandafter \@firstoftwo
 \else \expandafter \@secondoftwo
 \fi
}%
\providecommand \@ifx [1]{%
 \ifx #1\expandafter \@firstoftwo
 \else \expandafter \@secondoftwo
 \fi
}%
\providecommand \natexlab [1]{#1}%
\providecommand \enquote  [1]{``#1''}%
\providecommand \bibnamefont  [1]{#1}%
\providecommand \bibfnamefont [1]{#1}%
\providecommand \citenamefont [1]{#1}%
\providecommand \href@noop [0]{\@secondoftwo}%
\providecommand \href [0]{\begingroup \@sanitize@url \@href}%
\providecommand \@href[1]{\@@startlink{#1}\@@href}%
\providecommand \@@href[1]{\endgroup#1\@@endlink}%
\providecommand \@sanitize@url [0]{\catcode `\\12\catcode `\$12\catcode
  `\&12\catcode `\#12\catcode `\^12\catcode `\_12\catcode `\%12\relax}%
\providecommand \@@startlink[1]{}%
\providecommand \@@endlink[0]{}%
\providecommand \url  [0]{\begingroup\@sanitize@url \@url }%
\providecommand \@url [1]{\endgroup\@href {#1}{\urlprefix }}%
\providecommand \urlprefix  [0]{URL }%
\providecommand \Eprint [0]{\href }%
\providecommand \doibase [0]{http://dx.doi.org/}%
\providecommand \selectlanguage [0]{\@gobble}%
\providecommand \bibinfo  [0]{\@secondoftwo}%
\providecommand \bibfield  [0]{\@secondoftwo}%
\providecommand \translation [1]{[#1]}%
\providecommand \BibitemOpen [0]{}%
\providecommand \bibitemStop [0]{}%
\providecommand \bibitemNoStop [0]{.\EOS\space}%
\providecommand \EOS [0]{\spacefactor3000\relax}%
\providecommand \BibitemShut  [1]{\csname bibitem#1\endcsname}%
\let\auto@bib@innerbib\@empty
\bibitem [{\citenamefont {{Ginsparg}}(1988)}]{AppliedCFT}%
  \BibitemOpen
  \bibfield  {author} {\bibinfo {author} {\bibfnamefont {Paul}\ \bibnamefont
  {{Ginsparg}}},\ }\bibfield  {title} {\enquote {\bibinfo {title} {{Applied
  Conformal Field Theory}},}\ }\href@noop {} {\  (\bibinfo {year} {1988})},\
  \Eprint {http://arxiv.org/abs/hep-th/9108028} {arXiv:hep-th/9108028}
  \BibitemShut {NoStop}%
\bibitem [{\citenamefont {{Gaberdiel}}(2000)}]{IntroCFT}%
  \BibitemOpen
  \bibfield  {author} {\bibinfo {author} {\bibfnamefont {Matthias~R.}\
  \bibnamefont {{Gaberdiel}}},\ }\bibfield  {title} {\enquote {\bibinfo {title}
  {{An introduction to conformal field theory}},}\ }\href {\doibase
  10.1088/0034-4885/63/4/203} {\bibfield  {journal} {\bibinfo  {journal}
  {Reports on Progress in Physics}\ }\textbf {\bibinfo {volume} {63}},\
  \bibinfo {pages} {607--667} (\bibinfo {year} {2000})},\ \Eprint
  {http://arxiv.org/abs/hep-th/9910156} {arXiv:hep-th/9910156} \BibitemShut
  {NoStop}%
\bibitem [{\citenamefont {Coldea}\ \emph {et~al.}(2010)\citenamefont {Coldea},
  \citenamefont {Tennant}, \citenamefont {Wheeler}, \citenamefont {Wawrzynska},
  \citenamefont {Prabhakaran}, \citenamefont {Telling}, \citenamefont
  {Habicht}, \citenamefont {Smeibidl},\ and\ \citenamefont
  {Kiefer}}]{ColdeaE8}%
  \BibitemOpen
  \bibfield  {author} {\bibinfo {author} {\bibfnamefont {R.}~\bibnamefont
  {Coldea}}, \bibinfo {author} {\bibfnamefont {D.~A.}\ \bibnamefont {Tennant}},
  \bibinfo {author} {\bibfnamefont {E.~M.}\ \bibnamefont {Wheeler}}, \bibinfo
  {author} {\bibfnamefont {E.}~\bibnamefont {Wawrzynska}}, \bibinfo {author}
  {\bibfnamefont {D.}~\bibnamefont {Prabhakaran}}, \bibinfo {author}
  {\bibfnamefont {M.}~\bibnamefont {Telling}}, \bibinfo {author} {\bibfnamefont
  {K.}~\bibnamefont {Habicht}}, \bibinfo {author} {\bibfnamefont
  {P.}~\bibnamefont {Smeibidl}}, \ and\ \bibinfo {author} {\bibfnamefont
  {K.}~\bibnamefont {Kiefer}},\ }\bibfield  {title} {\enquote {\bibinfo {title}
  {{Quantum Criticality in an Ising Chain: Experimental Evidence for Emergent
  E8 Symmetry}},}\ }\href {\doibase 10.1126/science.1180085} {\bibfield
  {journal} {\bibinfo  {journal} {Science}\ }\textbf {\bibinfo {volume}
  {327}},\ \bibinfo {pages} {177--180} (\bibinfo {year} {2010})},\ \Eprint
  {http://arxiv.org/abs/1103.3694} {arXiv:1103.3694} \BibitemShut {NoStop}%
\bibitem [{\citenamefont {{Fava}}\ \emph {et~al.}(2020)\citenamefont {{Fava}},
  \citenamefont {{Coldea}},\ and\ \citenamefont {{Parameswaran}}}]{FavaE8}%
  \BibitemOpen
  \bibfield  {author} {\bibinfo {author} {\bibfnamefont {Michele}\ \bibnamefont
  {{Fava}}}, \bibinfo {author} {\bibfnamefont {Radu}\ \bibnamefont {{Coldea}}},
  \ and\ \bibinfo {author} {\bibfnamefont {S.~A.}\ \bibnamefont
  {{Parameswaran}}},\ }\bibfield  {title} {\enquote {\bibinfo {title} {{Glide
  symmetry breaking and Ising criticality in the quasi-1D magnet CoNb2O6}},}\
  }\href {\doibase 10.1073/pnas.2007986117} {\bibfield  {journal} {\bibinfo
  {journal} {Proceedings of the National Academy of Science}\ }\textbf
  {\bibinfo {volume} {117}},\ \bibinfo {pages} {25219--25224} (\bibinfo {year}
  {2020})},\ \Eprint {http://arxiv.org/abs/2004.04169} {arXiv:2004.04169}
  \BibitemShut {NoStop}%
\bibitem [{\citenamefont {Moore}\ and\ \citenamefont {Read}(1991)}]{MooreRead}%
  \BibitemOpen
  \bibfield  {author} {\bibinfo {author} {\bibfnamefont {Gregory}\ \bibnamefont
  {Moore}}\ and\ \bibinfo {author} {\bibfnamefont {Nicholas}\ \bibnamefont
  {Read}},\ }\bibfield  {title} {\enquote {\bibinfo {title} {Nonabelions in the
  fractional quantum {Hall} effect},}\ }\href {\doibase
  10.1016/0550-3213(91)90407-O} {\bibfield  {journal} {\bibinfo  {journal}
  {Nuclear Physics B}\ }\textbf {\bibinfo {volume} {360}},\ \bibinfo {pages}
  {362--396} (\bibinfo {year} {1991})}\BibitemShut {NoStop}%
\bibitem [{\citenamefont {{Browaeys}}\ and\ \citenamefont
  {{Lahaye}}(2020)}]{Browaeys2020}%
  \BibitemOpen
  \bibfield  {author} {\bibinfo {author} {\bibfnamefont {Antoine}\ \bibnamefont
  {{Browaeys}}}\ and\ \bibinfo {author} {\bibfnamefont {Thierry}\ \bibnamefont
  {{Lahaye}}},\ }\bibfield  {title} {\enquote {\bibinfo {title} {{Many-body
  physics with individually controlled Rydberg atoms}},}\ }\href {\doibase
  10.1038/s41567-019-0733-z} {\bibfield  {journal} {\bibinfo  {journal} {Nature
  Physics}\ }\textbf {\bibinfo {volume} {16}},\ \bibinfo {pages} {132--142}
  (\bibinfo {year} {2020})},\ \Eprint {http://arxiv.org/abs/2002.07413}
  {arXiv:2002.07413} \BibitemShut {NoStop}%
\bibitem [{\citenamefont {{Morgado}}\ and\ \citenamefont
  {{Whitlock}}(2021)}]{Morgado2021}%
  \BibitemOpen
  \bibfield  {author} {\bibinfo {author} {\bibfnamefont {M.}~\bibnamefont
  {{Morgado}}}\ and\ \bibinfo {author} {\bibfnamefont {S.}~\bibnamefont
  {{Whitlock}}},\ }\bibfield  {title} {\enquote {\bibinfo {title} {{Quantum
  simulation and computing with Rydberg-interacting qubits}},}\ }\href
  {\doibase 10.1116/5.0036562} {\bibfield  {journal} {\bibinfo  {journal} {AVS
  Quantum Science}\ }\textbf {\bibinfo {volume} {3}},\ \bibinfo {eid} {023501}
  (\bibinfo {year} {2021})},\ \Eprint {http://arxiv.org/abs/2011.03031}
  {arXiv:2011.03031} \BibitemShut {NoStop}%
\bibitem [{\citenamefont {{Bernien}}\ \emph {et~al.}(2017)\citenamefont
  {{Bernien}}, \citenamefont {{Schwartz}}, \citenamefont {{Keesling}},
  \citenamefont {{Levine}}, \citenamefont {{Omran}}, \citenamefont {{Pichler}},
  \citenamefont {{Choi}}, \citenamefont {{Zibrov}}, \citenamefont {{Endres}},
  \citenamefont {{Greiner}}, \citenamefont {{Vuleti{\'c}}},\ and\ \citenamefont
  {{Lukin}}}]{Lukin2017}%
  \BibitemOpen
  \bibfield  {author} {\bibinfo {author} {\bibfnamefont {Hannes}\ \bibnamefont
  {{Bernien}}}, \bibinfo {author} {\bibfnamefont {Sylvain}\ \bibnamefont
  {{Schwartz}}}, \bibinfo {author} {\bibfnamefont {Alexander}\ \bibnamefont
  {{Keesling}}}, \bibinfo {author} {\bibfnamefont {Harry}\ \bibnamefont
  {{Levine}}}, \bibinfo {author} {\bibfnamefont {Ahmed}\ \bibnamefont
  {{Omran}}}, \bibinfo {author} {\bibfnamefont {Hannes}\ \bibnamefont
  {{Pichler}}}, \bibinfo {author} {\bibfnamefont {Soonwon}\ \bibnamefont
  {{Choi}}}, \bibinfo {author} {\bibfnamefont {Alexander~S.}\ \bibnamefont
  {{Zibrov}}}, \bibinfo {author} {\bibfnamefont {Manuel}\ \bibnamefont
  {{Endres}}}, \bibinfo {author} {\bibfnamefont {Markus}\ \bibnamefont
  {{Greiner}}}, \bibinfo {author} {\bibfnamefont {Vladan}\ \bibnamefont
  {{Vuleti{\'c}}}}, \ and\ \bibinfo {author} {\bibfnamefont {Mikhail~D.}\
  \bibnamefont {{Lukin}}},\ }\bibfield  {title} {\enquote {\bibinfo {title}
  {{Probing many-body dynamics on a 51-atom quantum simulator}},}\ }\href
  {\doibase 10.1038/nature24622} {\bibfield  {journal} {\bibinfo  {journal}
  {\nat}\ }\textbf {\bibinfo {volume} {551}},\ \bibinfo {pages} {579--584}
  (\bibinfo {year} {2017})},\ \Eprint {http://arxiv.org/abs/1707.04344}
  {arXiv:1707.04344} \BibitemShut {NoStop}%
\bibitem [{\citenamefont {{Scholl}}\ \emph {et~al.}(2021)\citenamefont
  {{Scholl}}, \citenamefont {{Schuler}}, \citenamefont {{Williams}},
  \citenamefont {{Eberharter}}, \citenamefont {{Barredo}}, \citenamefont
  {{Schymik}}, \citenamefont {{Lienhard}}, \citenamefont {{Henry}},
  \citenamefont {{Lang}}, \citenamefont {{Lahaye}}, \citenamefont
  {{L{\"a}uchli}},\ and\ \citenamefont {{Browaeys}}}]{Scholl2021}%
  \BibitemOpen
  \bibfield  {author} {\bibinfo {author} {\bibfnamefont {Pascal}\ \bibnamefont
  {{Scholl}}}, \bibinfo {author} {\bibfnamefont {Michael}\ \bibnamefont
  {{Schuler}}}, \bibinfo {author} {\bibfnamefont {Hannah~J.}\ \bibnamefont
  {{Williams}}}, \bibinfo {author} {\bibfnamefont {Alexander~A.}\ \bibnamefont
  {{Eberharter}}}, \bibinfo {author} {\bibfnamefont {Daniel}\ \bibnamefont
  {{Barredo}}}, \bibinfo {author} {\bibfnamefont {Kai-Niklas}\ \bibnamefont
  {{Schymik}}}, \bibinfo {author} {\bibfnamefont {Vincent}\ \bibnamefont
  {{Lienhard}}}, \bibinfo {author} {\bibfnamefont {Louis-Paul}\ \bibnamefont
  {{Henry}}}, \bibinfo {author} {\bibfnamefont {Thomas~C.}\ \bibnamefont
  {{Lang}}}, \bibinfo {author} {\bibfnamefont {Thierry}\ \bibnamefont
  {{Lahaye}}}, \bibinfo {author} {\bibfnamefont {Andreas~M.}\ \bibnamefont
  {{L{\"a}uchli}}}, \ and\ \bibinfo {author} {\bibfnamefont {Antoine}\
  \bibnamefont {{Browaeys}}},\ }\bibfield  {title} {\enquote {\bibinfo {title}
  {{Quantum simulation of 2D antiferromagnets with hundreds of Rydberg
  atoms}},}\ }\href {\doibase 10.1038/s41586-021-03585-1} {\bibfield  {journal}
  {\bibinfo  {journal} {\nat}\ }\textbf {\bibinfo {volume} {595}},\ \bibinfo
  {pages} {233--238} (\bibinfo {year} {2021})},\ \Eprint
  {http://arxiv.org/abs/2012.12268} {arXiv:2012.12268} \BibitemShut {NoStop}%
\bibitem [{\citenamefont {{Ebadi}}\ \emph {et~al.}(2021)\citenamefont
  {{Ebadi}}, \citenamefont {{Wang}}, \citenamefont {{Levine}}, \citenamefont
  {{Keesling}}, \citenamefont {{Semeghini}}, \citenamefont {{Omran}},
  \citenamefont {{Bluvstein}}, \citenamefont {{Samajdar}}, \citenamefont
  {{Pichler}}, \citenamefont {{Ho}}, \citenamefont {{Choi}}, \citenamefont
  {{Sachdev}}, \citenamefont {{Greiner}}, \citenamefont {{Vuleti{\'c}}},\ and\
  \citenamefont {{Lukin}}}]{Lukin256}%
  \BibitemOpen
  \bibfield  {author} {\bibinfo {author} {\bibfnamefont {Sepehr}\ \bibnamefont
  {{Ebadi}}}, \bibinfo {author} {\bibfnamefont {Tout~T.}\ \bibnamefont
  {{Wang}}}, \bibinfo {author} {\bibfnamefont {Harry}\ \bibnamefont
  {{Levine}}}, \bibinfo {author} {\bibfnamefont {Alexander}\ \bibnamefont
  {{Keesling}}}, \bibinfo {author} {\bibfnamefont {Giulia}\ \bibnamefont
  {{Semeghini}}}, \bibinfo {author} {\bibfnamefont {Ahmed}\ \bibnamefont
  {{Omran}}}, \bibinfo {author} {\bibfnamefont {Dolev}\ \bibnamefont
  {{Bluvstein}}}, \bibinfo {author} {\bibfnamefont {Rhine}\ \bibnamefont
  {{Samajdar}}}, \bibinfo {author} {\bibfnamefont {Hannes}\ \bibnamefont
  {{Pichler}}}, \bibinfo {author} {\bibfnamefont {Wen~Wei}\ \bibnamefont
  {{Ho}}}, \bibinfo {author} {\bibfnamefont {Soonwon}\ \bibnamefont {{Choi}}},
  \bibinfo {author} {\bibfnamefont {Subir}\ \bibnamefont {{Sachdev}}}, \bibinfo
  {author} {\bibfnamefont {Markus}\ \bibnamefont {{Greiner}}}, \bibinfo
  {author} {\bibfnamefont {Vladan}\ \bibnamefont {{Vuleti{\'c}}}}, \ and\
  \bibinfo {author} {\bibfnamefont {Mikhail~D.}\ \bibnamefont {{Lukin}}},\
  }\bibfield  {title} {\enquote {\bibinfo {title} {{Quantum phases of matter on
  a 256-atom programmable quantum simulator}},}\ }\href {\doibase
  10.1038/s41586-021-03582-4} {\bibfield  {journal} {\bibinfo  {journal}
  {\nat}\ }\textbf {\bibinfo {volume} {595}},\ \bibinfo {pages} {227--232}
  (\bibinfo {year} {2021})},\ \Eprint {http://arxiv.org/abs/2012.12281}
  {arXiv:2012.12281} \BibitemShut {NoStop}%
\bibitem [{\citenamefont {Fendley}\ \emph {et~al.}(2004)\citenamefont
  {Fendley}, \citenamefont {Sengupta},\ and\ \citenamefont
  {Sachdev}}]{Fendley2004}%
  \BibitemOpen
  \bibfield  {author} {\bibinfo {author} {\bibfnamefont {Paul}\ \bibnamefont
  {Fendley}}, \bibinfo {author} {\bibfnamefont {K.}~\bibnamefont {Sengupta}}, \
  and\ \bibinfo {author} {\bibfnamefont {Subir}\ \bibnamefont {Sachdev}},\
  }\bibfield  {title} {\enquote {\bibinfo {title} {Competing density-wave
  orders in a one-dimensional hard-boson model},}\ }\href {\doibase
  10.1103/PhysRevB.69.075106} {\bibfield  {journal} {\bibinfo  {journal} {Phys.
  Rev. B}\ }\textbf {\bibinfo {volume} {69}},\ \bibinfo {pages} {075106}
  (\bibinfo {year} {2004})},\ \Eprint {http://arxiv.org/abs/cond-mat/0309438}
  {arXiv:cond-mat/0309438} \BibitemShut {NoStop}%
\bibitem [{\citenamefont {{Lesanovsky}}\ and\ \citenamefont
  {{Katsura}}(2012)}]{Lesanovsky2012}%
  \BibitemOpen
  \bibfield  {author} {\bibinfo {author} {\bibfnamefont {Igor}\ \bibnamefont
  {{Lesanovsky}}}\ and\ \bibinfo {author} {\bibfnamefont {Hosho}\ \bibnamefont
  {{Katsura}}},\ }\bibfield  {title} {\enquote {\bibinfo {title} {{Interacting
  Fibonacci anyons in a Rydberg gas}},}\ }\href {\doibase
  10.1103/PhysRevA.86.041601} {\bibfield  {journal} {\bibinfo  {journal}
  {\pra}\ }\textbf {\bibinfo {volume} {86}},\ \bibinfo {eid} {041601(R)}
  (\bibinfo {year} {2012})},\ \Eprint {http://arxiv.org/abs/1204.0903}
  {arXiv:1204.0903 [cond-mat.quant-gas]} \BibitemShut {NoStop}%
\bibitem [{\citenamefont {Rader}\ and\ \citenamefont
  {L\"{a}uchli}(2019)}]{Rader2019}%
  \BibitemOpen
  \bibfield  {author} {\bibinfo {author} {\bibfnamefont {Michael}\ \bibnamefont
  {Rader}}\ and\ \bibinfo {author} {\bibfnamefont {Andreas~M.}\ \bibnamefont
  {L\"{a}uchli}},\ }\bibfield  {title} {\enquote {\bibinfo {title} {Floating
  phases in one-dimensional rydberg ising chains},}\ }\href@noop {} {\
  (\bibinfo {year} {2019})},\ \Eprint {http://arxiv.org/abs/1908.02068}
  {arXiv:1908.02068} \BibitemShut {NoStop}%
\bibitem [{\citenamefont {Samajdar}\ \emph {et~al.}(2018)\citenamefont
  {Samajdar}, \citenamefont {Choi}, \citenamefont {Pichler}, \citenamefont
  {Lukin},\ and\ \citenamefont {Sachdev}}]{Samajdar2018}%
  \BibitemOpen
  \bibfield  {author} {\bibinfo {author} {\bibfnamefont {Rhine}\ \bibnamefont
  {Samajdar}}, \bibinfo {author} {\bibfnamefont {Soonwon}\ \bibnamefont
  {Choi}}, \bibinfo {author} {\bibfnamefont {Hannes}\ \bibnamefont {Pichler}},
  \bibinfo {author} {\bibfnamefont {Mikhail~D.}\ \bibnamefont {Lukin}}, \ and\
  \bibinfo {author} {\bibfnamefont {Subir}\ \bibnamefont {Sachdev}},\
  }\bibfield  {title} {\enquote {\bibinfo {title} {Numerical study of the
  chiral ${Z}_{3}$ quantum phase transition in one spatial dimension},}\ }\href
  {\doibase 10.1103/PhysRevA.98.023614} {\bibfield  {journal} {\bibinfo
  {journal} {Phys. Rev. A}\ }\textbf {\bibinfo {volume} {98}},\ \bibinfo
  {pages} {023614} (\bibinfo {year} {2018})}\BibitemShut {NoStop}%
\bibitem [{\citenamefont {Whitsitt}\ \emph {et~al.}(2018)\citenamefont
  {Whitsitt}, \citenamefont {Samajdar},\ and\ \citenamefont
  {Sachdev}}]{Whitsitt2018}%
  \BibitemOpen
  \bibfield  {author} {\bibinfo {author} {\bibfnamefont {Seth}\ \bibnamefont
  {Whitsitt}}, \bibinfo {author} {\bibfnamefont {Rhine}\ \bibnamefont
  {Samajdar}}, \ and\ \bibinfo {author} {\bibfnamefont {Subir}\ \bibnamefont
  {Sachdev}},\ }\bibfield  {title} {\enquote {\bibinfo {title} {Quantum field
  theory for the chiral clock transition in one spatial dimension},}\ }\href
  {\doibase 10.1103/PhysRevB.98.205118} {\bibfield  {journal} {\bibinfo
  {journal} {Phys. Rev. B}\ }\textbf {\bibinfo {volume} {98}},\ \bibinfo
  {pages} {205118} (\bibinfo {year} {2018})}\BibitemShut {NoStop}%
\bibitem [{\citenamefont {Kibble}(1976)}]{Kibble1976}%
  \BibitemOpen
  \bibfield  {author} {\bibinfo {author} {\bibfnamefont {T~W~B}\ \bibnamefont
  {Kibble}},\ }\bibfield  {title} {\enquote {\bibinfo {title} {Topology of
  cosmic domains and strings},}\ }\href {\doibase 10.1088/0305-4470/9/8/029}
  {\bibfield  {journal} {\bibinfo  {journal} {Journal of Physics A:
  Mathematical and General}\ }\textbf {\bibinfo {volume} {9}},\ \bibinfo
  {pages} {1387--1398} (\bibinfo {year} {1976})}\BibitemShut {NoStop}%
\bibitem [{\citenamefont {Zurek}(1985)}]{Zurek1985}%
  \BibitemOpen
  \bibfield  {author} {\bibinfo {author} {\bibfnamefont {W.~H.}\ \bibnamefont
  {Zurek}},\ }\bibfield  {title} {\enquote {\bibinfo {title} {Cosmological
  experiments in superfluid helium?}}\ }\href {\doibase 10.1038/317505a0}
  {\bibfield  {journal} {\bibinfo  {journal} {Nature}\ }\textbf {\bibinfo
  {volume} {317}},\ \bibinfo {pages} {505--508} (\bibinfo {year}
  {1985})}\BibitemShut {NoStop}%
\bibitem [{\citenamefont {{Keesling}}\ \emph {et~al.}(2019)\citenamefont
  {{Keesling}}, \citenamefont {{Omran}}, \citenamefont {{Levine}},
  \citenamefont {{Bernien}}, \citenamefont {{Pichler}}, \citenamefont {{Choi}},
  \citenamefont {{Samajdar}}, \citenamefont {{Schwartz}}, \citenamefont
  {{Silvi}}, \citenamefont {{Sachdev}}, \citenamefont {{Zoller}}, \citenamefont
  {{Endres}}, \citenamefont {{Greiner}}, \citenamefont {{Vuleti{\'c}}},
  \citenamefont {{}},\ and\ \citenamefont {{Lukin}}}]{RydbergKibbleZurek}%
  \BibitemOpen
  \bibfield  {author} {\bibinfo {author} {\bibfnamefont {Alexander}\
  \bibnamefont {{Keesling}}}, \bibinfo {author} {\bibfnamefont {Ahmed}\
  \bibnamefont {{Omran}}}, \bibinfo {author} {\bibfnamefont {Harry}\
  \bibnamefont {{Levine}}}, \bibinfo {author} {\bibfnamefont {Hannes}\
  \bibnamefont {{Bernien}}}, \bibinfo {author} {\bibfnamefont {Hannes}\
  \bibnamefont {{Pichler}}}, \bibinfo {author} {\bibfnamefont {Soonwon}\
  \bibnamefont {{Choi}}}, \bibinfo {author} {\bibfnamefont {Rhine}\
  \bibnamefont {{Samajdar}}}, \bibinfo {author} {\bibfnamefont {Sylvain}\
  \bibnamefont {{Schwartz}}}, \bibinfo {author} {\bibfnamefont {Pietro}\
  \bibnamefont {{Silvi}}}, \bibinfo {author} {\bibfnamefont {Subir}\
  \bibnamefont {{Sachdev}}}, \bibinfo {author} {\bibfnamefont {Peter}\
  \bibnamefont {{Zoller}}}, \bibinfo {author} {\bibfnamefont {Manuel}\
  \bibnamefont {{Endres}}}, \bibinfo {author} {\bibfnamefont {Markus}\
  \bibnamefont {{Greiner}}}, \bibinfo {author} {\bibnamefont {{Vuleti{\'c}}}},
  \bibinfo {author} {\bibfnamefont {Vladan}\ \bibnamefont {{}}}, \ and\
  \bibinfo {author} {\bibfnamefont {Mikhail~D.}\ \bibnamefont {{Lukin}}},\
  }\bibfield  {title} {\enquote {\bibinfo {title} {{Quantum Kibble-Zurek
  mechanism and critical dynamics on a programmable Rydberg simulator}},}\
  }\href {\doibase 10.1038/s41586-019-1070-1} {\bibfield  {journal} {\bibinfo
  {journal} {\nat}\ }\textbf {\bibinfo {volume} {568}},\ \bibinfo {pages}
  {207--211} (\bibinfo {year} {2019})},\ \Eprint
  {http://arxiv.org/abs/1809.05540} {arXiv:1809.05540} \BibitemShut {NoStop}%
\bibitem [{\citenamefont {Ebadi}\ \emph {et~al.}(2021)\citenamefont {Ebadi},
  \citenamefont {Wang}, \citenamefont {Levine}, \citenamefont {Keesling},
  \citenamefont {Semeghini}, \citenamefont {Omran}, \citenamefont {Bluvstein},
  \citenamefont {Samajdar}, \citenamefont {Pichler}, \citenamefont {Ho},
  \citenamefont {Choi}, \citenamefont {Sachdev}, \citenamefont {Greiner},
  \citenamefont {Vuletic},\ and\ \citenamefont {Lukin}}]{Ebadi2021}%
  \BibitemOpen
  \bibfield  {author} {\bibinfo {author} {\bibfnamefont {Sepehr}\ \bibnamefont
  {Ebadi}}, \bibinfo {author} {\bibfnamefont {Tout~T.}\ \bibnamefont {Wang}},
  \bibinfo {author} {\bibfnamefont {Harry}\ \bibnamefont {Levine}}, \bibinfo
  {author} {\bibfnamefont {Alexander}\ \bibnamefont {Keesling}}, \bibinfo
  {author} {\bibfnamefont {Giulia}\ \bibnamefont {Semeghini}}, \bibinfo
  {author} {\bibfnamefont {Ahmed}\ \bibnamefont {Omran}}, \bibinfo {author}
  {\bibfnamefont {Dolev}\ \bibnamefont {Bluvstein}}, \bibinfo {author}
  {\bibfnamefont {Rhine}\ \bibnamefont {Samajdar}}, \bibinfo {author}
  {\bibfnamefont {Hannes}\ \bibnamefont {Pichler}}, \bibinfo {author}
  {\bibfnamefont {Wen~Wei}\ \bibnamefont {Ho}}, \bibinfo {author}
  {\bibfnamefont {Soonwon}\ \bibnamefont {Choi}}, \bibinfo {author}
  {\bibfnamefont {Subir}\ \bibnamefont {Sachdev}}, \bibinfo {author}
  {\bibfnamefont {Markus}\ \bibnamefont {Greiner}}, \bibinfo {author}
  {\bibfnamefont {Vladan}\ \bibnamefont {Vuletic}}, \ and\ \bibinfo {author}
  {\bibfnamefont {Mikhail~D.}\ \bibnamefont {Lukin}},\ }\bibfield  {title}
  {\enquote {\bibinfo {title} {Quantum phases of matter on a 256-atom
  programmable quantum simulator},}\ }\href {\doibase
  10.1038/s41586-021-03582-4} {\bibfield  {journal} {\bibinfo  {journal}
  {Nature}\ }\textbf {\bibinfo {volume} {595}},\ \bibinfo {pages} {227--232}
  (\bibinfo {year} {2021})}\BibitemShut {NoStop}%
\bibitem [{\citenamefont {{Kane}}\ \emph {et~al.}(2002)\citenamefont {{Kane}},
  \citenamefont {{Mukhopadhyay}},\ and\ \citenamefont
  {{Lubensky}}}]{wiresKane}%
  \BibitemOpen
  \bibfield  {author} {\bibinfo {author} {\bibfnamefont {C.~L.}\ \bibnamefont
  {{Kane}}}, \bibinfo {author} {\bibfnamefont {Ranjan}\ \bibnamefont
  {{Mukhopadhyay}}}, \ and\ \bibinfo {author} {\bibfnamefont {T.~C.}\
  \bibnamefont {{Lubensky}}},\ }\bibfield  {title} {\enquote {\bibinfo {title}
  {{Fractional Quantum Hall Effect in an Array of Quantum Wires}},}\ }\href
  {\doibase 10.1103/PhysRevLett.88.036401} {\bibfield  {journal} {\bibinfo
  {journal} {\prl}\ }\textbf {\bibinfo {volume} {88}},\ \bibinfo {eid} {036401}
  (\bibinfo {year} {2002})},\ \Eprint {http://arxiv.org/abs/cond-mat/0108445}
  {arXiv:cond-mat/0108445} \BibitemShut {NoStop}%
\bibitem [{\citenamefont {{Teo}}\ and\ \citenamefont
  {{Kane}}(2011)}]{wiresTeo}%
  \BibitemOpen
  \bibfield  {author} {\bibinfo {author} {\bibfnamefont {Jeffrey C.~Y.}\
  \bibnamefont {{Teo}}}\ and\ \bibinfo {author} {\bibfnamefont {C.~L.}\
  \bibnamefont {{Kane}}},\ }\bibfield  {title} {\enquote {\bibinfo {title}
  {{From Luttinger liquid to non-Abelian quantum Hall states}},}\ }\href@noop
  {} {\  (\bibinfo {year} {2011})},\ \Eprint {http://arxiv.org/abs/1111.2617}
  {arXiv:1111.2617} \BibitemShut {NoStop}%
\bibitem [{\citenamefont {{Li}}\ \emph {et~al.}(2020)\citenamefont {{Li}},
  \citenamefont {{Ebisu}}, \citenamefont {{Sahoo}}, \citenamefont {{Oreg}},\
  and\ \citenamefont {{Franz}}}]{wireTCI}%
  \BibitemOpen
  \bibfield  {author} {\bibinfo {author} {\bibfnamefont {Chengshu}\
  \bibnamefont {{Li}}}, \bibinfo {author} {\bibfnamefont {Hiromi}\ \bibnamefont
  {{Ebisu}}}, \bibinfo {author} {\bibfnamefont {Sharmistha}\ \bibnamefont
  {{Sahoo}}}, \bibinfo {author} {\bibfnamefont {Yuval}\ \bibnamefont {{Oreg}}},
  \ and\ \bibinfo {author} {\bibfnamefont {Marcel}\ \bibnamefont {{Franz}}},\
  }\bibfield  {title} {\enquote {\bibinfo {title} {{Coupled wire construction
  of a topological phase with chiral tricritical Ising edge modes}},}\ }\href
  {\doibase 10.1103/PhysRevB.102.165123} {\bibfield  {journal} {\bibinfo
  {journal} {\prb}\ }\textbf {\bibinfo {volume} {102}},\ \bibinfo {eid}
  {165123} (\bibinfo {year} {2020})},\ \Eprint
  {http://arxiv.org/abs/2008.04438} {arXiv:2008.04438} \BibitemShut {NoStop}%
\bibitem [{\citenamefont {{Yao}}\ \emph {et~al.}(2021)\citenamefont {{Yao}},
  \citenamefont {{Pan}}, \citenamefont {{Liu}},\ and\ \citenamefont
  {{Zhai}}}]{YaoScarCriticality}%
  \BibitemOpen
  \bibfield  {author} {\bibinfo {author} {\bibfnamefont {Zhiyuan}\ \bibnamefont
  {{Yao}}}, \bibinfo {author} {\bibfnamefont {Lei}\ \bibnamefont {{Pan}}},
  \bibinfo {author} {\bibfnamefont {Shang}\ \bibnamefont {{Liu}}}, \ and\
  \bibinfo {author} {\bibfnamefont {Hui}\ \bibnamefont {{Zhai}}},\ }\bibfield
  {title} {\enquote {\bibinfo {title} {{Quantum Many-Body Scars and Quantum
  Criticality}},}\ }\href@noop {} {\  (\bibinfo {year} {2021})},\ \Eprint
  {http://arxiv.org/abs/2108.05113} {arXiv:2108.05113} \BibitemShut {NoStop}%
\bibitem [{\citenamefont {Turner}\ \emph {et~al.}(2018)\citenamefont {Turner},
  \citenamefont {Michailidis}, \citenamefont {Abanin}, \citenamefont {Serbyn},\
  and\ \citenamefont {Papi\'{c}}}]{TurnerScar}%
  \BibitemOpen
  \bibfield  {author} {\bibinfo {author} {\bibfnamefont {C.~J.}\ \bibnamefont
  {Turner}}, \bibinfo {author} {\bibfnamefont {A.~A.}\ \bibnamefont
  {Michailidis}}, \bibinfo {author} {\bibfnamefont {D.~A.}\ \bibnamefont
  {Abanin}}, \bibinfo {author} {\bibfnamefont {M.}~\bibnamefont {Serbyn}}, \
  and\ \bibinfo {author} {\bibfnamefont {Z.}~\bibnamefont {Papi\'{c}}},\
  }\bibfield  {title} {\enquote {\bibinfo {title} {Weak ergodicity breaking
  from quantum many-body scars},}\ }\href {\doibase 10.1038/s41567-018-0137-5}
  {\bibfield  {journal} {\bibinfo  {journal} {Nature Physics}\ }\textbf
  {\bibinfo {volume} {14}},\ \bibinfo {pages} {745--749} (\bibinfo {year}
  {2018})},\ \Eprint {http://arxiv.org/abs/1711.03528} {arXiv:1711.03528}
  \BibitemShut {NoStop}%
\bibitem [{\citenamefont {{Ovchinnikov}}\ \emph
  {et~al.}(2003{\natexlab{a}})\citenamefont {{Ovchinnikov}}, \citenamefont
  {{Dmitriev}}, \citenamefont {{Krivnov}},\ and\ \citenamefont
  {{Cheranovskii}}}]{OvchinnikovIsingXZ}%
  \BibitemOpen
  \bibfield  {author} {\bibinfo {author} {\bibfnamefont {A.~A.}\ \bibnamefont
  {{Ovchinnikov}}}, \bibinfo {author} {\bibfnamefont {D.~V.}\ \bibnamefont
  {{Dmitriev}}}, \bibinfo {author} {\bibfnamefont {V.~Ya.}\ \bibnamefont
  {{Krivnov}}}, \ and\ \bibinfo {author} {\bibfnamefont {V.~O.}\ \bibnamefont
  {{Cheranovskii}}},\ }\bibfield  {title} {\enquote {\bibinfo {title}
  {{Antiferromagnetic Ising chain in a mixed transverse and longitudinal
  magnetic field}},}\ }\href {\doibase 10.1103/PhysRevB.68.214406} {\bibfield
  {journal} {\bibinfo  {journal} {\prb}\ }\textbf {\bibinfo {volume} {68}},\
  \bibinfo {eid} {214406} (\bibinfo {year} {2003}{\natexlab{a}})},\ \Eprint
  {http://arxiv.org/abs/cond-mat/0306468} {arXiv:cond-mat/0306468} \BibitemShut
  {NoStop}%
\bibitem [{\citenamefont {Baxter}(1982)}]{Baxter1982}%
  \BibitemOpen
  \bibfield  {author} {\bibinfo {author} {\bibfnamefont {R.~J.}\ \bibnamefont
  {Baxter}},\ }\href@noop {} {\emph {\bibinfo {title} {{Exactly solved models
  in statistical mechanics}}}}\ (\bibinfo  {publisher} {Academic},\ \bibinfo
  {year} {1982})\BibitemShut {NoStop}%
\bibitem [{\citenamefont {Chepiga}\ and\ \citenamefont
  {Mila}(2019{\natexlab{a}})}]{Chepiga2019}%
  \BibitemOpen
  \bibfield  {author} {\bibinfo {author} {\bibfnamefont {Natalia}\ \bibnamefont
  {Chepiga}}\ and\ \bibinfo {author} {\bibfnamefont {Fr\'ed\'eric}\
  \bibnamefont {Mila}},\ }\bibfield  {title} {\enquote {\bibinfo {title}
  {Floating phase versus chiral transition in a 1d hard-boson model},}\ }\href
  {\doibase 10.1103/PhysRevLett.122.017205} {\bibfield  {journal} {\bibinfo
  {journal} {Phys. Rev. Lett.}\ }\textbf {\bibinfo {volume} {122}},\ \bibinfo
  {pages} {017205} (\bibinfo {year} {2019}{\natexlab{a}})}\BibitemShut
  {NoStop}%
\bibitem [{\citenamefont {Chepiga}\ and\ \citenamefont
  {Mila}(2019{\natexlab{b}})}]{Chepiga2019b}%
  \BibitemOpen
  \bibfield  {author} {\bibinfo {author} {\bibfnamefont {Natalia}\ \bibnamefont
  {Chepiga}}\ and\ \bibinfo {author} {\bibfnamefont {Fr\'ed\'eric}\
  \bibnamefont {Mila}},\ }\bibfield  {title} {\enquote {\bibinfo {title} {{DMRG
  investigation of constrained models: from quantum dimer and quantum loop
  ladders to hard-boson and Fibonacci anyon chains}},}\ }\href {\doibase
  10.21468/SciPostPhys.6.3.033} {\bibfield  {journal} {\bibinfo  {journal}
  {SciPost Phys.}\ }\textbf {\bibinfo {volume} {6}},\ \bibinfo {pages} {33}
  (\bibinfo {year} {2019}{\natexlab{b}})}\BibitemShut {NoStop}%
\bibitem [{\citenamefont {Giudici}\ \emph {et~al.}(2019)\citenamefont
  {Giudici}, \citenamefont {Angelone}, \citenamefont {Magnifico}, \citenamefont
  {Zeng}, \citenamefont {Giudice}, \citenamefont {Mendes-Santos},\ and\
  \citenamefont {Dalmonte}}]{Giudici2019}%
  \BibitemOpen
  \bibfield  {author} {\bibinfo {author} {\bibfnamefont {G.}~\bibnamefont
  {Giudici}}, \bibinfo {author} {\bibfnamefont {A.}~\bibnamefont {Angelone}},
  \bibinfo {author} {\bibfnamefont {G.}~\bibnamefont {Magnifico}}, \bibinfo
  {author} {\bibfnamefont {Z.}~\bibnamefont {Zeng}}, \bibinfo {author}
  {\bibfnamefont {G.}~\bibnamefont {Giudice}}, \bibinfo {author} {\bibfnamefont
  {T.}~\bibnamefont {Mendes-Santos}}, \ and\ \bibinfo {author} {\bibfnamefont
  {M.}~\bibnamefont {Dalmonte}},\ }\bibfield  {title} {\enquote {\bibinfo
  {title} {Diagnosing potts criticality and two-stage melting in
  one-dimensional hard-core boson models},}\ }\href {\doibase
  10.1103/PhysRevB.99.094434} {\bibfield  {journal} {\bibinfo  {journal} {Phys.
  Rev. B}\ }\textbf {\bibinfo {volume} {99}},\ \bibinfo {pages} {094434}
  (\bibinfo {year} {2019})}\BibitemShut {NoStop}%
\bibitem [{\citenamefont {{Ovchinnikov}}\ \emph
  {et~al.}(2003{\natexlab{b}})\citenamefont {{Ovchinnikov}}, \citenamefont
  {{Dmitriev}}, \citenamefont {{Krivnov}},\ and\ \citenamefont
  {{Cheranovskii}}}]{IsingLongitudinal}%
  \BibitemOpen
  \bibfield  {author} {\bibinfo {author} {\bibfnamefont {A.~A.}\ \bibnamefont
  {{Ovchinnikov}}}, \bibinfo {author} {\bibfnamefont {D.~V.}\ \bibnamefont
  {{Dmitriev}}}, \bibinfo {author} {\bibfnamefont {V.~Ya.}\ \bibnamefont
  {{Krivnov}}}, \ and\ \bibinfo {author} {\bibfnamefont {V.~O.}\ \bibnamefont
  {{Cheranovskii}}},\ }\bibfield  {title} {\enquote {\bibinfo {title}
  {{Antiferromagnetic Ising chain in a mixed transverse and longitudinal
  magnetic field}},}\ }\href {\doibase 10.1103/PhysRevB.68.214406} {\bibfield
  {journal} {\bibinfo  {journal} {\prb}\ }\textbf {\bibinfo {volume} {68}},\
  \bibinfo {eid} {214406} (\bibinfo {year} {2003}{\natexlab{b}})},\ \Eprint
  {http://arxiv.org/abs/cond-mat/0306468} {arXiv:cond-mat/0306468} \BibitemShut
  {NoStop}%
\bibitem [{\citenamefont {Baxter}(2008)}]{Baxbook}%
  \BibitemOpen
  \bibfield  {author} {\bibinfo {author} {\bibfnamefont {R.J.}\ \bibnamefont
  {Baxter}},\ }\href@noop {} {\emph {\bibinfo {title} {Exactly Solved Models in
  Statistical Mechanics}}}\ (\bibinfo  {publisher} {Dover},\ \bibinfo {year}
  {2008})\BibitemShut {NoStop}%
\bibitem [{\citenamefont {{Feiguin}}\ \emph {et~al.}(2007)\citenamefont
  {{Feiguin}}, \citenamefont {{Trebst}}, \citenamefont {{Ludwig}},
  \citenamefont {{Troyer}}, \citenamefont {{Kitaev}}, \citenamefont {{Wang}},\
  and\ \citenamefont {{Freedman}}}]{Feiguin2007}%
  \BibitemOpen
  \bibfield  {author} {\bibinfo {author} {\bibfnamefont {Adrian}\ \bibnamefont
  {{Feiguin}}}, \bibinfo {author} {\bibfnamefont {Simon}\ \bibnamefont
  {{Trebst}}}, \bibinfo {author} {\bibfnamefont {Andreas W.~W.}\ \bibnamefont
  {{Ludwig}}}, \bibinfo {author} {\bibfnamefont {Matthias}\ \bibnamefont
  {{Troyer}}}, \bibinfo {author} {\bibfnamefont {Alexei}\ \bibnamefont
  {{Kitaev}}}, \bibinfo {author} {\bibfnamefont {Zhenghan}\ \bibnamefont
  {{Wang}}}, \ and\ \bibinfo {author} {\bibfnamefont {Michael~H.}\ \bibnamefont
  {{Freedman}}},\ }\bibfield  {title} {\enquote {\bibinfo {title} {{Interacting
  Anyons in Topological Quantum Liquids: The Golden Chain}},}\ }\href {\doibase
  10.1103/PhysRevLett.98.160409} {\bibfield  {journal} {\bibinfo  {journal}
  {\prl}\ }\textbf {\bibinfo {volume} {98}},\ \bibinfo {eid} {160409} (\bibinfo
  {year} {2007})},\ \Eprint {http://arxiv.org/abs/cond-mat/0612341}
  {arXiv:cond-mat/0612341} \BibitemShut {NoStop}%
\bibitem [{\citenamefont {Cardy}(1984)}]{Cardy_1984}%
  \BibitemOpen
  \bibfield  {author} {\bibinfo {author} {\bibfnamefont {J~L}\ \bibnamefont
  {Cardy}},\ }\bibfield  {title} {\enquote {\bibinfo {title} {Conformal
  invariance and universality in finite-size scaling},}\ }\href {\doibase
  10.1088/0305-4470/17/7/003} {\bibfield  {journal} {\bibinfo  {journal}
  {Journal of Physics A: Mathematical and General}\ }\textbf {\bibinfo {volume}
  {17}},\ \bibinfo {pages} {L385--L387} (\bibinfo {year} {1984})}\BibitemShut
  {NoStop}%
\bibitem [{\citenamefont {Cardy}(1986)}]{CARDY1986186}%
  \BibitemOpen
  \bibfield  {author} {\bibinfo {author} {\bibfnamefont {John~L.}\ \bibnamefont
  {Cardy}},\ }\bibfield  {title} {\enquote {\bibinfo {title} {Operator content
  of two-dimensional conformally invariant theories},}\ }\href {\doibase
  10.1016/0550-3213(86)90552-3} {\bibfield  {journal} {\bibinfo  {journal}
  {Nuclear Physics B}\ }\textbf {\bibinfo {volume} {270}},\ \bibinfo {pages}
  {186--204} (\bibinfo {year} {1986})}\BibitemShut {NoStop}%
\bibitem [{Note1()}]{Note1}%
  \BibitemOpen
  \bibinfo {note} {When $u<0$, $T \protect \overline {T}$ is dangerously
  irrelevant in the sense that the long-distance IR physics is sensitive to the
  UV cutoff.}\BibitemShut {Stop}%
\bibitem [{Afo()}]{Afoot:deltaj}%
  \BibitemOpen
  \href@noop {} {}\bibinfo {note} {The use of $\delta j = \frac{L}{\pi} \sin
  \frac{\pi \Delta j}{L}$ in \figref{fig:sigma nnScaling} is a standard choice
  that reduces finite-size effects. For instance, this choice leads to an exact
  power-law correlation function for free fermions. Since $\delta j \approx
  \Delta j$ when $|\Delta j| \ll L$, the main advantage is improving
  extrapolation to $|\Delta j| \sim L$. In \figref{fig:gammaScaling}, we
  generalize this choice to $\delta_\epsilon j = \frac{L}{\pi} \sin
  \frac{\pi}{L} (\Delta j + \epsilon)$. For \figref{fig:muScaling} and
  \ref{fig:gammaDecay}, we choose an $\epsilon$ with a symmetry such that there
  is overlap between the $\Delta j \leq L/2$ and $\Delta j > L/2$ points. For
  \figref{fig:gammaPowerLaw}, we instead choose the $\epsilon$ that results in
  the best-looking data.}\BibitemShut {Stop}%
\bibitem [{Note2()}]{Note2}%
  \BibitemOpen
  \bibinfo {note} {The Hermitian operator $b_j + b_j^\dagger $ has identical
  symmetry properties to $n_j$, and thus exhibits a low-energy expansion of the
  same form (of course with different coefficients). The Hermitian operator
  $i(b_j-b_j^\dagger )$ is odd under time reversal but even under $R_x$. In the
  Heisenberg picture, we therefore obtain $i[b_j(t)-b_j^\dagger (t)] \sim
  (-1)^j \partial _t \sigma + \protect \cdots $, implying $i(b_j-b_j^\dagger
  )\sim i(-1)^j[\sigma ,H]$ for the Schrodinger picture that we typically
  employ in this paper. We focus on the number operator rather than creation
  and annihilation operators due to ease of measurement.}\BibitemShut {Stop}%
\bibitem [{\citenamefont {{O'Brien}}\ and\ \citenamefont
  {{Fendley}}(2018)}]{Fendley2018}%
  \BibitemOpen
  \bibfield  {author} {\bibinfo {author} {\bibfnamefont {Edward}\ \bibnamefont
  {{O'Brien}}}\ and\ \bibinfo {author} {\bibfnamefont {Paul}\ \bibnamefont
  {{Fendley}}},\ }\bibfield  {title} {\enquote {\bibinfo {title} {{Lattice
  Supersymmetry and Order-Disorder Coexistence in the Tricritical Ising
  Model}},}\ }\href {\doibase 10.1103/PhysRevLett.120.206403} {\bibfield
  {journal} {\bibinfo  {journal} {\prl}\ }\textbf {\bibinfo {volume} {120}},\
  \bibinfo {eid} {206403} (\bibinfo {year} {2018})},\ \Eprint
  {http://arxiv.org/abs/1712.06662} {arXiv:1712.06662} \BibitemShut {NoStop}%
\bibitem [{\citenamefont {Aasen}\ \emph {et~al.}(2020)\citenamefont {Aasen},
  \citenamefont {Mong}, \citenamefont {Hunt}, \citenamefont {Mandrus},\ and\
  \citenamefont {Alicea}}]{Aasen2020}%
  \BibitemOpen
  \bibfield  {author} {\bibinfo {author} {\bibfnamefont {David}\ \bibnamefont
  {Aasen}}, \bibinfo {author} {\bibfnamefont {Roger S.~K.}\ \bibnamefont
  {Mong}}, \bibinfo {author} {\bibfnamefont {Benjamin~M.}\ \bibnamefont
  {Hunt}}, \bibinfo {author} {\bibfnamefont {David}\ \bibnamefont {Mandrus}}, \
  and\ \bibinfo {author} {\bibfnamefont {Jason}\ \bibnamefont {Alicea}},\
  }\bibfield  {title} {\enquote {\bibinfo {title} {Electrical probes of the
  non-abelian spin liquid in kitaev materials},}\ }\href {\doibase
  10.1103/PhysRevX.10.031014} {\bibfield  {journal} {\bibinfo  {journal} {Phys.
  Rev. X}\ }\textbf {\bibinfo {volume} {10}},\ \bibinfo {pages} {031014}
  (\bibinfo {year} {2020})},\ \Eprint {http://arxiv.org/abs/2002.01944}
  {arXiv:2002.01944} \BibitemShut {NoStop}%
\bibitem [{\citenamefont {Rahmani}\ and\ \citenamefont
  {Franz}(2019)}]{Rahmani2019}%
  \BibitemOpen
  \bibfield  {author} {\bibinfo {author} {\bibfnamefont {Armin}\ \bibnamefont
  {Rahmani}}\ and\ \bibinfo {author} {\bibfnamefont {Marcel}\ \bibnamefont
  {Franz}},\ }\bibfield  {title} {\enquote {\bibinfo {title} {Interacting
  {M}ajorana fermions},}\ }\href {\doibase 10.1088/1361-6633/ab28ef} {\bibfield
   {journal} {\bibinfo  {journal} {Reports on Progress in Physics}\ }\textbf
  {\bibinfo {volume} {82}},\ \bibinfo {pages} {084501} (\bibinfo {year}
  {2019})}\BibitemShut {NoStop}%
\bibitem [{Note3()}]{Note3}%
  \BibitemOpen
  \bibinfo {note} {Technically, $c^{\protect \rm eff}_\varepsilon =
  c_{\varepsilon } \kappa $ varies with $V_2$ due to a combination of changes
  in $\kappa $ \protect \emph {and} $c_\varepsilon $. Variation in
  $c_\varepsilon $ can have a trivial origin unrelated to interactions, e.g.,
  the lattice operator $\protect \hat \varepsilon ^\protect \text
  {bare}_{j+1/2}$ can have a smaller overlap with the CFT field $\varepsilon $
  as $V_2$ increase simply due to curvature in the phase boundary of Fig.~\ref
  {fig:phases}. We expect, however, that the latter effect is $O(1)$, in
  contrast to the dramatic change in $c^{\protect \rm eff}_\varepsilon $
  (again, by more than a factor of 30!) evident in Fig.~\ref
  {cVsV}.}\BibitemShut {Stop}%
\bibitem [{\citenamefont {{Sandvik}}(2010)}]{SandvikComputational}%
  \BibitemOpen
  \bibfield  {author} {\bibinfo {author} {\bibfnamefont {Anders~W.}\
  \bibnamefont {{Sandvik}}},\ }\bibfield  {title} {\enquote {\bibinfo {title}
  {{Computational Studies of Quantum Spin Systems}},}\ }in\ \href {\doibase
  10.1063/1.3518900} {\emph {\bibinfo {booktitle} {Lectures on the Physics of
  Strongly Correlated Systems Xiv: Fourteenth Training Course in the Physics of
  Strongly Correlated Systems}}},\ \bibinfo {series} {American Institute of
  Physics Conference Series}, Vol.\ \bibinfo {volume} {1297},\ \bibinfo
  {editor} {edited by\ \bibinfo {editor} {\bibfnamefont {Adolfo}\ \bibnamefont
  {{Avella}}}\ and\ \bibinfo {editor} {\bibfnamefont {Ferdinando}\ \bibnamefont
  {{Mancini}}}}\ (\bibinfo {year} {2010})\ pp.\ \bibinfo {pages} {135--338},\
  \Eprint {http://arxiv.org/abs/1101.3281} {arXiv:1101.3281} \BibitemShut
  {NoStop}%
\bibitem [{\citenamefont {Friedan}\ \emph {et~al.}(1985)\citenamefont
  {Friedan}, \citenamefont {Qiu},\ and\ \citenamefont {Shenker}}]{Friedan1984}%
  \BibitemOpen
  \bibfield  {author} {\bibinfo {author} {\bibfnamefont {Daniel}\ \bibnamefont
  {Friedan}}, \bibinfo {author} {\bibfnamefont {Zong-an}\ \bibnamefont {Qiu}},
  \ and\ \bibinfo {author} {\bibfnamefont {Stephen~H.}\ \bibnamefont
  {Shenker}},\ }\bibfield  {title} {\enquote {\bibinfo {title} {{Superconformal
  Invariance in Two-Dimensions and the Tricritical Ising Model}},}\ }\href
  {\doibase 10.1016/0370-2693(85)90819-6} {\bibfield  {journal} {\bibinfo
  {journal} {Phys. Lett. B}\ }\textbf {\bibinfo {volume} {151}},\ \bibinfo
  {pages} {37--43} (\bibinfo {year} {1985})}\BibitemShut {NoStop}%
\bibitem [{\citenamefont {Lassig}\ \emph {et~al.}(1991)\citenamefont {Lassig},
  \citenamefont {Mussardo},\ and\ \citenamefont {Cardy}}]{Lassig1990}%
  \BibitemOpen
  \bibfield  {author} {\bibinfo {author} {\bibfnamefont {Michael}\ \bibnamefont
  {Lassig}}, \bibinfo {author} {\bibfnamefont {Giuseppe}\ \bibnamefont
  {Mussardo}}, \ and\ \bibinfo {author} {\bibfnamefont {John~L.}\ \bibnamefont
  {Cardy}},\ }\bibfield  {title} {\enquote {\bibinfo {title} {{The scaling
  region of the tricritical Ising model in two-dimensions}},}\ }\href {\doibase
  10.1016/0550-3213(91)90206-D} {\bibfield  {journal} {\bibinfo  {journal}
  {Nucl. Phys. B}\ }\textbf {\bibinfo {volume} {348}},\ \bibinfo {pages}
  {591--618} (\bibinfo {year} {1991})}\BibitemShut {NoStop}%
\bibitem [{\citenamefont {Bl{\"o}te}\ \emph {et~al.}(1986)\citenamefont
  {Bl{\"o}te}, \citenamefont {Cardy},\ and\ \citenamefont
  {Nightingale}}]{Bloete1986}%
  \BibitemOpen
  \bibfield  {author} {\bibinfo {author} {\bibfnamefont {H.~W.~J.}\
  \bibnamefont {Bl{\"o}te}}, \bibinfo {author} {\bibfnamefont {John~L.}\
  \bibnamefont {Cardy}}, \ and\ \bibinfo {author} {\bibfnamefont {M.~P.}\
  \bibnamefont {Nightingale}},\ }\bibfield  {title} {\enquote {\bibinfo {title}
  {{Conformal Invariance, the Central Charge, and Universal Finite Size
  Amplitudes at Criticality}},}\ }\href {\doibase 10.1103/PhysRevLett.56.742}
  {\bibfield  {journal} {\bibinfo  {journal} {Phys. Rev. Lett.}\ }\textbf
  {\bibinfo {volume} {56}},\ \bibinfo {pages} {742--745} (\bibinfo {year}
  {1986})}\BibitemShut {NoStop}%
\bibitem [{\citenamefont {Affleck}(1986)}]{Affleck1986}%
  \BibitemOpen
  \bibfield  {author} {\bibinfo {author} {\bibfnamefont {Ian}\ \bibnamefont
  {Affleck}},\ }\bibfield  {title} {\enquote {\bibinfo {title} {{Universal Term
  in the Free Energy at a Critical Point and the Conformal Anomaly}},}\ }\href
  {\doibase 10.1103/PhysRevLett.56.746} {\bibfield  {journal} {\bibinfo
  {journal} {Phys. Rev. Lett.}\ }\textbf {\bibinfo {volume} {56}},\ \bibinfo
  {pages} {746--748} (\bibinfo {year} {1986})}\BibitemShut {NoStop}%
\bibitem [{Note4()}]{Note4}%
  \BibitemOpen
  \bibinfo {note} {We show results for the bond-centered CDW order parameter
  rather than Eq.~\protect \textup {\hbox {\mathsurround \z@ \protect
  \normalfont (\ignorespaces \ref {sigma_lattice}\unskip \@@italiccorr )}}
  since the latter exhibits a pronounced even-odd effect that muddies somewhat
  the power-law correlations arising from the $c = 7/10$ $\sigma $
  field.}\BibitemShut {Stop}%
\bibitem [{\citenamefont {{Kim}}\ \emph {et~al.}(2016)\citenamefont {{Kim}},
  \citenamefont {{Lee}}, \citenamefont {{Lee}}, \citenamefont {{Jo}},
  \citenamefont {{Song}},\ and\ \citenamefont {{Ahn}}}]{arrayJaewook}%
  \BibitemOpen
  \bibfield  {author} {\bibinfo {author} {\bibfnamefont {Hyosub}\ \bibnamefont
  {{Kim}}}, \bibinfo {author} {\bibfnamefont {Woojun}\ \bibnamefont {{Lee}}},
  \bibinfo {author} {\bibfnamefont {Han-Gyeol}\ \bibnamefont {{Lee}}}, \bibinfo
  {author} {\bibfnamefont {Hanlae}\ \bibnamefont {{Jo}}}, \bibinfo {author}
  {\bibfnamefont {Yunheung}\ \bibnamefont {{Song}}}, \ and\ \bibinfo {author}
  {\bibfnamefont {Jaewook}\ \bibnamefont {{Ahn}}},\ }\bibfield  {title}
  {\enquote {\bibinfo {title} {{In situ single-atom array synthesis using
  dynamic holographic optical tweezers}},}\ }\href {\doibase
  10.1038/ncomms13317} {\bibfield  {journal} {\bibinfo  {journal} {Nature
  Communications}\ }\textbf {\bibinfo {volume} {7}},\ \bibinfo {eid} {13317}
  (\bibinfo {year} {2016})},\ \Eprint {http://arxiv.org/abs/1601.03833}
  {arXiv:1601.03833} \BibitemShut {NoStop}%
\bibitem [{\citenamefont {{Barredo}}\ \emph {et~al.}(2016)\citenamefont
  {{Barredo}}, \citenamefont {{de L{\'e}s{\'e}leuc}}, \citenamefont
  {{Lienhard}}, \citenamefont {{Lahaye}},\ and\ \citenamefont
  {{Browaeys}}}]{arrayBrowaeys}%
  \BibitemOpen
  \bibfield  {author} {\bibinfo {author} {\bibfnamefont {Daniel}\ \bibnamefont
  {{Barredo}}}, \bibinfo {author} {\bibfnamefont {Sylvain}\ \bibnamefont {{de
  L{\'e}s{\'e}leuc}}}, \bibinfo {author} {\bibfnamefont {Vincent}\ \bibnamefont
  {{Lienhard}}}, \bibinfo {author} {\bibfnamefont {Thierry}\ \bibnamefont
  {{Lahaye}}}, \ and\ \bibinfo {author} {\bibfnamefont {Antoine}\ \bibnamefont
  {{Browaeys}}},\ }\bibfield  {title} {\enquote {\bibinfo {title} {{An
  atom-by-atom assembler of defect-free arbitrary two-dimensional atomic
  arrays}},}\ }\href {\doibase 10.1126/science.aah3778} {\bibfield  {journal}
  {\bibinfo  {journal} {Science}\ }\textbf {\bibinfo {volume} {354}},\ \bibinfo
  {pages} {1021--1023} (\bibinfo {year} {2016})},\ \Eprint
  {http://arxiv.org/abs/1607.03042} {arXiv:1607.03042} \BibitemShut {NoStop}%
\bibitem [{\citenamefont {{Huang}}\ \emph {et~al.}(2020)\citenamefont
  {{Huang}}, \citenamefont {{Kueng}},\ and\ \citenamefont
  {{Preskill}}}]{classicalShadow}%
  \BibitemOpen
  \bibfield  {author} {\bibinfo {author} {\bibfnamefont {Hsin-Yuan}\
  \bibnamefont {{Huang}}}, \bibinfo {author} {\bibfnamefont {Richard}\
  \bibnamefont {{Kueng}}}, \ and\ \bibinfo {author} {\bibfnamefont {John}\
  \bibnamefont {{Preskill}}},\ }\bibfield  {title} {\enquote {\bibinfo {title}
  {{Predicting many properties of a quantum system from very few
  measurements}},}\ }\href {\doibase 10.1038/s41567-020-0932-7} {\bibfield
  {journal} {\bibinfo  {journal} {Nature Physics}\ }\textbf {\bibinfo {volume}
  {16}},\ \bibinfo {pages} {1050--1057} (\bibinfo {year} {2020})},\ \Eprint
  {http://arxiv.org/abs/2002.08953} {arXiv:2002.08953} \BibitemShut {NoStop}%
\bibitem [{\citenamefont {{Cotler}}\ \emph {et~al.}(2021)\citenamefont
  {{Cotler}}, \citenamefont {{Mark}}, \citenamefont {{Huang}}, \citenamefont
  {{Hernandez}}, \citenamefont {{Choi}}, \citenamefont {{Shaw}}, \citenamefont
  {{Endres}},\ and\ \citenamefont {{Choi}}}]{rydbergDesigns}%
  \BibitemOpen
  \bibfield  {author} {\bibinfo {author} {\bibfnamefont {Jordan~S.}\
  \bibnamefont {{Cotler}}}, \bibinfo {author} {\bibfnamefont {Daniel~K.}\
  \bibnamefont {{Mark}}}, \bibinfo {author} {\bibfnamefont {Hsin-Yuan}\
  \bibnamefont {{Huang}}}, \bibinfo {author} {\bibfnamefont {Felipe}\
  \bibnamefont {{Hernandez}}}, \bibinfo {author} {\bibfnamefont {Joonhee}\
  \bibnamefont {{Choi}}}, \bibinfo {author} {\bibfnamefont {Adam~L.}\
  \bibnamefont {{Shaw}}}, \bibinfo {author} {\bibfnamefont {Manuel}\
  \bibnamefont {{Endres}}}, \ and\ \bibinfo {author} {\bibfnamefont {Soonwon}\
  \bibnamefont {{Choi}}},\ }\bibfield  {title} {\enquote {\bibinfo {title}
  {{Emergent quantum state designs from individual many-body wavefunctions}},}\
  }\href@noop {} {\  (\bibinfo {year} {2021})},\ \Eprint
  {http://arxiv.org/abs/2103.03536} {arXiv:2103.03536} \BibitemShut {NoStop}%
\bibitem [{\citenamefont {Burkhardt}\ and\ \citenamefont
  {Guim}(1993)}]{Burkhardt1993}%
  \BibitemOpen
  \bibfield  {author} {\bibinfo {author} {\bibfnamefont {Theodore~W.}\
  \bibnamefont {Burkhardt}}\ and\ \bibinfo {author} {\bibfnamefont {Ihnsouk}\
  \bibnamefont {Guim}},\ }\bibfield  {title} {\enquote {\bibinfo {title}
  {Conformal theory of the two-dimensional ising model with homogeneous
  boundary conditions and with disordred boundary fields},}\ }\href {\doibase
  10.1103/PhysRevB.47.14306} {\bibfield  {journal} {\bibinfo  {journal} {Phys.
  Rev. B}\ }\textbf {\bibinfo {volume} {47}},\ \bibinfo {pages} {14306--14311}
  (\bibinfo {year} {1993})}\BibitemShut {NoStop}%
\end{thebibliography}%

\end{document}